\documentclass[aps,twocolumn,preprintnumbers,floats,nofootinbib]{revtex4}\def\@cite#1#2{\textsuperscript{[{#1\if@tempswa , #2\fi}]}}

\usepackage{graphicx}
\usepackage{amsmath}
\usepackage{bm}
\usepackage{amsfonts}
\usepackage{amssymb}
\usepackage{color}
\usepackage{subfigure}
\usepackage{epsfig}
\usepackage{morefloats}
\usepackage{multirow}
\usepackage{graphicx,booktabs}
\usepackage{mathrsfs}
\usepackage{txfonts}
\usepackage{footnote}
\usepackage{pifont}
\usepackage{indentfirst}
\usepackage{graphicx,booktabs}
\usepackage{longtable,lscape}
\usepackage[figuresright]{rotating}

\usepackage[normalem]{ulem}

\usepackage[colorlinks, citecolor=blue,anchorcolor=red,menucolor=red, linkcolor=red,filecolor=red,urlcolor=blue,frenchlinks=red]{hyperref}

\newcommand{\vxi}{\mbox{\boldmath$\xi$\unboldmath}}

\begin{document}

\title{Singly heavy tetraquarks}
	
	\author{Jun-Jie Liu$^{1}$ , Zhi-Biao Liang$^{1}$ , Feng-Xiao Liu$^{2}$ ~\footnote {E-mail: lfx@usc.edu.cn}, Mu-Yang Chen$^{1,5}$~\footnote {E-mail: muyang@hunnu.edu.cn},
		Xian-Hui Zhong$^{1,5}$ ~\footnote {E-mail: zhongxh@hunnu.edu.cn}, Qiang Zhao$^{3,4}$ ~\footnote {E-mail: zhaoq@ihep.ac.cn}}
	\affiliation{ 1) Department of Physics, Hunan Normal University, and Key Laboratory of Low-Dimensional Quantum Structures and Quantum Control of Ministry of Education, Changsha 410081, China }
\affiliation{2) School of Nuclear Science and Technology, University of South China, 421001 Hengyang, China}
\affiliation{ 3) Institute of High Energy Physics, Chinese Academy of Sciences, Beijing 100049, China}
\affiliation{ 4) University of Chinese Academy of Sciences, Beijing 100049, China}
	\affiliation{5) Synergetic Innovation Center for Quantum Effects and Applications (SICQEA),
		Hunan Normal University, Changsha 410081, China}
	%
	
\begin{abstract}

In this work, we carry out a systematic study of the spectra of the $1S$-wave states for
the whole singly-heavy tetraquark systems within a semi-relativistic hybrid quark potential model,
in which both the one-gluon exchange (OGE) and one-boson exchange (OBE) interactions are included. Furthermore,
the fall-apart decays are evaluated with the quark exchange model by combining the obtained spectra.
It is found that besides the OGE potentials, the OBE potentials play crucial roles for describing the spectrum.
All of our obtained states lie far above the lowest dissociation meson-meson threshold.
They are compact states with relatively narrow fall-apart widths $\sim 1-120$~MeV.
The $D_{s0}(2317)$, $D_{s1}(2460)$, $T_{b\bar{s}}(5568)$, and $T_{c\bar{s}}(2327)$ resonances reported
from experiments cannot be explained as compact
tetraquarks. While the $T_{\bar{c}\bar{s}0}(2870)$ and $T_{c\bar{s}0}(2900)$ favor the tetraquark states with
$IJ^P=00^+$ and $10^+$, i.e. $T_{(\bar{c}\bar{s}[ud])0^+}^0(2919)$ and $T_{(cn\{\bar{s}\bar{n}\})0^+}^{1}(2922)$, respectively.
More singly-heavy tetraquark states have good potentials to be observed in some of their
dominant decay channels in experiments.

\end{abstract}
	
	
	\maketitle

\section{Introduction}

In the quark model, except for the conventional meson ($q\bar{q}$) and baryon ($qqq$) states, there also
exist ``exotic" hadrons composed of multiquarks, such as tetraquarks ($qq\bar{q}\bar{q}$),
and pentaquarks ($qqqq\bar{q}$). Since the discovery of $X(3872)$ by the Belle Collaboration
in 2003 \cite{Belle:2003nnu}, many candidates of exotic hadrons have been observed in experiments~\cite{ParticleDataGroup:2024cfk}.
These exotic hadrons serve as a valuable probe of QCD in the non-perturbative regime,
offering insights into quark confinement mechanisms and the emergence of complex hadronic
structures beyond the conventional quark model. The status of
experimental and theoretical studies can be found in recent reviews~\cite{Chen:2016qju,Chen:2016spr,Esposito:2016noz,Guo:2017jvc,Olsen:2017bmm,Lebed:2016hpi,Ali:2017jda,
Liu:2019zoy,Brambilla:2019esw,Chen:2022asf,Dai:2026fkg}

The singly-heavy tetraquarks have received considerable attention since the discovery of
$D_{s0}(2317)$ and $D_{s1}(2460)$ in 2003~\cite{Belle:2003guh,CLEO:2003ggt,BaBar:2003oey}.
Due to the anomalously small masses beyond the conventional quark model
expectation, they may be tetraquark states composed of $cn\bar{s}\bar{n}$ ($n=u,d$) as
discussed in the literature at that time~\cite{Jovanovic:2007bz,Cheng:2003kg,Chen:2004dy,Maiani:2004vq,Terasaki:2003qa,Hayashigaki:2004st,Terasaki:2004yx,
Dmitrasinovic:2004cu,Dmitrasinovic:2005gc,Bracco:2005kt,Nielsen:2005zr,Kim:2005gt,Wang:2006uba,Terasaki:2006qd}.
In addition, stimulated by the exotic nature of $D_{s0}(2317)$ and $D_{s1}(2460)$,
some other studies of mass spectra for the singly-heavy tetraquarks were carried out within various quark models before
2016 by several groups~\cite{Ebert:2010af,Gerasyuta:2008ps,Zhang:2006hv}. However, these studies do not
support $D_{s0}(2317)$ and $D_{s1}(2460)$ as tetraquark assignments.

In 2016, the D0 Collaboration reported a narrow structure $X(5568)$ (also known as $T_{b\bar{s}}(5568)$~\cite{ParticleDataGroup:2024cfk}) in the $B_s^0\pi^{\pm}$
final state~\cite{D0:2016mwd}. This structure is most likely contributed by genuine tetraquarks composed of
$bu\bar{s}\bar{d}$ and $bd\bar{s}\bar{u}$, which reignited interests in the study of singly-heavy tetraquarks.
Although most of the studies focus on the $bn\bar{s}\bar{n}$ system related to $X(5568)$~\cite{Wang:2016tsi,Chen:2016mqt,Agaev:2016mjb,Wang:2016mee,Zanetti:2016wjn,
Agaev:2016ijz,Liu:2016ogz,Wang:2016wkj,Stancu:2016sfd,Tang:2016pcf,Lang:2016jpk,Goerke:2016hxf,Agaev:2016ifn,
Lucha:2018dzq,Zhang:2017xwc,Mutuk:2019uez,Yang:2016sws,Dias:2016dme},
some other systems with open charm and/or bottom were also studied from
the aspects of mass spectra~\cite{Cheng:2020nho,Agaev:2016lkl,Sungu:2019ybf,Agamaliev:2016wtt,Agaev:2017oay,
Chen:2018hts,Huang:2019otd,Chen:2017rhl}, decays~\cite{Agaev:2016lkl,Cheng:2020nho,Agaev:2019wkk,Xing:2019hjg}, and production~\cite{Ali:2016gdg,He:2016xvd,Yu:2017pmn}.
Unfortunately, the $X(5568)$ structure has not been confirmed by the subsequent
experiments carried out by the LHCb, CMS, CDF, and ATLAS Collaborations~\cite{LHCb:2016dxl,CMS:2017hfy,ATLAS:2018udc,CDF:2017dwr},
while its existence is also questioned in theory~\cite{Guo:2016nhb,Lang:2016jpk}.

The experimental explorations of the singly-heavy tetraquarks have not stopped.
In 2020, the LHCb Collaboration observed two resonances in the $D^-K^+$ final state
with determined spin-parity numbers $J^P=0^+$ and $1^-$ from the decay $B^+ \to D^+ D^- K^+$~\cite{LHCb:2020bls,LHCb:2020pxc},
which are denoted by $T_{\bar{c}\bar{s}0}(2870)^0$ and $T_{\bar{c}\bar{s}1}(2900)^0$
by the Particle Data Group (PDG)~\cite{ParticleDataGroup:2024cfk}. Their existence was confirmed in the same final state
from another decay $B^+ \to D^{*+} D^- K^+$~\cite{LHCb:2024vfz}. These two resonances should be exotic states with
minimal quark content $\bar{c}\bar{s}ud$. In 2022, the LHCb Collaboration observed another two
charmed-strange tetraquark candidates, $T_{c\bar{s}0}^a(2900)^{++}$ and $T_{c\bar{s}0}^a(2900)^{0}$,
in the $D_s^+ \pi^+$ and $D_s^+ \pi^-$ invariant mass spectra from the decays $B^+ \to D^- D_s^+ \pi^+$ and $B^0 \to \bar{D}^0 D_s^+ \pi^-$,
respectively \cite{LHCb:2022sfr,LHCb:2022lzp}. Their isospin and spin-parity quantum numbers have been determined to be $IJ^P = 10^+$, identifying them as two members of an isospin triplet with minimal quark content of $c u \bar{s} \bar{d}$ and $c d \bar{s} \bar{u}$, respectively.
Recently, from the $B^+ \to D^0 D_s^+ \pi^+\pi^-$ and $B^0 \to D^- D_s^+ \pi^+\pi^-$ decays
the LHCb Collaboration found that there might exist additional exotic contributions
decaying to $D_s^+ \pi^{\pm}$, denoted as $T_{c\bar{s}}(2327)^{++,0}$~\cite{LHCb:2024iuo}.
The significant experimental progress at LHCb stimulated more extensive research interests in the singly-heavy tetraquarks,
for understanding the nature of these newly observed resonances, and/or giving predictions for more states.
Many studies have been carried out within various methods and models~\cite{Meng:2023jqk,Ortega:2023azl,Jalili:2023kmw,Chen:2023syh,Wu:2023hhk,Zheng:2025uzy,
Gordillo:2025caj,Mutuk:2025hql,Chen:2026wrh,Karliner:2020vsi,He:2020jna,Wang:2020prk,Lu:2020qmp,Tan:2020cpu,
Yang:2021izl,Liu:2022hbk,Guo:2021mja,Wei:2022wtr,Xue:2020vtq,Dmitrasinovic:2023eei,Wang:2020xyc,
Zhang:2020oze,Albuquerque:2020ugi,Agaev:2022eeh,Mutuk:2020igv,Lian:2023cgs,Yang:2023evp,
Chen:2020aos,Agaev:2022eyk,Agaev:2022duz,Agaev:2020nrc,Yu:2025xip,Liu:2026qon}, such as constituent quark models~\cite{Meng:2023jqk,Ortega:2023azl,Jalili:2023kmw,Chen:2023syh,Wu:2023hhk,Zheng:2025uzy,
Gordillo:2025caj,Mutuk:2025hql,Chen:2026wrh,Karliner:2020vsi,He:2020jna,Wang:2020prk,Lu:2020qmp,Tan:2020cpu,
Yang:2021izl,Liu:2022hbk,Guo:2021mja,Wei:2022wtr,Xue:2020vtq,Dmitrasinovic:2023eei},
QCD sum rules~\cite{Wang:2020xyc,Zhang:2020oze,Albuquerque:2020ugi,Agaev:2022eeh,Mutuk:2020igv,Lian:2023cgs,Yang:2023evp,
Chen:2020aos,Agaev:2022eyk,Agaev:2022duz,Agaev:2020nrc}, and so on.

From the theoretical works existing in the literature, one can find (i)
most of the works focus on the masses, there are only a few works caring about the decays together with the masses~\cite{Agaev:2016mjb,Wang:2016wkj,Cheng:2020nho,Agaev:2019wkk,Wang:2020prk,Liu:2022hbk,Ortega:2023azl,Chen:2023syh,Guo:2021mja,Zheng:2025uzy}.
(ii) Most of the works focus on some special singly-heavy tetraquark systems,
there are only a few systematic studies of the whole singly-heavy tetraquark systems~\cite{Lu:2020qmp,Guo:2021mja}.
(iii) Most of the studies focus on the $Qs\bar{n}\bar{n}$ and $Qn\bar{s}\bar{n}$ ($Q=c,b$) systems
containing one strange quark, there are only a few studies for the nonstrange system $Qn\bar{n}\bar{n}$~\cite{Terasaki:2004yx,Wu:2023hhk,Guo:2021mja,Jovanovic:2007bz,Lu:2016zhe,Ebert:2010af,Lu:2020qmp,Chen:2017rhl,Jalili:2023kmw}, and
the multistrange systems $Qs\bar{s}\bar{n}$, $Qn\bar{s}\bar{s}$, and $Qs\bar{s}\bar{s}$~\cite{Guo:2021mja,Lu:2016zhe,Ebert:2010af,Lu:2020qmp,Jalili:2023kmw,Gerasyuta:2008ps,Zheng:2025uzy}.
(iv) There is significant model dependency in the predictions, which result in
contradictory explanations for the exotic candidates observed in experiments.
For example, the $T_{\bar{c}\bar{s}0}(2870)^0$ resonance is explained as compact tetraquarks~\cite{Karliner:2020vsi,He:2020jna,Wang:2020xyc,Zhang:2020oze,Wang:2020prk,Albuquerque:2020ugi,
Yang:2021izl,Liu:2022hbk,Guo:2021mja,Mutuk:2020igv,Wei:2022wtr,Dmitrasinovic:2023eei}, hadronic molecular states~\cite{Xue:2020vtq,Chen:2023syh,Mutuk:2020igv,Chen:2020aos,Huang:2020ptc,Xiao:2020ltm,Ke:2022ocs,Molina:2020hde,
He:2020btl,Liu:2020nil,Wang:2021lwy,Agaev:2020nrc}, or threshold/kinematical effects~\cite{Liu:2020orv,Burns:2020epm}.

In this work, we carry out a unified study of the spectra and fall-apart decays of the $1S$-wave states for the
whole singly-heavy tetraquark systems, $Qn\bar{n}\bar{n}$, $Qs\bar{n}\bar{n}$, $Qn\bar{s}\bar{n}$, $Qs\bar{s}\bar{n}$, $Qn\bar{s}\bar{s}$,
and $Qs\bar{s}\bar{s}$ within a constituent quark model. 
Compared to our previous study of the $cs\bar{n}\bar{n}$ and $cn\bar{s}\bar{n}$ systems~\cite{Liu:2022hbk},
several significant improvements have been made. (i) First, to properly consider the contributions of the light meson exchanges we
adopt a hybrid potential model which incorporates not only the
OGE potentials but also the OBE potentials
from the pseudoscalar mesons ($\pi, K,\eta, \eta'$), scalar meson ($\sigma$), and vector mesons ($\rho,\omega,K^*, \phi$).
(ii) Second, to properly account for relativistic effects of light quarks,
we replace the nonrelativistic kinetic energy term of the Hamiltonian with a relativistic form. (iii) Third, to enhance the reliability of the decay width predictions, the decay amplitudes are computed by using the genuine tetraquark wave functions
obtained from the mass spectrum calculations, rather than a single Gaussian approximate
form as that adopted in the previous work~\cite{Liu:2022hbk}.
	
This paper is organized as follows. The framework is given in Sec.~\ref{frame}.
Our results and discussions are presented in Sec.~\ref{sec:results}. Finally, a summary is given in Sec.~\ref{sec:summary}.

\section{Framework}\label{frame}

In this part we first give an introduction of the tetraquark configurations classified in the quark model based on the flavor, spin, and color symmetries. Then, we give a review of the potential model for the mass spectrum calculations.
Finally, a brief review of the quark-exchange model for dealing with the fall-apart decays of the tetraquark states is given.

\subsection{Configurations}
	
According to the strangeness quantum numbers, the singly heavy tetraquarks can be categorized into \( Qn\bar{n}\bar{n} \),
\( Qn\bar{s}\bar{n} \), \( Qs\bar{n}\bar{n} \), \( Qs\bar{s}\bar{n} \), \( Qn\bar{s}\bar{s} \) and \( Qs\bar{s}\bar{s} \) ($Q=c,b$; $n=u,d$).
To calculate the spectroscopy of tetraquarks, firstly we construct the configurations in the space of flavor $\otimes$ color $\otimes$ spin according to symmetry.

In the flavor space, for a tetraquark system $Q_1q_2\bar{q}_3\bar{q}_4$ containing a heavy quark $Q$ and three light quarks $q\bar{q}\bar{q}$ $(q=u, d$, or $ s)$, the light antiquark pair $\bar{q}_3\bar{q}_4$ should satisfy the SU(3) flavor symmetry. Therefore, the $Q_1q_2\bar{q}_3\bar{q}_4$ system can form two different SU(3) flavor representations: the symmetric sextet $\mathbf{6}_F$ and antisymmetric antitriplet $\bar{\mathbf{3}}_F$. Furthermore, isospin symmetry is required for the $Qn\bar{n}\bar{n}$ and $Qn\bar{s}\bar{n}$ systems. The obtained flavor functions of the $Q_1q_2\bar{q}_3\bar{q}_4$ system with different isospin quantum numbers are given in Table~\ref{tab:flavor_wavefunctions}.

In the spin space, according to the SU(2) Clebsch-Gordan coefficients,
the spin wave functions are constructed as follows:
	\begin{align}
		\chi_{00}^{00} & =\frac{1}{2}(\uparrow\downarrow\uparrow\downarrow-\uparrow\downarrow\downarrow\uparrow-\downarrow\uparrow\uparrow\downarrow+\downarrow\uparrow\downarrow\uparrow),\\
		\chi_{00}^{11} & =\sqrt{\frac{1}{12}}(2\uparrow\uparrow\downarrow\downarrow-\uparrow\downarrow\uparrow\downarrow\nonumber\\	 &-\uparrow\downarrow\downarrow\uparrow-\downarrow\uparrow\uparrow\downarrow-\downarrow\uparrow\downarrow\uparrow+2\downarrow\downarrow\uparrow\uparrow),\\
		\chi_{11}^{01} & =\sqrt{\frac{1}{2}}(\uparrow\downarrow\uparrow\uparrow-\downarrow\uparrow\uparrow\uparrow),\\
		\chi_{11}^{10} & =\sqrt{\frac{1}{2}}(\uparrow\uparrow\uparrow\downarrow-\uparrow\uparrow\downarrow\uparrow),\\
		\chi_{11}^{11} & =\frac{1}{2}(\uparrow\uparrow\uparrow\downarrow+\uparrow\uparrow\downarrow\uparrow-\uparrow\downarrow\uparrow\uparrow-\downarrow\uparrow\uparrow\uparrow),\\
		\chi_{22}^{11} & =\uparrow\uparrow\uparrow\uparrow.
	\end{align}
In the spin wave function $\chi ^{S_{12}S_{34}}_{SS_z}$, the $S_{12}$ and $S_{34}$ represent the spin quantum numbers of the subsystems $Q_1q_2$ and $\bar{q}_3\bar{q}_4$, respectively. $S$ is quantum number of the total spin of the tetraquarks, and $S_z$ is the quantum number
of the third component of the total spin.

In the color space, due to color confinement, the color wave functions of the tetraquark states must be singlets. Starting from the diquark-antidiquark structure, according to the symmetry of the SU(3) group, one can obtain two different color singlets for a tetraquark state, i.e.,
	\begin{equation}
		\label{eq:color_singlets}
		\left\{
		\begin{aligned}
			\left | 6\bar{6}   \right \rangle =(Q_1q_2)^6(\bar{q}_3 \bar{q}_4 )^{\bar{6} }, \\
			\left | \bar{3} 3  \right \rangle= (Q_1q_2)^{\bar{3} }(\bar{q}_3 \bar{q}_4 )^{3},
		\end{aligned}
		\right.
	\end{equation}
which can be further explicitly expressed as
	\begin{eqnarray}
			\left | 6\bar{6}   \right \rangle =& \frac{1}{2\sqrt{6}} \Big[ (rb + br)(\bar{r}\bar{b} + \bar{b}\bar{r}) +(gr + rg)(\bar{g}\bar{r} + \bar{r}\bar{g}) +\nonumber\\
			&+ (gb + bg)(\bar{g}\bar{b} + \bar{b}\bar{g})  \nonumber\\
			&+ 2(rr)(\bar{r}\bar{r}) + 2(gg)(\bar{g}\bar{g}) +
			2(bb)(\bar{b}\bar{b}) \Big],\\
			\left | \bar{3} 3  \right \rangle =& \frac{1}{2\sqrt{3}} \Big[ (br - rb)(\bar{b}\bar{r} - \bar{r}\bar{b}) +
			(rg - gr)(\bar{r}\bar{g} -\bar{g}\bar{r}) \nonumber\\
			&+(bg - gb)(\bar{b}\bar{g} - \bar{g}\bar{b}) \Big].
	\end{eqnarray}

Considering the Pauli principle and the requirement of color confinement in the four-quark system, the total wave function must be completely antisymmetric under the exchange of identical quarks or identical antiquarks. It is worth noting that since we consider only the $1S$-wave states, the spatial wave function is known to be symmetric. Therefore, it is sufficient to ensure the wave function of flavor $\otimes$ color $\otimes$ spin is overall antisymmetric under an exchange of any two identical particles. We represent all possible configurations of the $Q_1q_2\bar{q}_3\bar{q}_4$ systems in Table~\ref{tab:1S_configurations}.

\begin{table}[hptb]
	\caption{\label{tab:flavor_wavefunctions}	Flavor wave functions of the singly-heavy tetraquarks $Q_1q_2\bar{q}_3\bar{q}_4$ containing a heavy quark $Q ~(Q=b,c)$ and three light quarks ($u, d$, or $s$). In the table, we define $\left \{ a b  \right \}= \sqrt[]{\frac{1}{2} }(ab+ba) $ and $\left [ ab \right ]= \sqrt[]{\frac{1}{2} }(ab-ba) $.}
	\tabcolsep=0.07cm
	\renewcommand\arraystretch{1.30}
	\begin{tabular}{ccccc}
		\hline\hline
\textrm{$~$}&	
\textrm{$I$}&
\textrm{$I_3$}&
\textrm{$6_F$}&
\textrm{$\bar{3}_F$}
\\
\hline
$~$&$\frac{3}{2}$&~$+\frac{3}{2}$ &~$ Qu \bar{d}\bar{d}   $&~ $ ~ $
\\
$\multirow{5}{*}{\( Qn\bar{n}\bar{n} \)}$&$\frac{3}{2}$&~$+\frac{1}{2}$ &~$ \sqrt[]{\frac{1}{3} }Qd\bar{d}\bar{d}   -\sqrt[]{\frac{2}{3} }Qu\left \{ \bar{d}\bar{u}   \right \} $&~ $~ $
\\
$~$&$\frac{3}{2}$&~$-\frac{1}{2}$ &~$ \sqrt[]{\frac{1}{3} }Qu \bar{u}\bar{u}  -\sqrt[]{\frac{2}{3} }Qd\left \{ \bar{d}\bar{u}   \right \} $&~ $~ $
\\
$~$&$\frac{3}{2}$&~$-\frac{3}{2}$ &~$Qd \bar{u}\bar{u} $&~ $~ $
\\
$~$&$\frac{1}{2}$&~$+\frac{1}{2}$ &~$ -\sqrt[]{\frac{2}{3} }Qd\bar{d}\bar{d}   -\sqrt[]{\frac{1}{3} }Qu\left \{ \bar{d}\bar{u} \right \}  $&~ $-Qu \left[\bar{d}\bar{u}\right] $
\\
$~$&$\frac{1}{2}$&~$-\frac{1}{2}$ &~$ \sqrt[]{\frac{2}{3} }Qu \bar{u}\bar{u}  +\sqrt[]{\frac{1}{3} }Qd\left \{ \bar{d}\bar{u}   \right \} $&~ $-Qd\left[\bar{d}\bar{u}\right] $
\\
\hline\hline
$~$&$0$&~$0$ &~$ \sqrt[]{\frac{1}{2} }Qd\left \{ \bar{s}\bar{d}   \right \} +\sqrt[]{\frac{1}{2} }Qu\left \{ \bar{s}\bar{u}   \right \} $&~ $ \sqrt[]{\frac{1}{2} }Qd\left [ \bar{s}\bar{d}   \right ] +\sqrt[]{\frac{1}{2} }Qu\left [ \bar{s}\bar{u}   \right ] $
\\
$\multirow{2}{*}{\( Qn\bar{s}\bar{n} \)}$&$1$&~$+1$ &~$ -Qu\left \{ \bar{s}\bar{d}   \right \} $&~ $-Qu\left [ \bar{s}\bar{d}   \right ] $
\\
$~$&$1$&~$0$ &~$ \sqrt[]{\frac{1}{2} }Qd\left \{ \bar{s}\bar{d}   \right \} -\sqrt[]{\frac{1}{2} }Qu\left \{ \bar{s}\bar{u}   \right \} $&~ $\sqrt[]{\frac{1}{2} }Qd\left [ \bar{s}\bar{d}   \right ] +\sqrt[]{\frac{1}{2} }Qu\left [ \bar{s}\bar{u}   \right ] $
\\
$~$&$1$&~$-1$ &~$ Qd\left \{ \bar{s}\bar{u}   \right \} $&~ $Qd\left [ \bar{s}\bar{u}   \right ]    $
\\
\hline\hline
$~$&$0$&~$0$ &~$ ~ $&~ $ -Qs\left [ \bar{d}\bar{u}   \right ]  $
\\
$\multirow{2}{*}{\( Qs\bar{n}\bar{n} \)}$&$1$&~$+1$ &~$ Qs \bar{d}\bar{d} $&~ $~ $
\\
$~$&$1$&~$0$ &~$- Qs\left \{ \bar{d}\bar{u}   \right \}  $&~ $~ $
\\
$~$&$1$&~$-1$ &~$ Qs\bar{u}\bar{u}   $& $~  $
\\
\hline\hline
$\multirow{2}{*}{\( Qs\bar{s}\bar{n} \)}$&$\frac{1}{2}$&~$+\frac{1}{2}$ &~$ -Qs\left \{ \bar{s}\bar{d}   \right \}$&~ $-Qs\left [ \bar{s}\bar{d}   \right ]  $
\\
$~$&$\frac{1}{2}$&~$-\frac{1}{2}$ &~$ Qs\left \{ \bar{s}\bar{u}   \right \} $&~ $Qs\left [ \bar{s}\bar{u}   \right ] $
\\
\hline\hline
$\multirow{2}{*}{\( Qn\bar{s}\bar{s} \)}$&$\frac{1}{2}$&~$+\frac{1}{2}$ &~$ Qu \bar{s}\bar{s}   $&~ $~  $
\\
$~$&$\frac{1}{2}$&~$-\frac{1}{2}$ &~$ Qd \bar{s}\bar{s}   $&~ $~$
\\
\hline\hline			
$\multirow{1}{*}{\( Qs\bar{s}\bar{s} \)}$&$0$&~$0$ &~$ Qs \bar{s}\bar{s}    $&~ $~ $
\\
\hline\hline
	\end{tabular}
\end{table}	

\begin{table}[htpb]
	\setlength{\tabcolsep}{5pt}
\renewcommand\arraystretch{1.50}
	\begin{center}
		\caption{\label{tab:1S_configurations}	
The $1S$-wave configurations of the $Q_1q_2\bar{q}_3\bar{q}_4$ systems constructed in the quark model.
The subscripts and superscripts of each configuration are the spin quantum numbers and representations of the color SU(3) group, respectively.}
		\begin{tabular}{clcc}
\hline\hline
\multirow{2}{*}{$~(I)J^P$}&
\multicolumn{3}{c}{\textrm{Configuration}}\\
\cline{2-4}					\rule{0pt}{0.4cm}
~&\multicolumn{3}{c}{\textrm{$Qn\bar{n}\bar{n}$ ~System}}\\
\cline{2-4}					\rule{0pt}{0.5cm}
$(\frac{1}{2})0^+$&$\left |  (Qn)^{\bar{3}}_0\left [ \bar{n}\bar{n}   \right ]^{3}_0  \right \rangle _0$&~$\left |  (Qn)^{6}_1\left [ \bar{n}\bar{n}   \right ] ^{\bar{6} }_1  \right \rangle _0$ &~$\sim  $\\
$~(\frac{1}{2})1^+$&$\left |  (Qn)^{6}_0\left [ \bar{n}\bar{n}   \right ]^{\bar{6}}_1  \right \rangle _1$&~$\left |  (Qn)^{\bar{3}}_1\left [ \bar{n}\bar{n}   \right ]^{3}_0  \right \rangle _1$ &~$ \left |  (Qn)^{6}_1\left [ \bar{n}\bar{n}   \right ]^{\bar{6}}_1  \right \rangle _1 $\\
$~(\frac{1}{2})2^+$&$\left |  (Qn)^{6}_1\left [ \bar{n}\bar{n}   \right ]^{\bar{6}}_1  \right \rangle _2$&~$\sim  $ &~$\sim   $\\
$~(\frac{1}{2},\frac{3}{2})0^+$&$	\left |  (Qn)^6_0 \left\{\bar{n}\bar{n}  \right\} ^{\bar{6} }_0  \right \rangle _0	$&~$\left |  (Qn)^{\bar{3}}_1\left \{ \bar{n}\bar{n}   \right \} ^{3}_1  \right \rangle _0$ &~$\sim  $\\
$~(\frac{1}{2},\frac{3}{2})1^+$&$\left |  (Qn)^{6}_1\left \{ \bar{n}\bar{n}   \right \}^{\bar{6}}_0  \right \rangle _1$&~$\left |  (Qn)^{\bar{3}}_1\left \{ \bar{n}\bar{n}   \right \} ^{3 }_1  \right \rangle _1$ &~$\left |  (Qn)^{\bar{3}}_0\left \{ \bar{n}\bar{n}   \right \} ^{3 }_1  \right \rangle _1$\\
$~(\frac{1}{2},\frac{3}{2})2^+$&$\left |  (Qn)^{\bar{3}}_1\left \{ \bar{n}\bar{n}   \right \}^{3}_1  \right \rangle _2$&~$\sim  $ &~$\sim  $\\
\hline\hline	
\rule{0pt}{0.4cm}	
~&\multicolumn{3}{c}{\textrm{$Qn\bar{s}\bar{n}$ ~System}}\\
\cline{2-4}					\rule{0pt}{0.5cm}	
$(0,1)0^+$&$\left |  (Qn)^{\bar{3}}_0\left [ \bar{s}\bar{n}   \right ]^{3}_0  \right \rangle _0$&~$\left |  (Qn)^{6}_1\left [ \bar{s}\bar{n}   \right ] ^{\bar{6} }_1  \right \rangle _0$ &~$\sim  $\\	
$~(0,1)1^+$&$\left |  (Qn)^{6}_0\left [ \bar{s}\bar{n}   \right ]^{\bar{6}}_1  \right \rangle _1$&~$\left |  (Qn)^{\bar{3}}_1\left [ \bar{s}\bar{n}   \right ]^{3}_0  \right \rangle _1$ &~$ \left |  (Qn)^{6}_1\left [ \bar{s}\bar{n}   \right ]^{\bar{6}}_1  \right \rangle _1 $\\	
$~(0,1)2^+$&$\left |  (Qn)^{6}_1\left [ \bar{s}\bar{n}   \right ]^{\bar{6}}_1  \right \rangle _2$&~$\sim  $ &~$\sim   $\\		

$~(0,1)0^+$&$\left |  (Qn)^{6}_0\left \{ \bar{s}\bar{n}   \right \}^{\bar{6}}_0  \right \rangle _0$&~$\left |  (Qn)^{\bar{3}}_1\left \{ \bar{s}\bar{n}   \right \} ^{3 }_1  \right \rangle _0$ &~$\sim  $\\
$~(0,1)1^+$&$\left |  (Qn)^{6}_1\left \{ \bar{s}\bar{n}   \right \}^{\bar{6}}_0  \right \rangle _1$&~$\left |  (Qn)^{\bar{3}}_1\left \{ \bar{s}\bar{n}   \right \} ^{3 }_1  \right \rangle _1$ &~$\left |  (Qn)^{\bar{3}}_0\left \{ \bar{s}\bar{n}   \right \} ^{3 }_1  \right \rangle _1$\\
$~(0,1)2^+$&$\left |  (Qn)^{\bar{3}}_1\left \{ \bar{s}\bar{n}   \right \}^{3}_1  \right \rangle _2$&~$\sim  $ &~$\sim  $\\			
\hline\hline
\rule{0pt}{0.4cm}
~&\multicolumn{3}{c}{\textrm{$Qs\bar{n}\bar{n}$ ~System}}\\
\cline{2-4}						\rule{0pt}{0.5cm}
$(0)0^+$&$\left | (Qs)^{\bar{3}}_0\left [ \bar{n}\bar{n}   \right ]^{3}_0  \right \rangle _0$&~$\left |  (Qs)^{6}_1\left [ \bar{n}\bar{n}   \right ] ^{\bar{6} }_1  \right \rangle _0$ &~$\sim  $\\	
$~(0)1^+$&$\left | (Qs)^{6}_0\left [ \bar{n}\bar{n}  \right ]^{\bar{6}}_1  \right \rangle _1$&~$\left |  (Qs)^{\bar{3}}_1\left [ \bar{n}\bar{n}  \right ]^{3}_0  \right \rangle _1$ &~$ \left |  (Qs)^{6}_1\left [ \bar{n}\bar{n}   \right ]^{\bar{6}}_1  \right \rangle _1 $\\	
$~(0)2^+$&$\left | (Qs)^{6}_1\left [ \bar{n}\bar{n}   \right ]^{\bar{6}}_1  \right \rangle _2$&~$\sim  $ &~$\sim   $\\		

$~(1)0^+$&$\left |  (Qs)^{6}_0\left \{ \bar{n}\bar{n}   \right \}^{\bar{6}}_0  \right \rangle _0$&~$\left |  (Qs)^{\bar{3}}_1\left \{ \bar{n}\bar{n}   \right \} ^{3 }_1  \right \rangle _0$ &~$\sim  $\\
$~(1)1^+$&$\left |  (Qs)^{6}_1\left \{ \bar{n}\bar{n}   \right \}^{\bar{6}}_0  \right \rangle _1$&~$\left |  (Qs)^{\bar{3}}_1\left \{ \bar{n}\bar{n}   \right \} ^{3 }_1  \right \rangle _1$ &~$\left |  (Qs)^{\bar{3}}_0\left \{ \bar{n}\bar{n}   \right \} ^{3 }_1  \right \rangle _1$\\
$~(1)2^+$&$\left |  (Qs)^{\bar{3}}_1\left \{ \bar{n}\bar{n}   \right \}^{3}_1  \right \rangle _2$&~$\sim  $ &~$\sim  $\\		
\hline\hline
\rule{0pt}{0.4cm}
~&\multicolumn{3}{c}{\textrm{$Qs\bar{s}\bar{n}$ ~System}}\\
\cline{2-4}						\rule{0pt}{0.5cm}
$(\frac{1}{2})0^+$&$\left |  (Qs)^{\bar{3}}_0\left [ \bar{s}\bar{n}   \right ]^{3}_0  \right \rangle _0$&~$\left |  (Qs)^{6}_1\left [ \bar{s}\bar{n}   \right ] ^{\bar{6} }_1  \right \rangle _0$ &~$\sim  $\\
$~(\frac{1}{2})1^+$&$\left |  (Qs)^{6}_0\left [ \bar{s}\bar{n}   \right ]^{\bar{6}}_1  \right \rangle _1$&~$\left |  (Qs)^{\bar{3}}_1\left [ \bar{s}\bar{n}   \right ]^{3}_0  \right \rangle _1$ &~$ \left |  (Qs)^{6}_1\left [ \bar{s}\bar{n}   \right ]^{\bar{6}}_1  \right \rangle _1 $\\
$~(\frac{1}{2})2^+$&$\left |  (Qs)^{6}_1\left [ \bar{s}\bar{n}   \right ]^{\bar{6}}_1  \right \rangle _2$&~$\sim  $ &~$\sim   $\\

$~(\frac{1}{2})0^+$&$\left |  (Qs)^{6}_0\left \{ \bar{s}\bar{n}   \right \}^{\bar{6}}_0  \right \rangle _0$&~$\left |  (Qs)^{\bar{3}}_1\left \{ \bar{s}\bar{n}   \right \} ^{3}_1  \right \rangle _0$ &~$\sim  $\\
$~(\frac{1}{2})1^+$&$\left |  (Qs)^{6}_1\left \{ \bar{s}\bar{n}   \right \}^{\bar{6}}_0  \right \rangle _1$&~$\left |  (Qs)^{\bar{3}}_1\left \{ \bar{s}\bar{n}   \right \} ^{3 }_1  \right \rangle _1$ &~$\left |  (Qs)^{\bar{3}}_0\left \{ \bar{s}\bar{n}   \right \} ^{3 }_1  \right \rangle _1$\\
$~(\frac{1}{2})2^+$&$\left |  (Qs)^{\bar{3}}_1\left \{ \bar{s}\bar{n}   \right \}^{3}_1  \right \rangle _2$&~$\sim  $ &~$\sim  $\\		
\hline\hline
\rule{0pt}{0.4cm}
~&\multicolumn{3}{c}{\textrm{$Qn\bar{s}\bar{s}$ ~System}}\\
\cline{2-4}					\rule{0pt}{0.5cm}
$(\frac{1}{2})0^+$&$\left |  (Qn)^{6}_0\left \{ \bar{s}\bar{s}   \right \}^{\bar{6}}_0  \right \rangle _0$&~$\left |  (Qn)^{\bar{3}}_1\left \{ \bar{s}\bar{s}   \right \} ^{3}_1  \right \rangle _0$ &~$\sim  $\\
$~(\frac{1}{2})1^+$&$\left |  (Qn)^{6}_1\left \{ \bar{s}\bar{s}   \right \}^{\bar{6}}_0  \right \rangle _1$&~$\left |  (Qn)^{\bar{3}}_1\left \{ \bar{s}\bar{s}   \right \} ^{3 }_1  \right \rangle _1$ &~$\left |  (Qn)^{\bar{3}}_0\left \{ \bar{s}\bar{s}   \right \} ^{3 }_1  \right \rangle _1$\\
$~(\frac{1}{2})2^+$&$\left |  (Qn)^{\bar{3}}_1\left \{ \bar{s}\bar{s}   \right \}^{3}_1  \right \rangle _2$&~$\sim  $ &~$\sim  $\\	
\hline\hline
\rule{0pt}{0.4cm}
~&\multicolumn{3}{c}{\textrm{$Qs\bar{s}\bar{s}$ ~System}}\\
\cline{2-4}					\rule{0pt}{0.5cm}
$(0)0^+$&$\left |  (Qs)^{6}_0\left \{ \bar{s}\bar{s}   \right \}^{\bar{6}}_0  \right \rangle _0$&~$\left |  (Qs)^{\bar{3}}_1\left \{ \bar{s}\bar{s}   \right \} ^{3}_1  \right \rangle _0$ &~$\sim  $\\
$~(0)1^+$&$\left |  (Qs)^{6}_1\left \{ \bar{s}\bar{s}   \right \}^{\bar{6}}_0  \right \rangle _1$&~$\left |  (Qs)^{\bar{3}}_1\left \{ \bar{s}\bar{s}   \right \} ^{3 }_1  \right \rangle _1$ &~$\left |  (Qs)^{\bar{3}}_0\left \{ \bar{s}\bar{s}   \right \} ^{3 }_1  \right \rangle _1$\\
$~(0)2^+$&$\left |  (Qs)^{\bar{3}}_1\left \{ \bar{s}\bar{s}   \right \}^{3}_1  \right \rangle _2$&~$\sim  $ &~$\sim  $\\		
\hline	\hline
		\end{tabular}
	\end{center}
\end{table}

\subsection{Potential model}

In this work, we calculate the mass spectrum of the singly heavy tetraquarks within a
semi-relativistic quark potential model by including both the OGE and OBE potentials.
In the following, we briefly review the model Hamiltonian, parameter determination, and numerical method.

\subsubsection{\label{sec:hamiltonian}Hamiltonian}

To describe a singly-heavy tetraquark system, we adopt a semi-relativistic Hamiltonian, i.e.,	
\begin{equation}\label{eq:hamiltonian}
H=\sum\limits_{i=1}^{4}\sqrt{p_i^2+m_i^2}+\sum\limits_{i< j}V_{ij}(r_{ij}) ,
\end{equation}
where $\sqrt{\boldsymbol{p}_i^2+m_i^2}$ represents the relativistic kinetic term of the $i$-th constituent quark with mass $m_i$ and momentum $\boldsymbol{p}_{i}$. $V_{ij}(r_{ij})$ stands for the effective potentials between the $i$-th and $j$-th quarks with a distance $r_{ij}\equiv \left | \boldsymbol{r}_i-\boldsymbol{r}_j \right | $.

In hadronic systems containing light quarks, the spontaneous breaking of chiral symmetry gives rise to
Goldstone bosons, whose exchange is expected to play a significant role in the quark-quark interaction dynamics.
To achieve a more reliable description of the mass spectrum of the singly heavy tetraquark states,
in this work the effective potential is adopted as
	\begin{equation}
		\label{eq:Vij_OGE_OBE}
		\begin{split}
			V_{ij}(r_{ij})&=V_{ij}^{OGE}(r_{ij})+V_{ij}^{OBE}(r_{ij}),
		\end{split}
	\end{equation}	
which not only includes the potential $V_{ij}^{OGE}(r_{ij})$ widely adopted in the OGE potential models,
but also includes the OBE potential $V_{ij}^{OBE}(r_{ij})$ from the light pseudoscalar meson ($\pi$, $K$, $\eta$, $\eta'$) exchanges, the scalar meson ($\sigma$) exchange, and light vector meson ($\rho$, $\omega$, $K^*$, $\phi$) exchanges.
This hybrid quark potential model including both the OGE and OBE potentials has been widely applied in the study of the hadron spectra and hadron-hadron interactions in the literature, e.g., Refs.~\cite{Huang:2015nja,He:2023ucd,Zhong:2024mnt,Lu:2024dtb,Dai:2003dz,Huang:2005hy,Vijande:2004he,Brauer:1990kt,Zhang:1994pp,Yu:1995ag,Valcarce:2005em,Valcarce:2005rr,Valcarce:2008dr,
Wang:2011rga,Huang:2004sj,Yang:2017qan,Huang:2017dwn}, which have demonstrated that the chiral dynamics is important for some special
hadron systems containing light quarks. For example, our recent study of nucleon and $\Delta$ baryon spectra shows that
the $\pi$ meson exchange plays a crucial role in the Roper resonance $N(1440)1/2^+$, with which the mass reversal
between $N(1440)1/2^+$ and $N(1535)1/2^-$ can be naturally explained~\cite{Zhong:2024mnt}. Moreover, the recent analysis of the $S$-wave phase shifts of nucleon-nucleon scattering shows that vector meson ($\rho,\omega$) exchanges between $u/d$ quarks are crucial in the short range~\cite{Lu:2024dtb}.

The $V_{ij}^{OGE}(r_{ij})$ part is explicitly expressed as
\begin{eqnarray}\label{eq:conf}
	V_{ij}^{OGE}(r_{ij}) &&= -\frac{3}{16}(\boldsymbol{\lambda}_i^c \cdot \boldsymbol{\lambda}_j^c)\biggl[(b_{ij} r_{ij} + C_{ij})-\frac{4}{3}\frac{\alpha_{ij}}{r_{ij}}\nonumber\\
       && +
	\frac{32\alpha_{ij}\pi}{9m_im_j}\frac{e^{-r_{ij}^2/r_0^2}}{\pi^{3/2}r_0^3}
		(\boldsymbol{S}_i\cdot\boldsymbol{S}_j)\biggr],
\end{eqnarray}
where the first, second, and last terms stand for the linear confinement potential,
color-Coulomb potential, and color-magnetic spin-spin interaction, respectively.
In the above equation, the parameter $b_{ij}$ denotes the strength of the confinement potential between the $i$-th and $j$-th quarks, while $C_{ij}$ is the zero point energy. $\boldsymbol{\lambda}_{i/j}^c$ and $\boldsymbol{S}_{i/j}$ represent the color and spin operators acting on the $i/j$-th quark, respectively.
The effective strong coupling constant $\alpha_{ij} $ is expressed as a product of the individual OGE coupling
constants $g_i$ and $g_j$ for the $i$-th and $j$-th quarks,
respectively, i.e., $\alpha_{ij} \equiv g_i g_j$ as adopted in Refs.~\cite{Dai:2003dz,Huang:2015nja}. This flavor-dependent parameterization allows the effective coupling strength to vary with different quark
pairs (e.g., light--light, heavy--light, and heavy--heavy).
For $r_0$, we choose a constituent quark mass dependent form as suggested in Ref.~\cite{Silvestre-Brac:1996myf}:
\begin{equation}
	r_0(m_i, m_j) = A\left(\frac{2m_im_j}{m_i+m_j}\right)^{-B},
\end{equation}
where $A$ and $B$ are two free parameters. Finally, it should be mentioned that for the quark-quark (antiquark-antiquark) interaction, $\boldsymbol{\lambda}_i^c \cdot \boldsymbol{\lambda}_j^c \equiv \sum_{a=1}^{8} \lambda_i^a \cdot \lambda_j^a$, while for the quark-antiquark interaction  $\boldsymbol{\lambda}_i^c \cdot \boldsymbol{\lambda}_j^c \equiv -\sum_{a=1}^{8} \lambda_i^a \cdot \lambda_j^{a*}$, where $\lambda^{a*}$ is the complex conjugate of the Gell-Mann
matrices $\lambda^{a}$ ($a=1,\cdots,8$) associated with the quark color wave function.

The center part of the one-boson-exchange (OBE) quark-quark (antiquark-antiquark) potentials by the exchanges of $\pi$, $K$, $\eta$, $\eta'$, $\sigma$, $\rho$, $K^*$, $\phi$ and $\omega$ mesons is
given by
\begin{equation}
	V_{ij}^{\mathrm{OBE}}(r_{ij})= \sum_{\chi=\pi,K,\eta,\eta'} V_{\chi}^{c}(r_{ij})+ V_{\sigma}^{c}(r_{ij})+ \sum_{v=\rho,\omega,K^*,\phi} V_{v}^{c}(r_{ij}),
\end{equation}
with
\begin{eqnarray}\label{eq:Vchi_central}
		\begin{split}	
V_{\chi}^{c}(r_{ij})=&\frac{g_{\chi}^2}{4\pi}\frac{\Lambda^2_{\chi}}{\Lambda^2_{\chi}-m_{\chi}^2}\frac{m_{\chi}^3}{3m_im_j}\left [ Y(m_{\chi }r_{ij})-\frac{\Lambda^3_{\chi}}{m_{\chi}^3}Y(\Lambda_{\chi} r_{ij})  \right ] \\
			&\cdot(\boldsymbol{S} _i\cdot \boldsymbol{S}_j) \mathcal{I}_{\chi},
	\end{split}
\end{eqnarray}
\begin{eqnarray}\label{eq:Vsigma_central}
		\begin{split}	
&V_{\sigma}^{c}(r_{ij})=-\frac{g_{\sigma}^2}{4\pi}\frac{\Lambda_{\sigma}^2}{\Lambda_{\sigma}^2-m_{\sigma}^2}m_{\sigma}\left [ Y(m_{\sigma }r_{ij})-\frac{\Lambda_{\sigma}}{m_{\sigma }}Y(\Lambda_{\sigma} r_{ij})  \right ]\mathcal{I}_{\sigma},
	\end{split}
\end{eqnarray}
\begin{eqnarray}\label{eq:Vv_central}
\begin{split}
V_{v}^{c}(r_{ij})
		&={}  \frac{g_{v}^{2}}{4\pi}
		\frac{\Lambda_{v}^{2}}
		{\Lambda_{v}^{2}-m_{v}^{2}}\,
		m_{v}\biggl\{\biggl[Y(m_{v}r_{ij})
		-\frac{\Lambda_{v}}{m_{v}}
		Y(\Lambda_{v}r_{ij})\biggr] \\
		&+ \frac{2m_{v}^{2}}{3m_{i}m_{j}} \biggl[Y(m_{v}r_{ij})
		-\Bigl(\frac{\Lambda_{v}}{m_{v}}\Bigr)^{\!3}
		Y(\Lambda_{v}r_{ij})\biggr](\boldsymbol{S} _i\cdot \boldsymbol{S}_j)\biggl\}\mathcal{I}_{v},
	\end{split}
\end{eqnarray}
where $Y(x)$ is the standard Yukawa function $Y(x)=\frac{e^{-x}}{x}$, and $g_{\chi/\sigma/v}$ are the quark-chiral-field coupling constants. $m_{\chi}$, $m_{\sigma}$, and $m_{v}$ stand for the masses of the pseudoscalar mesons ($\chi=\pi$, $K$, $\eta$, $\eta'$), the $\sigma$ meson, and the vector mesons ($v=\rho$, $K^*$, $\omega$, $\phi$), respectively. $\Lambda$ stands for the cutoff parameters of the meson fields, which characterize the energy scale of spontaneous chiral symmetry breaking. $\mathcal{I}_{\chi}$, $\mathcal{I}_{v}$, and $I_\sigma$ stand for the flavor operators of the pseudoscalar, vector, and scalar $\sigma$ mesons, respectively.
They are explicitly expressed as follows,
\begin{eqnarray}
	\mathcal{I}_{\pi,\rho} = \sum_{a=1}^{3}
	(\lambda_{i}^{a}\cdot\lambda_{j}^{a}),~
	\mathcal{I}_{K,K^*} = \sum_{a=4}^{7}
	(\lambda_{i}^{a}\cdot\lambda_{j}^{a}),~\mathcal{I}_{\sigma}= \lambda_{i}^{0}\cdot\lambda_{j}^{0},\label{eq:flavor_ops_pi_K_sigma}\\
	\mathcal{I}_{\eta}= \lambda_{i}^{\eta}\cdot\lambda_{j}^{\eta},
\mathcal{I}_{\eta'}= \lambda_{i}^{\eta'}\cdot\lambda_{j}^{\eta'},\mathcal{I}_{\omega}
=\lambda_{i}^{\omega}\cdot\lambda_{j}^{\omega},\mathcal{I}_{\phi}
=\lambda_{i}^{\phi}\cdot\lambda_{j}^{\phi}\label{eq:flavor_ops_eta_eta_omega_phi}.
\end{eqnarray}
In Eq.~(\ref{eq:flavor_ops_pi_K_sigma}), $\lambda^{a}$ ($a=1,\cdots,8$) are the Gell-Mann matrices, while $\lambda^{0}$ is defined as $\lambda^0=\sqrt{\frac{2}{3}}\,\mathbb{I}$, where $\mathbb{I}$ is the $3\times3$ identity matrix.
The other matrices $\lambda^{\eta,\eta',\omega,\phi}$ appearing in Eq.~(\ref{eq:flavor_ops_eta_eta_omega_phi}) are defined as
$\lambda^{\eta}=\lambda^{8}\cos\theta_{P}- \lambda^{0}\sin\theta_{P}$, $\lambda^{\eta'}=\lambda^{8}\sin\theta_{P}+\lambda^{0}\cos\theta_{P}$, $\lambda^{\omega}=\lambda^{8}\sin\theta_{V}+ \lambda^{0}\cos\theta_{V}$, and $\lambda^{\phi}=\lambda^{8}\cos\theta_{V}- \lambda^{0}\sin\theta_{V}$, where $\theta_P$ and $\theta_V$ are the mixing angles for the pseudoscalar and vector mesons in the $SU(3)$ flavor basis, respectively. They are taken as the standard values of PDG, i.e., $\theta_P = -15^{\circ}$ and $\theta_V = 35.3^{\circ}$~\cite{ParticleDataGroup:2024cfk}.
Note that for the quark-antiquark potentials, the flavor operators in Eqs.~(\ref{eq:flavor_ops_pi_K_sigma}) and (\ref{eq:flavor_ops_eta_eta_omega_phi}) should be replaced
by
	\begin{equation}
		\label{eq:qqbar_operator_rule}
\lambda_i\cdot\lambda_j\to	
\left\{
		\begin{aligned}
			\lambda_i\cdot\lambda_j^{*}, ~\mathrm{for~scalar~and~pseudoscalar ~mesons},\\
			-\lambda_i\cdot\lambda_j^{*}, ~\mathrm{for ~vector ~mesons,~~~~~~~~~~~~~~~~~~~~~~~~~~~}
		\end{aligned}\nonumber
		\right.
	\end{equation}
where $\lambda^{*}_j$ is the transformation matrix acting on the $j$-th antiquark.

Finally, it should be pointed out that in this work we only focus on the low-lying $1S$-wave tetraquark states,
neither tensor nor spin-orbit interactions are incorporated into the OGE and OBE potentials.
The details of the tensor and spin-orbit potentials originating from  OGE and OBE can be found in the literature, e.g. Refs.~\cite{Vijande:2004he,Huang:2005hy,Dai:2003dz,Ni:2023lvx,Zhong:2024mnt}.

\subsubsection{\label{sec:parameters}Parameters}

To be consistent with the previous work of our group~\cite{Zhong:2024mnt},
the coupling constants, $g_{\chi}$ ($\chi=\pi,K,\eta$), for the pseudoscalar
meson exchanges are determined by
\begin{equation}
	g_{\chi} = \delta\,\frac{m_u}{f_\chi},
\end{equation}
where $\delta$ and $f_\chi$ are a global parameter accounting for the strength of
the quark-pseudoscalar-meson couplings and the corresponding
meson decay constant, respectively, which are associated with the chiral Lagrangian $\mathcal{L}_{ps}=\frac{\delta}{\sqrt{2}f_{\chi }}\bar{\psi}\gamma^{\mu}\gamma^{5}\psi \partial ^{\mu}\phi_{m}$, where $\psi$ stands for the light quark field with $\psi^{\top}\equiv(u,~ d,~s)$, and $\phi_{m}$ is the matrix representation of the pseudoscalar-meson fields. For the parameter $\delta$, we take the same value, $\delta=0.576$, as that determined by the strong decays of light strange baryon resonances in
Refs.~\cite{Xiao:2013xi,Xiao:2018pwe,Liu:2019wdr}.
For the $\pi$, $K$, $\eta$ and $\eta^{\prime}$ mesons, the decay constants are fixed with
$f_{\pi} = 93$~MeV, $f_{K} = f_{\eta/\eta^{\prime} } = 113$~MeV. For the
vector mesons $\rho$, $\omega$, $K^*$ and $\phi$, the coupling constants
are fixed with $g_{\rho/K^*/\omega/\phi}= 1.7$, which are associated
with the effective interaction $\mathcal{L}_{v}=g_{v}\bar{\psi}\gamma_{\mu}V^{\mu}\psi$,
where $V^{\mu}$ is the matrix representation of the vector-meson fields.
These coupling constants are determined by the vector meson photoproduction processes~\cite{Zhao:1998fn}.
In this work, we take $g_{\sigma}=g_{\pi}$ as an approximation.

The cutoff parameter $\Lambda$ characterizes the energy
scale of spontaneous chiral symmetry breaking. For the
pseudoscalar ($\pi$, $K$, $\eta$, $\eta^{\prime}$) and scalar ($\sigma$)
mesons, to be consistent with the previous work of our group,
the cutoff parameters are fixed with
$\Lambda_{\pi/K/\eta/\eta^{\prime}}= \Lambda_{\sigma} = 0.66$~GeV~\cite{Zhong:2024mnt}. While for the
vector mesons, we take $\Lambda_{\rho/\omega/K^*/\phi} = 0.85$~GeV,
which are determined by fitting the light meson spectrum.
The masses of the exchange mesons in the OBE potentials are taken as their physical values:
$m_{\pi} = 138$~MeV, $m_{K} = 495$~MeV,
$m_{\eta} = 548$~MeV, $m_{\eta^{\prime}} = 958$~MeV, $m_{\sigma} = 458$~MeV,
$m_{\rho} = 770$~MeV, $m_{K^*} = 892$~MeV, $m_{\phi} = 1020$~MeV and $m_{\omega} = 782$~MeV~\cite{ParticleDataGroup:2024cfk}.
It should be mentioned that at short range of the quark-(anti)quark distance
the masses of $\omega$ meson and $\Lambda_c$ baryon
are sensitive to $\pi$ meson exchange and vector mesons $\rho$ and $\omega$ exchanges, respectively,
the divergent behavior at short range will lead to instability of numerical results,
thus, we introduce cutoff distances $r_{ij}^{\pi/\rho/\omega}=0.30$~fm
for the $\pi$- and $\rho/\omega$-exchange potentials,
which are determined by the measured masses of the $\omega$ meson and $\Lambda_c$ baryon.

The constituent quark masses and the parameters $b$, $g_{i/j}$, $C_{ij}$, $A$,
and $B$ appearing in the confinement and OGE potentials
adopted in this work have been collected in Table~\ref{tab:optimal_params},
which are determined by fitting the $1S$- and $2S$-wave meson states observed in experiments.
The fitted meson spectrum compared with the data is presented in Table~\ref{tab:mesons}.

\begin{table}[htbp]
	\centering
	\caption{Parameters of the quark model.}
	\label{tab:optimal_params}
		\begin{tabular}{cccccc}
			\hline
			\hline
			{Parameter}~~~ & {Value}~~~
			& {Parameter}~~~ & {Value}
			& {Parameter}  & {Value (GeV)} \\ \hline
			$m_n$\,(GeV)                    & 0.3221
			& $m_s$\,(GeV)                  & 0.4747
			& $C_{nn}$  & -0.6530 \\
			$m_c$\,(GeV)                    & 1.4736
			& $m_b$\,(GeV)                  & 4.7976
			& $C_{ns}$  & -0.6680 \\
			$g_n$                           & 0.7100
			& $g_s$                         & 0.6271
			& $C_{nc}$  & -0.4374 \\
			$g_c$                           & 0.5600
			& $g_b$                         & 0.5000
			& $C_{nb}$  & -0.3460 \\
			$A$\,(GeV$^{B-1}$)              & 1.0627
			& $B$            & 0.4966
			& $C_{ss}$  & -0.6535 \\
			$\Lambda$\,(GeV)                & 0.6600
			& $\Lambda_v$\,(GeV)            & 0.8500
			& $C_{sc}$  & -0.4663 \\
			$r_{ij}^{\pi/\rho/\omega}$\,(fm)                  & 0.3000
			& $b$\,(GeV$^{2}$)              & 0.2134
			& $C_{sb}$  & -0.3773 \\
			\hline\hline
		\end{tabular}%
\end{table}
	
\begin{table}[hptb]
	\caption{Mass spectra (MeV), RMS radii (fm), and model parameter $\alpha$ (GeV) for mesons. Experimental data from PDG~\cite{ParticleDataGroup:2024cfk}.}
	\label{tab:mesons}
	\centering
		\begin{tabular}{lcccc}
			\hline\hline
			{State}
			& {$M_{\text{th}}$ (MeV)}
			& {$M_{\text{exp}}$ (MeV)}
			& {$\sqrt{\langle r^{2}\rangle}$ (fm)}
			& {$\alpha$ (GeV)} \\
			\hline
			\multicolumn{5}{l}{\textit{Ground states ($1S$)}} \\[2pt]
			$\pi$                  &  148 &      140 & 0.356 & 0.679 \\
			$\rho$                 &  818 &      775 & 0.594 & 0.407 \\
			$\omega$               &  731 &      783 & 0.560 & 0.432 \\
			$K$                    &  485 &      498 & 0.392 & 0.616 \\
			$K^{*}$                &  906 &      892 & 0.568 & 0.426 \\
			$\phi$                 & 1012 &     1019 & 0.541 & 0.447 \\[3pt]
			$D$                    & 1831 &     1865 & 0.405 & 0.597 \\
			$D^{*}$                & 1998 &     2007 & 0.482 & 0.501 \\
			$D_{s}$                & 1969 &     1969 & 0.410 & 0.590 \\
			$D_{s}^{*}$            & 2092 &     2112 & 0.472 & 0.512 \\[3pt]
			$B$                    & 5263 &     5280 & 0.386 & 0.626 \\
			$B^{*}$                & 5323 &     5325 & 0.414 & 0.583 \\
			$B_{s}$                & 5370 &     5367 & 0.383 & 0.630 \\
			$B_{s}^{*}$            & 5416 &     5416 & 0.407 & 0.593 \\[3pt]
			\hline
			\multicolumn{5}{l}{\textit{Radially excited states}} \\[2pt]
			$\rho(1450)$           & 1529 & 1465(25) & 0.974 & 0.248 \\
			$\omega(1420)$         & 1478 & 1410(60) & 0.949 & 0.255 \\
			$\phi(1680)$           & 1724 & 1680(20) & 0.932 & 0.259 \\
			$D_{0}(2550)$          & 2560 & 2549(19) & 0.807 & 0.300 \\
			$D_{1}^{*}(2600)$      & 2651 & 2627(10) & 0.849 & 0.285 \\
			$D_{s1}^{*}(2700)$     & 2734 &  2714(5) & 0.837 & 0.289 \\[3pt]
			\hline\hline
		\end{tabular}%
\end{table}

\begin{figure*}[htbp]
\centering
\includegraphics[width=1.50\columnwidth]{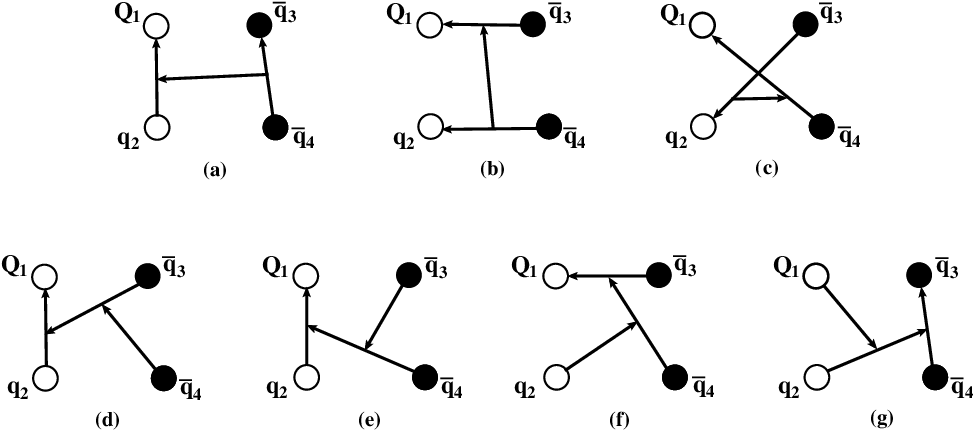}
\caption{The illustration of the Jacobi coordinates taken for the tetraquarks $Q_1q_2\bar{q}_3\bar{q}_4$.}
\label{fig:jacobi_coordinates}
\end{figure*}

\subsubsection{\label{sec:level4}Numerical method}

To accurately solve the four-body system, we adopt the explicitly correlated
Gaussian (ECG) method~\cite{Varga:1995dm,Mitroy:2013eom}, in which the trial spatial wave function is expanded with the most
general nondiagonal Gaussian basis functions. Such a
basis function can be expressed as
	\begin{equation}
		\psi (\boldsymbol{r}_1,\boldsymbol{r}_2,\boldsymbol{r}_3,\boldsymbol{r}_4)=\exp\left({-\sum\limits_{i< j}^{4}a_{ij}r_{ij}^2 }\right),
	\end{equation}
where $a_{ij}$ are variational parameters. Given that we are investigating a singly-heavy tetraquark system $Q_1q_2\bar{q}_3\bar{q}_4$, in the diquark-antidiquark configuration given in Table~\ref{tab:flavor_wavefunctions}, the $\bar{q}_3\bar{q}_4$ can be treated as identical particles in the SU(3) symmetry limit. Thus, we have four independent variational parameters, i.e., $a_{12}=a$, $a_{34}=b$, \( a_{13} = a_{14} \equiv  c \),  and  \( a_{23} =  a_{24} \equiv  d \).

To facilitate the calculations, we adopt a set of Jacobi coordinates constructed as shown in Fig.~\ref{fig:jacobi_coordinates}, rather than using the relative distance vectors $\boldsymbol{r}_{ij}=$($\boldsymbol{r}_i-\boldsymbol{r}_j$). For instance, one can take a set of Jacobi coordinates as that shown in Fig.~\ref{fig:jacobi_coordinates} (a), i.e.,
	\begin{equation}
		\begin{aligned}
			\vxi_1 &= \boldsymbol{r}_1 - \boldsymbol{r}_2, \\
			\vxi_2 &= \boldsymbol{r}_3 - \boldsymbol{r}_4, \\
			\vxi_3 &=  \frac{m_1 \boldsymbol{r}_1 + m_2 \boldsymbol{r}_2 }{m_1 + m_2 } - \frac{m_3 \boldsymbol{r}_3 + m_4 \boldsymbol{r}_4}{m_3 + m_4}, \\
			\boldsymbol{R} &=  \frac{m_1 \boldsymbol{r}_1 + m_2 \boldsymbol{r}_2 +m_3 \boldsymbol{r}_3 + m_4 \boldsymbol{r}_4}{m_1 + m_2 +m_3 + m_4}.
		\end{aligned}
	\end{equation}

Removing the coordinate of center of mass, then, the correlated Gaussian basis function $\psi (\boldsymbol{r}_1,\boldsymbol{r}_2,\boldsymbol{r}_3,\boldsymbol{r}_4)$ can be rewritten as
	\begin{equation}
		\label{eq:correlated_gaussian}
		G(\xi, A) = \exp \left( -\sum_{i,j} A_{ij}\vxi_i \cdot \vxi_j \right) =\exp \left[ -(\xi^T A \xi) \right ]
	\end{equation}
where $\xi^T=(\vxi_1,\vxi_2,\vxi_3)$, and $A$ is a $3\times 3$ symmetric positive-definite matrix whose elements are composed of variational parameters and the constituent quark masses.

The trial spatial wave function is expanded with a set of correlated Gaussians,
	\begin{equation}
		\label{eq:trial_wavefunction_expansion}
		\psi(\xi, A) = \sum_{n=1}^{n_{max}} C_n G(\xi, A_n),
	\end{equation}
where the $n_{max}$ denotes the number of Gaussian basis functions. The accuracy of the trial function is governed
by $n_{max }$ and the nonlinear parameter matrix $A_{n}$. In our calculations, following the procedure in Ref.~\cite{Hiyama:2003cu},
the variational parameters are chosen to form a geometric progression.
For example, for the variational parameter $a$, we take
	\[
	a_n = \frac{1}{(r_{a_1} q_a^{n-1})^2} ~~(n = 1, \ldots, n_{\text{max}}).
	\]
The Gaussian size parameters $\{r_{a_1},q_a, n_{\text{max}}\}$ will be determined by the variation method. In the calculations, the final
results should be stable and independent of these
parameters.	

For a given tetraquark configuration, one can work out the Hamiltonian matrix elements,
\begin{eqnarray}
	H_{nn'}=\langle \psi_{CS}G(\boldsymbol{\xi},A_n) | H |\psi_{CS}G(\boldsymbol{\xi},A_{n'}) \rangle,
\end{eqnarray}
where $\psi_{CS}$ is the spin-color wave function.
Then, by solving the generalized matrix eigenvalue problem,
\begin{eqnarray}
	\sum_{n'=1}^{N}(H_{nn'}-EN_{nn'})C_{n'}=0,
\end{eqnarray}
one can obtain the eigenenergy $E$, and the expansion coefficients
$\{C_n\}$. The $N_{nn'}$ is an overlap factor defined by $N_{nn'}=\langle G(\boldsymbol{\xi},A_n)|G(\boldsymbol{\xi},A_{n'})\rangle$.

Finally, it should be mentioned that in the calculations we can obtain stable solutions when we take $N=n_{max}^a\times n_{max}^b\times n_{max}^c\times n_{max}^d=6\times6\times6\times6=6^4$ for the singly heavy tetraquark states.

\begin{figure}[htbp]
\centering
\includegraphics[width=0.90\columnwidth]{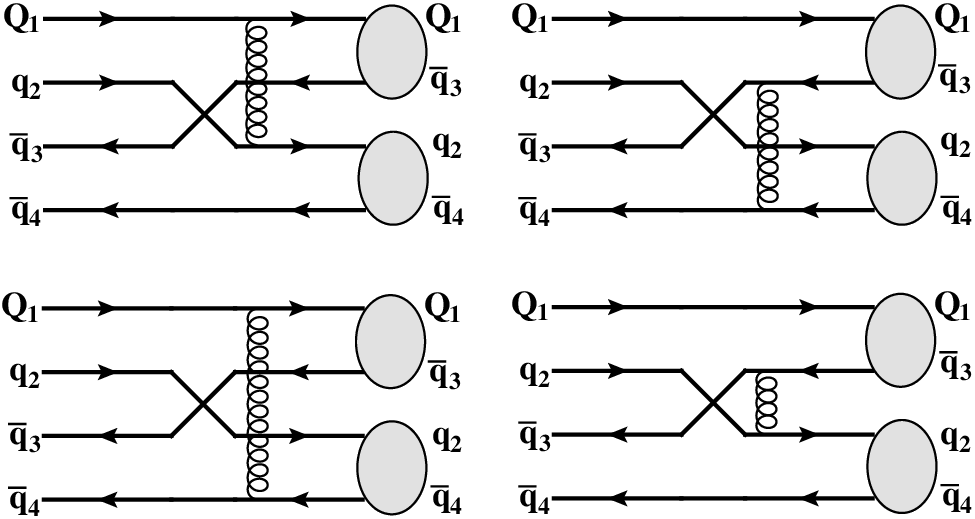}
\caption{The fall-apart decays of a singly-heavy tetraquark state induced by the interactions
$V_{ij} (ij\neq 13, 24$ or $ij\neq 14, 23)$ between inner quarks of final hadrons
$B$ and $C$.}
\label{fig:fall_apart_decay}
\end{figure}

\subsection{\label{sec:citeref}Fall-apart decays}
	
In this work, we also investigate the decay properties of singly heavy tetraquark systems within the framework of the quark-exchange model~\cite{Barnes:1991em,Barnes:2000hu}. Specifically, we analyze the decay processes of $Q_1q_2\bar{q}_3\bar{q}_4$ system, such as
	\[
	A(Q_1q_2\bar{q}_3\bar{q}_4  )\longrightarrow B(Q_1\bar{q}_3 )+C(q_2\bar{q}_4 ) ~\mathrm{or}~ B(Q_1\bar{q}_4 )+C(q_2\bar{q}_3 ),
	\]
which proceed via a quark-exchange mechanism. The exchange process is assumed to be induced by the OGE potential as shown in Fig. \ref{fig:fall_apart_decay}. In these processes, constituent quarks (or antiquarks) from the initial compact tetraquark state are rearranged to form two color-singlet mesons in the final states. The decay amplitude for the decay process $A\to  BC$ is expressed as
	\begin{equation}
		\mathcal{M}(A\to  BC)=-\sqrt{(2\pi)^3}\sqrt{8M_AE_BE_C}\left \langle BC  \Bigg|\sum\limits_{i\in B,j\in C} V_{ij}^{OGE} \Bigg|A  \right \rangle,
	\end{equation}
where $V_{ij}^{OGE}$ is the OGE potential given by Eq.~(\ref{eq:conf}). $M_A$ represents the mass of the initial tetraquark state, while $E_B$ and $E_C$ denote the energies of the final hadrons $B$ and $C$, respectively, in the rest frame of the initial hadron.
With the decay amplitudes, the partial decay width of $A\longrightarrow BC$ is determined by
	\begin{equation}
		\Gamma =\frac{1}{2J_A+1}\frac{|\boldsymbol{q}|}{8\pi M_A^2}\Big |\mathcal{M}(A\to BC)\Big |^2
	\end{equation}	
where $\boldsymbol{q}$ is the three-vector momentum of the final state $B$ or $C$ in the initial-hadron-rest frame.
	
This phenomenological model has achieved a good description of the low-energy $S$-wave phase shift for the $\pi\pi$ scattering at the quark level~\cite{Barnes:1991em,Barnes:2000hu}. Recently, this model has been extensively employed in the study of fall-apart decays of the multiquark systems in the literature~\cite{Wang:2018pwi,Wang:2020prk,Wang:2019spc,Xiao:2019spy,Han:2022fup,Zhou:2019swr,Liu:2026ljb,
Liang:2024met,Liu:2022hbk,Liu:2024fnh,liu:2020eha,An:2026xlu}. Considering the complexity of the initial tetraquark
states, we take the numerical wave functions obtained from our potential-model calculations.
For simplicity, the wave functions for the relatively simple final $B$ and $C$ meson states take a single harmonic oscillator (SHO) form. Their SHO parameters are determined by fitting the root mean square radii, which are obtained from our potential model calculations
with the same Hamiltonian given in Eq. (\ref{eq:hamiltonian}). Our determined root mean square (RMS) radii and SHO parameters for the
final meson and baryon states are collected in Table~\ref{tab:mesons}.
For the well-established meson states in the final states, the masses are taken from the PDG averaged
values~\cite{ParticleDataGroup:2024cfk}, which are collected Table~\ref{tab:mesons}. For the initial tetraquark states, the masses are taken from our potential model predictions.

\section{Results and Discussions}\label{sec:results}
	
In this section, we present and discuss the mass spectra and fall-apart decay properties of the $1S$-wave states
for the singly heavy tetraquark systems, $Qn\bar{n}\bar{n}$, $Qs\bar{n}\bar{n}$, $Qn\bar{s}\bar{n}$,
$Qs\bar{s}\bar{n}$, $Qn\bar{s}\bar{s}$, $Qs\bar{s}\bar{s}$.
The predicted mass spectra are given in Tables~\ref{tab:table_Qnnn_radii_merged}-\ref{tab:table_Qsss_radii_merged},
while the predicted fall-apart decay properties are given in Tables~\ref{tab:table_Qnnn_decay_merged}-\ref{tab:table_Qsss_decay_merged}.

Several global characters of the obtained $1S$-wave singly-heavy tetraquark states can be seen:
(i) they are compact states and lie far above the lowest dissociation meson-meson threshold. The root-mean-square distances
between any two quarks are predicted to be in the range of $\sim 0.4-0.7$~fm. (ii)
The $J^P=2^+$ states have extremely narrow fall-apart widths of $\mathcal{O}(1)$~MeV,
while most of the $J^P=0^+,1^+$ states also have narrow widths of a few tens MeV. It indicates
that these singly-heavy tetraquarks may be stable enough to be observed in future experiments.
(iii) For the $J^P=0^+$ and $1^+$ states,
there is significant mixing between different configurations.
The off-diagonal Hamiltonian matrix elements, which induce the configuration mixing,
are predominantly contributed by the color-magnetic spin-spin interaction.
(iv) For the mixed doublet with $J^P = 0^+$, there is a fairly
large mass splitting, $\sim 200-600$~MeV, which is caused by the large off-diagonal Hamiltonian matrix elements between
the two color configurations $| \mathbf{6}\bar{\mathbf{6}} \rangle$ and $| \bar{\mathbf{3}}\mathbf{3} \rangle$.
These large off-diagonal matrix elements arise from the constructive contributions of the color-magnetic interactions between any two quarks, i.e., the same signs of the color-spin factors $\langle \boldsymbol{\lambda}_i^c \cdot \boldsymbol{\lambda}_j^c\rangle  \langle\boldsymbol{S}_i \cdot \boldsymbol{S}_j \rangle$ ($i<j$).

Furthermore, to see the dynamic roles of various interactions from OGE and OBE in
different configurations of each tetraquark system, we present their average values in Tables~\ref{tab:merged_cnnn_bnnn}-\ref{tab:merged_csss_bsss} as an appendix. For the OGE sector, the interactions have few dependencies on the quark flavors,
and play similar roles in different tetraquark systems. Concretely, from Tables~\ref{tab:merged_cnnn_bnnn}-\ref{tab:merged_csss_bsss} one can see that (i) the color-Coulomb interaction plays a dominant role in quark-(anti)quark interactions; it provides a strong attractive potential, $\langle V^{Coul}\rangle \sim -(700\pm 150)$~MeV, for each configuration.
(ii) The spin-spin color-magnetic interaction plays a crucial role in the configurations belonging to $\bar{\mathbf{3}}_F$, $\left| (Q_1q_2)^{\bar{\mathbf{3}}}_0 [\bar{q}_3\bar{q}_4]^{\mathbf{3}}_0 \right\rangle_0$, $\left| (Q_1q_2)^{\mathbf{6}}_1 [\bar{q}_3\bar{q}_4]^{\bar{\mathbf{6}}}_{1} \right\rangle_{0,1}$, and $\left| (Q_1q_2)^{\bar{\mathbf{3}}}_1 [\bar{q}_3\bar{q}_4]^{\mathbf{3}}_0 \right\rangle_1$,
the contributions to these configurations can reach up to $\langle V^{SS}\rangle\sim(-500,-150)$~MeV.
(iii) Except the configurations mentioned above, the magnitude of the linear potential,
$V^{conf}\propto \sum_{i<j} \boldsymbol{\lambda}_i^c \cdot \boldsymbol{\lambda}_j^c(b_{ij}r_{ij}+C_{ij})$, is often no more than 100~MeV due
to the large cancellation between the confinement term $b_{ij}r_{ij}$ and zero energy $C_{ij}$.

While for the OBE sector, from Tables~\ref{tab:merged_cnnn_bnnn}-\ref{tab:merged_csss_bsss}
one can see that (i) the $\sigma$ exchanges contribute a large background $\langle V^{\sigma}\rangle\sim -(60\pm 10)$~MeV,
which is a relatively stable value for each configuration, and can be absorbed in the other parameters,
such as constituent quark mass and zero energy. Thus, the presence or absence of the $\sigma$ exchange has
few effects on the spectral properties. (ii) The $\eta$ and $\eta'$ exchanges are less important; the magnitude of the potentials is often
no more than 10~MeV. (iii) The contributions from the other meson ($\pi$, $K$, $\omega$, $\rho$, $K^*$, $\phi$) exchanges strongly depend
on the light quark flavors, their roles will be discussed by combining the specific tetraquark systems.	
			
More special and detailed discussions about the $Qn\bar{n}\bar{n}$, $Qs\bar{n}\bar{n}$, $Qn\bar{s}\bar{n}$,
$Qs\bar{s}\bar{n}$, $Qn\bar{s}\bar{s}$, $Qs\bar{s}\bar{s}$ systems are given as follows one by one.

\begin{figure}[htbp]
 \centering \epsfxsize=7.0 cm \epsfbox{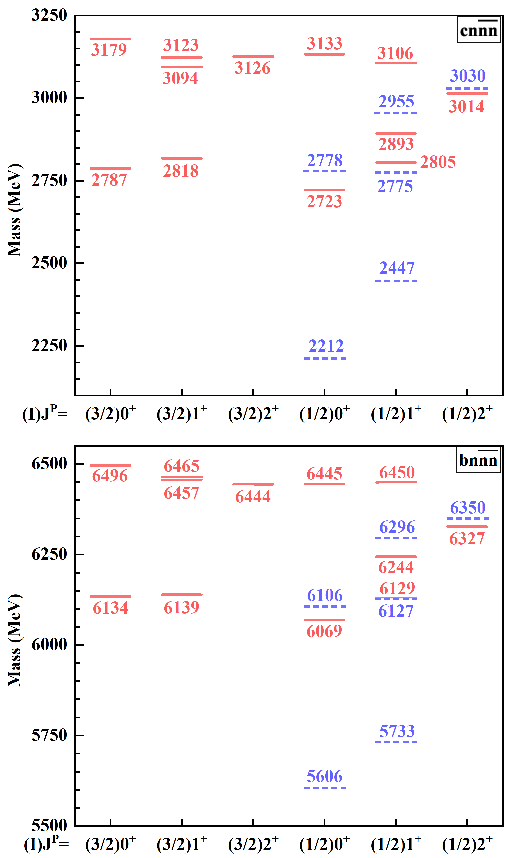}
 \caption{Mass spectra of the $1S$-wave states for the $cn\bar{n}\bar{n}$ (upper panel) and $bn\bar{n}\bar{n}$ (lower panel) systems.
 The red solid and blue dashed lines represent the states with flavor symmetries belonging to $\mathbf{6}_F$ and $\bar{\mathbf{3}}_F$, respectively.
}\label{fig:mass_Qnnn}
\end{figure}

\subsection{$Qn\bar{n}\bar{n}$ system}\label{Q3n}

For the $Qn\bar{n}\bar{n}$ system, according to the quark model classification as shown in Table~\ref{tab:1S_configurations}
there are eighteen $1S$-wave configurations: six ones with $I=3/2$ belonging
to $\mathbf{6}_F$, six ones with $I=1/2$ belonging to $\mathbf{6}_F$,
and six ones with $I=1/2$ belonging to $\bar{\mathbf{3}}_F$.


From Table~\ref{tab:merged_cnnn_bnnn}, one can see
the dynamic roles of the OBE potentials from the $\pi$, $\omega$, $\rho$ exchanges,
which are summarized as follows: (i) the $\pi$-exchange term plays an important role in several
special configurations with the $\bar{3} 3$ color structure, $\left| (Qn)^{\bar{\mathbf{3}}}_1 \{\bar{n}\bar{n}\}^{\mathbf{3}}_1 \right\rangle_0$, $\left| (Qn)^{\bar{\mathbf{3}}}_0 [\bar{n}\bar{n}]^{\mathbf{3}}_0 \right\rangle_0$ and $\left| (Qn)^{\bar{\mathbf{3}}}_1 [\bar{n}\bar{n}]^{\mathbf{3}}_0 \right\rangle_1$,
the magnitude of the contributions can reach up to $|\langle V^{\pi}\rangle|\sim 160$~MeV.
For the other configurations, the contribution of the $\pi$ exchange is less important.
(ii) For the $I=3/2$ configurations, due to significant cancelation among the $\omega$- and $\rho$-exchange terms,
the total contributions from the vector meson exchanges are less important. However, for the
$I=1/2$ ones, the $\omega$- and $\rho$-exchange terms play important roles.
For most of the configurations, they contribute a large attractive potential $\langle V^{\rho}\rangle+\langle V^{\omega}\rangle\sim (-200,-100)$~MeV.

It should be mentioned that the various OGE potential models widely employed in the literature (e.g. Refs.~\cite{Guo:2021mja,Lu:2020qmp}),
due to a lack of flavor-dependent interactions, yield a degenerate mass spectrum for
the $\mathbf{6}_F$ representation with different isospins, $I=3/2$ and $1/2$.
In the present work, we incorporate the OBE potentials whose flavor-dependent operators $\mathcal{I}_{\chi}$ and $\mathcal{I}_{v}$ give rise to distinct matrix elements for different isospin configurations.
As a result, we obtain two completely different spectra for the $\mathbf{6}_F$ representation with
isospin $I=3/2$ and $I=1/2$, which have been shown in Fig.~\ref{fig:mass_Qnnn}.

\subsubsection{$cn\bar{n}\bar{n}$ system}

For the $cn\bar{n}\bar{n}$ system, our predicted masses for the twelve states belonging to
$\mathbf{6}_F$ with $I=3/2$ and $I=1/2$ scatter in the range
of $\sim 2800-3200$~MeV, and $\sim 2700-3100$~MeV, respectively. While the predicted masses for the six states with $I=1/2$
belonging to $\bar{\mathbf{3}}_F$ scatter in a lower mass range
of $\sim 2200-3000$~MeV. The detailed behavior of the spectrum can be seen in Table~\ref{tab:table_Qnnn_radii_merged} and Fig.~\ref{fig:mass_Qnnn}.
A few studies of the $Qn\bar{n}\bar{n}$ system can be found in the literature~\cite{Terasaki:2004yx,Wu:2023hhk,Guo:2021mja,Jovanovic:2007bz,Lu:2016zhe,Ebert:2010af,Lu:2020qmp,Chen:2017rhl,Jalili:2023kmw}.
There are strong model dependencies in the predictions. For example, our predicted spectrum lies
$\sim 100-400$~MeV below that predicted within the GI model~\cite{Lu:2020qmp}, while $\sim 200-300$~MeV
above that predicted within the improved chromomagnetic interaction (ICMI) model~\cite{Guo:2021mja}.

It should be emphasized that in the isospin $I=1/2$ states belonging to $\bar{\mathbf{3}}_F$, there are two low-mass states with $I=1/2$,
$T_{(cn[\bar{n}\bar{n}])0^+}^{1/2}(2212)$ and $T_{(cn[\bar{n}\bar{n}])1^+}^{1/2}(2447)$, whose masses are about 400~MeV smaller
than the results of similar dynamic calculations based on the OGE potentials~\cite{Lu:2020qmp}. Their low mass nature is mainly caused by
the strong attractive interactions from the $\pi$ and $\rho$ exchanges together with a
large contribution of the color-magnetic interactions. Furthermore, the configuration mixing also
significantly lowers their masses. The mass of the scalar state $T_{(cn[\bar{n}\bar{n}])0^+}^{1/2}(2212)$ predicted in the present
work is consistent with that obtained by using the Glozman-Riska hyperfine interaction in Ref.~\cite{Jovanovic:2007bz}.
It is interesting to find that the tetraquark states $T_{(cn[\bar{n}\bar{n}])0^+}^{1/2}(2212)$ and $T_{(cn[\bar{n}\bar{n}])1^+}^{1/2}(2447)$ predicted in the present work significantly overlap with the observed resonances, $D_0(2300)$ and $D_1(2420,2430)$ listed in RPP~\cite{ParticleDataGroup:2024cfk}.
Our finding is consistent with that obtained in the framework of the relativistic quark model~\cite{Ebert:2010af}.


To provide useful information for future observations in experiments,
the fall-apart decay properties of the obtained states for the $cn\bar{n}\bar{n}$ system are studied.
Our results are given in Table~\ref{tab:table_Qnnn_decay_merged}. It is found that the scalar tetraquark state
$T_{(cn[\bar{n}\bar{n}])0^+}^{1/2}(2212)$ has a relatively narrow fall-apart width of $\sim 41$~MeV, which is saturated by the
$D\pi$ channel. Although the predicted mass and decay mode of $T_{(cn[\bar{n}\bar{n}])0^+}^{1/2}(2212)$ are
consistent with the observations of $D_0(2300)$, the predicted decay width is too narrow to be comparable
with the observed one, $\Gamma_{exp}=229\pm 16$~MeV~\cite{ParticleDataGroup:2024cfk}.
Thus, our results do not support the $D_0(2300)$ resonance as a tetraquark assignment as suggested in Ref.~\cite{Jovanovic:2007bz}.
It may be interesting to search for the narrow $0^+$ state $T_{(cn[\bar{n}\bar{n}])0^+}^{1/2}(2212)$ in the $D^+\pi^-$ and $D^0\pi^+$ final states around the mass range of $2.1-2.3$~GeV.

While for the axial vector state $T_{(cn[\bar{n}\bar{n}])1^+}^{1/2}(2447)$, the fall-apart decay is saturated by the
$D^*\pi$ channel. Both its decay width $\Gamma\simeq 45$~MeV and mass $M\simeq 2447$~MeV
are comparable with the observations of the well-established resonance $D_1(2420)$~\cite{ParticleDataGroup:2024cfk}.
Thus, it raises a puzzle that whether the $D_1(2420)$ is conventional meson state
in the $D$ meson family, or a compact tetraquark state, or an admixture between both of them.
To clarify the nature of $D_1(2420)$, more studies are needed to be carried out in both theory and experiments.

Furthermore, some other high-lying tetraquark states may have good discovery potentials
in their dominant decay channels as shown in Table~\ref{tab:table_Qnnn_decay_merged}.
For example, the isospin $I=3/2$ scalar state $T_{(cn\{\bar{n}\bar{n}\})0^+}^{3/2}(2787)$ has narrow fall-apart
decay width of $\sim 49$~MeV, which is saturated by the $D\pi$ and $D^*\rho$ channels with a partial width ratio
\begin{eqnarray}
	\frac{\Gamma[T_{(cn\{\bar{n}\bar{n}\})0^+}^{3/2}(2787)\to D\pi]}{\Gamma[T_{(cn\{\bar{n}\bar{n}\})0^+}^{3/2}(2787)\to D^*\rho]}\simeq 4.7.
\end{eqnarray}
The $T_{(cn\{\bar{n}\bar{n}\})0^+}^{3/2}(2787)$ composed of $cu\bar{d}\bar{d}/cd\bar{u}\bar{u}$ may have good potentials to be observed
in the $D^+\pi^+/D^0\pi^-$ final state. While for the $I=1/2$ scalar states $T_{(cn\{\bar{n}\bar{n}\})0^+}^{1/2}(2723)$ and
$T_{(cn[\bar{n}\bar{n}])0^+}^{1/2}(2778)$, the fall-apart decay widths are predicted to be $\sim 32$~MeV and
$\sim 17$~MeV, respectively. They have large decay rates into the $D\pi$ and $D\eta$ channels
with partial width ratios
\begin{eqnarray}
	\frac{\Gamma[T_{(cn\{\bar{n}\bar{n}\})0^+}^{1/2}(2723)\to D\pi]}{\Gamma[T_{(cn\{\bar{n}\bar{n}\})0^+}^{1/2}(2723)\to D\eta]}&\simeq & 0.53,\\
    \frac{\Gamma[T_{(cn[\bar{n}\bar{n}])0^+}^{1/2}(2778)\to D\pi]}{\Gamma[T_{(cn[\bar{n}\bar{n}])0^+}^{1/2}(2778)\to D\eta]}&\simeq & 5.0.
\end{eqnarray}
The $T_{(cn\{\bar{n}\bar{n}\})0^+}^{1/2}(2723)$ and $T_{(cn[\bar{n}\bar{n}])0^+}^{1/2}(2778)$ may have potentials to be observed
in the $D\pi$ and $D\eta$ channels.

Finally, it should be mentioned that the tensor states of the $cn\bar{n}\bar{n}$ system lie in the high mass range of $\sim 3.0-3.2$~GeV.
They may be extremely narrow states with fall-apart decay widths of $\sim 1$~MeV.
They dominantly decay into the final states containing two vector mesons, such as $D^*\rho$.

\subsubsection{$bn\bar{n}\bar{n}$ system}


For the $bn\bar{n}\bar{n}$ system, the predicted masses of various states belonging to
the $\mathbf{6}_F$ representation with $I=3/2$ and $I=1/2$ scatter in the range
of $\sim 6100-6500$~MeV, and $\sim 6100-6450$~MeV, respectively. While the predicted masses of the $I=1/2$
states belonging to $\bar{\mathbf{3}}_F$ scatter in a lower mass range
of $\sim 5600-6350$~MeV. The detailed behavior of the spectrum can be seen in Table~\ref{tab:table_Qnnn_radii_merged} and Fig.~\ref{fig:mass_Qnnn}.
There are a few studies of the $bn\bar{n}\bar{n}$ system in the literature~\cite{Wu:2023hhk,Guo:2021mja,Lu:2016zhe,Ebert:2010af,Lu:2020qmp,Chen:2017rhl,Jalili:2023kmw}.
Significant model dependencies exist in the predictions. For example, our predicted spectrum lies
$\sim 200-400$~MeV below that predicted within the GI model~\cite{Lu:2020qmp}, while $\sim 200-300$~MeV
above that predicted within the improved ICMI model~\cite{Guo:2021mja}.

There are two low mass states with $I=1/2$,
$T_{(bn[\bar{n}\bar{n}])0^+}^{1/2}(5606)$ and $T_{(bn[\bar{n}\bar{n}])1^+}^{1/2}(5733)$, as the partners of $T_{(cn[\bar{n}\bar{n}])0^+}^{1/2}(2212)$ and $T_{(cn[\bar{n}\bar{n}])1^+}^{1/2}(2447)$ of the charmed sector, whose masses overlap significantly with conventional orbitally excited $B$ meson states, $B_0(5573)$ and $B_1(5731)$ predicted within an unquenched quark model~\cite{Ni:2023lvx}. The $B_0(5573)$, as a broad state predicted within the quark model, has not been established, while the $B_1(5731)$ is often assigned to the observed narrow resonance $B_1(5721)$ listed in the Review of Particle Physics (RPP)~\cite{ParticleDataGroup:2024cfk}.
According to our predictions as shown in Table~\ref{tab:table_Qnnn_decay_merged}, both the $T_{(bn[\bar{n}\bar{n}])0^+}^{1/2}(5606)$ and $T_{(bn[\bar{n}\bar{n}])1^+}^{1/2}(5733)$ may have
narrow fall-apart decay widths of $\sim 20-50$~MeV, which are nearly saturated by the
$B\pi$ and $B^*\pi$ channels, respectively. It may be interesting to search for the narrow
scalar state $T_{(bn[\bar{n}\bar{n}])0^+}^{1/2}(5606)$ in the $B^-\pi^+$ and $B^0\pi^-$ final states.
However, for the axial vector state $T_{(bn[\bar{n}\bar{n}])1^+}^{1/2}(5733)$, our predicted mass, width, and decay mode
are also consistent with the observations of $B_1(5721)$~\cite{ParticleDataGroup:2024cfk}. Thus, it is still questionable
about the nature of the $1^+$ resonance $B_1(5721)$ usually classified as the $1P$ wave state in the $B$ meson family.
The $B_1(5721)$ may be an admixture of conventional $q\bar{q}$ excitations and four-quark configurations.
To clarify the nature of $B_1(5721)$, more observations, such as the radiative decays, are
needed to be carried out in future experiments.


From Table~\ref{tab:table_Qnnn_decay_merged}, it is seen that all of the $1S$-wave states
have narrow fall-apart decay widths, which scatter in the range of $\sim 1-82$~MeV.
Except for the two low-lying states $T_{(bn[\bar{n}\bar{n}])0^+}^{1/2}(5606)$ and $T_{(bn[\bar{n}\bar{n}])1^+}^{1/2}(5733)$,
several states in the $bn\bar{n}\bar{n}$ system may have good discovery potentials
in future experiments. Such as, the scalar state $T_{(bn\{\bar{n}\bar{n}\})0^+}^{3/2}(6134)$ with $I=3/2$,
as the bottom partner of $T_{(cn\{\bar{n}\bar{n}\})0^+}^{3/2}(2787)$,
has a fall-apart width of $\sim 66$~MeV, and dominantly decays into the $B\pi$ and $B^*\rho$ channels
with a partial width ratio
\begin{eqnarray}
	\frac{\Gamma[T_{(bn\{\bar{n}\bar{n}\})0^+}^{3/2}(6134)\to B\pi]}{\Gamma[T_{(bn\{\bar{n}\bar{n}\})0^+}^{3/2}(6134)\to B^*\rho]}&\simeq & 1.0.
\end{eqnarray}
This state composed of $bd\bar{u}\bar{u}/bu\bar{d}\bar{d}$ may be likely
established in the $B^-\pi^-/B^0\pi^+$ final state.
Furthermore, two scalar states $T_{(bn\{\bar{n}\bar{n}\})0^+}^{1/2}(6069)$
and $T_{(bn[\bar{n}\bar{n}])0^+}^{1/2}(6106)$ with $I=1/2$, as bottom partners of $T_{(cn\{\bar{n}\bar{n}\})0^+}^{1/2}(2723)$ and $T_{(cn[\bar{n}\bar{n}])0^+}^{1/2}(2778)$, are predicted to be narrow states with widths of
$\sim 26$~MeV and $\sim 42$~MeV, respectively. They have large decay rates into the $B\pi$ and $B\eta$ channels.
The partial width ratios are predicted to be
\begin{eqnarray}
	\frac{\Gamma[T_{(bn\{\bar{n}\bar{n}\})0^+}^{1/2}(6069)\to B\pi]}{\Gamma[T_{(bn\{\bar{n}\bar{n}\})0^+}^{1/2}(6069)\to B\eta]}&\simeq & 0.53,\\
    \frac{\Gamma[T_{(bn[\bar{n}\bar{n}])0^+}^{1/2}(6106)\to B\pi]}{\Gamma[T_{(bn[\bar{n}\bar{n}])0^+}^{1/2}(6106)\to B\eta]}&\simeq & 5.0.
\end{eqnarray}
The $T_{(bn\{\bar{n}\bar{n}\})0^+}^{1/2}(6069)$ and $T_{(bn[\bar{n}\bar{n}])0^+}^{1/2}(6106)$ may have potentials to be observed in the $B\pi$ and $B\eta$ channels.

Finally, it should be mentioned that the tensor states of the $bn\bar{n}\bar{n}$ system lie in the high mass range of $\sim 6.3-6.4$~GeV.
Their fall-apart decay widths are predicted to be very narrow, the values are only a few MeV.
They dominantly decay into the final states containing two vector mesons, such as $B^*\rho$.

\begin{figure*}[htbp]
 \centering \epsfxsize=13.6 cm \epsfbox{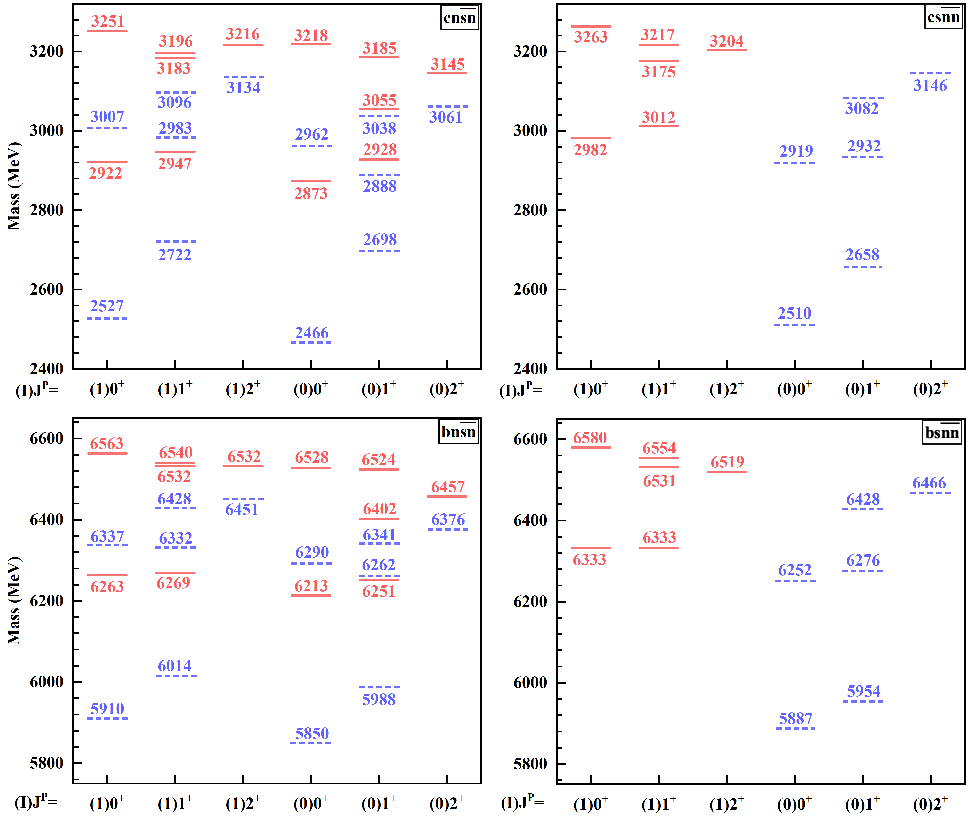}
 \caption{Mass spectra of the $1S$-wave states for the $Qn\bar{s}\bar{n}$ (left panel) and $Qs\bar{n}\bar{n}$ (right panel) systems. The red solid and blue dashed lines represent the states with flavor symmetries belonging to $\mathbf{6}_F$ and $\bar{\mathbf{3}}_F$, respectively.}\label{fig:Mass_Qnsn_Qsnn}
\end{figure*}

\subsection{$Qn\bar{s}\bar{n}$ system}

For the $Qn\bar{s}\bar{n}$ system, there are six $1S$-wave configurations with isospin $I=1$ and six ones with isospin $I=0$ belonging to the $\mathbf{6}_F$ representation. There are also six $1S$-wave configurations with isospin $I=1$ and six ones with isospin $I=0$ belonging to the $\bar{\mathbf{3}}_F$ representation.

From Table~\ref{tab:merged_cnsn_bnsn} given in the appendix, one can see the dynamic roles of the OBE potentials from the $\pi$, $K$, $\omega$, $\rho$, and $K^*$ exchanges, which are summarized as follows: (i) the $\pi$ exchange is less important compared with that in the $Qn\bar{n}\bar{n}$ system. It only has a sizable contribution, $|\langle V^{\pi}\rangle|\sim 20-50$~MeV, to several
special configurations, $\left| (Qn)^{\bar{\mathbf{3}}}_1 \{\bar{s}\bar{n}\}^{\mathbf{3}}_1 \right\rangle_{0,1,2}$, $\left| (Qn)^{\mathbf{6}}_1 [\bar{s}\bar{n}]^{\bar{\mathbf{6}}}_{1} \right\rangle_{0}$, and $\left| (cn)^{\mathbf{6}}_1 \{\bar{s}\bar{n}\}^{\bar{\mathbf{6}}}_{1} \right\rangle_{0,1,2}$. For the other configurations, the contribution of the $\pi$ exchange is negligibly small.
(ii) the $K$ exchange between $s$ and $u/d$ quarks also contributes a sizable attractive potential, $\langle V^{K}\rangle\sim -30$~MeV, to several
special configurations, $\left| (Qn)^{\bar{\mathbf{3}}}_0 [\bar{s}\bar{n}]^{\mathbf{3}}_0 \right\rangle_{0}$ and
$\left| (Qn)^{\bar{\mathbf{3}}}_1 [\bar{s}\bar{n}]^{\mathbf{3}}_0 \right\rangle_{1}$ with isospin $I=0$ and $1$.
For the other configurations, the contribution of the $K$ exchange is negligibly small.
(iii) For the $I=1$ states, due to a nearly exact cancelation among the $\omega$- and $\rho$-exchange terms,
the remaining vector meson contributions are about a few tens MeV, which come from the $K^*$ meson exchanges. However, in the
$I=0$ states the $\omega$-, $\rho$-, and $K$-exchange terms play important roles.
For most of the configurations belonging to $\bar{\mathbf{3}}_F$, there is a large attractive potential $\langle V^{\rho}\rangle+\langle V^{\omega}\rangle+\langle V^{K^*}\rangle\sim (-200,-100)$~MeV.

It should be mentioned that in the OGE potential models (e.g. Refs.~\cite{Liu:2022hbk,Lu:2020qmp}), the mass spectra
obtained for the same flavor representation ($\mathbf{6}_F$ or $\bar{\mathbf{3}}_F$)
with different isospins $I=0$ and $1$ are degenerate with each other, due to the lack of
isospin-dependent interactions. However, in the present work, we include the isospin-dependent
OBE potentials, thus, we obtain two completely different spectra for the same flavor representation with
isospin $I=0$ and $1$.

\subsubsection{$cn\bar{s}\bar{n}$ system}\label{Tcs2900}

The $cn\bar{s}\bar{n}$ is an interesting system which relates to several long standing controversial
states, such as $D_{s0}(2317)$ and $D_{s1}(2460)$ in the $D_s$ meson family~\cite{ParticleDataGroup:2024cfk}, as well as the newly
observed exotic states $T_{c\bar{s}0}^a(2900)^{++,0}$~\cite{LHCb:2022lzp,LHCb:2022sfr} and $T_{c\bar{s}}(2327)^{++,0}$~\cite{LHCb:2024iuo} at LHCb. These resonances may be candidates of the tetraquark states composed of $cn\bar{s}\bar{n}$ as discussed in the literature.
For the $cn\bar{s}\bar{n}$ system, our predicted masses of various states belonging to
the $\mathbf{6}_F$ representation scatter in a relatively narrow mass range
of $\sim 2900-3250$~MeV. While the predicted masses of the states belonging to $\bar{\mathbf{3}}_F$ scatter in a broad mass range
of $\sim 2400-3200$~MeV. The detailed behavior of the spectrum can be seen in Table~\ref{tab:table_Qnsn_radii_merged} and Fig.~\ref{fig:Mass_Qnsn_Qsnn}.

In the $cn\bar{s}\bar{n}$ system, the lowest state is the scalar state belonging to $\bar{\mathbf{3}}_F$ with $I=0$,
$T_{(cn[\bar{s}\bar{n}])0^+}^{0}(2466)$, which is a $|(cn)_1^6 [\bar{s}\bar{d}]_1^{\bar{6}}\rangle_0$
dominant state ($\sim 65\%$) with a sizeable $|(cn)_0^{\bar{3}} [\bar{s}\bar{d}]_0^3\rangle_0$ mixing ($\sim 35\%$).
The mass, $\sim 2466$~MeV, predicted in the present work is about $400$~MeV lower than our
previous prediction with the OGE potential model~\cite{Liu:2022hbk}, which is mainly caused by the strong
attractive interactions from the vector meson ($\rho$, $\omega$, and $K^*$) exchanges.
In the literature, the $D_{s0}(2317)$ was suggested to be a scalar $cn\bar{s}\bar{n}$ state based on various
phenomenological analysis~\cite{Zhang:2018mnm,Chen:2026wrh,Cheng:2003kg,Chen:2004dy,Maiani:2004vq,Terasaki:2003qa}.
Obviously, our study does not
support this assignment. Compared with our prediction in the present work,
a larger mass, $\sim 2.6-2.9$~GeV, for the lowest scalar $cn\bar{s}\bar{n}$ state
is predicted within the other quark models~\cite{Ebert:2010af,Wei:2022wtr,Lu:2020qmp,Xue:2020vtq} and QCD sum rules~\cite{Yang:2023evp,Chen:2017rhl,Albuquerque:2020ugi}. It should be mentioned that
an extremely small mass $\sim 2.2$~GeV was obtained within the chromo-magnetic interaction model~\cite{Cheng:2020nho,Guo:2021mja}.
We also study the fall-apart decay properties; our results show that
the $T_{(cn[\bar{s}\bar{n}])0^+}^{0}(2466)$ may be a narrow state with a width of $\sim 29$~MeV.
It may have good potentials to be observed in the dominant $DK$ ($D^+K^0+D^0K^+$) channel.

Another interesting state in the $cn\bar{s}\bar{n}$ system is the lowest axial state
with isospin $I=0$, which may be a candidate of $D_{s1}(2460)$ as suggested in the
literature, e.g. Refs.~\cite{Chen:2026wrh,Chen:2004dy}. In the present work, the lowest axial state
with isospin $I=0$ is predicted to be a mixed state with dominant components,
$|(cn)_1^6 [\bar{s}\bar{n}]_1^{\bar{6}}\rangle_1$ ($\sim 45\%$) and $|(cn)_1^{\bar{3}} [\bar{s}\bar{n}]_0^3\rangle_1$ ($\sim 45\%$),
denoted by $T_{(cn[\bar{s}\bar{n}])1^+}^{0}(2698)$. Whose mass, $\sim 2698$~MeV, is about $300$~MeV smaller than our
previous predictions only including the OGE potentials~\cite{Liu:2022hbk}. It is found that the
vector meson $\rho$ exchange plays crucial roles in this axial vector state.
Our analysis of the decay properties shows that the $T_{(cn[\bar{s}\bar{n}])1^+}^{0}(2698)$
may have a narrow fall-apart width of $\sim 36$~MeV, which is saturated by the $D^*K$ channel.
To establish the $T_{(cn[\bar{s}\bar{n}])1^+}^{0}(2698)$, the $D^0K^+$ channel is worth observing
in future experiments.

The $T_{c\bar{s}0}^a(2900)^{++,0}$ newly observed in $D_s^+\pi^{\pm}$ at LHCb~\cite{LHCb:2022lzp,LHCb:2022sfr}
seems to favor the lowest scalar tetraquarks belonging to $\mathbf{6}_F$ with isospin $I=1$, $T_{(cn\{\bar{s}\bar{n}\})0^+}^{1}(2922)$,
predicted in the present work. It is a $|(cn)_1^{\bar{3}} \{\bar{s}\bar{n}\}_1^3\rangle_0$
dominant state with a significant $|(cn)_0^6 \{\bar{s}\bar{n}\}_0^{\bar{6}}\rangle_0$ mixing.
We further analyze the decay properties, the results are given in Table~\ref{tab:table_Qnsn_decay_merged}. It is seen that the $T_{(cn\{\bar{s}\bar{n}\})0^+}^{1}(2922)$ has a
relatively narrow fall-apart width of $\sim 53$~MeV, which dominantly decays into the $DK$, $D_s\pi$, $D_s^*\rho$, and $D^*K^*$ channels. Our predicted partial width ratio between $DK$ and $D_s\pi$,
\begin{eqnarray}
R_{DK/D_s\pi}=\frac{\Gamma[T_{(cn\{\bar{s}\bar{n}\})0^+}^{1}(2922)\to DK]}{\Gamma[T_{(cn\{\bar{s}\bar{n}\})0^+}^{1}(2922)\to D_s\pi]}\simeq 1.6,
\end{eqnarray}
is slightly larger than the ratio $\sim 1.1$ predicted with QCD sum rules~\cite{Lian:2023cgs}.
The $T_{(cn\{\bar{s}\bar{n}\})0^+}^{1}(2922)$ as an assignment
of $T_{c\bar{s}0}^a(2900)$, except the predicted decay width is slightly narrower than the data $\mathcal{O}(100)$~MeV, the mass, decay mode, and quantum numbers are consistent with the observations. Our assignment is consistent with that of Refs.~\cite{Mutuk:2025hql,Gordillo:2025caj,Wei:2022wtr,Ortega:2023azl,Lian:2023cgs}.
To confirm the nature of $T_{c\bar{s}0}^a(2900)$, the $D^{0,+}K^+$ final states together with the partial width ratio
$R_{DK/D_s\pi}$ is worth observing in future experiments.

It should be mentioned that in our previous study based on the OGE model~\cite{Liu:2022hbk},
the $T_{c\bar{s}0}^a(2900)$ was assigned to the lowest scalar tetraquark state belonging to $\mathbf{\bar{3}}_F$ with isospin $I=1$,
where a higher mass, $\sim 2830$~MeV, was obtained due to the lack of OBE potentials. However, in the present work,
the mass of this state is predicted to be $\sim 2527$~MeV, denoted by $T_{(cn [\bar{s}\bar{n}])0^+}^{1}(2527)$.
It is a mixed state dominated by the $|(cn)_1^6 [\bar{s}\bar{n}]_1^{\bar{6}}\rangle_0$ component.
The low mass nature is mainly caused by the strong attractive interactions from the $\rho$ and $\omega$ exchanges.
While its high mass partner $T_{(cn [\bar{s}\bar{n}])0^+}^{1}(3007)$ is dominated by the
$|(cn)_0^{\bar{3}} \{\bar{s}\bar{n}\}_0^3\rangle_0$ component.
As shown in Table~\ref{tab:table_Qnsn_decay_merged}, both $T_{(cn [\bar{s}\bar{n}])0^+}^{1}(2527)$
and $T_{(cn [\bar{s}\bar{n}])0^+}^{1}(3007)$ are relatively broad states with widths
of $\sim 60$ and $\sim 117$~MeV, respectively.
For the $T_{(cn [\bar{s}\bar{n}])0^+}^{1}(2527)$, the fall-apart width is nearly saturated by the $DK$ and
$D_s\pi$ channels. The partial width ratio is predicted to be
\begin{eqnarray}
\frac{\Gamma[T_{(cn [\bar{s}\bar{n}])0^+}^{1}(2527)\to DK]}{\Gamma[T_{(cn [\bar{s}\bar{n}])0^+}^{1}(2527)\to D_s\pi]}\simeq 1.3.
\end{eqnarray}
While the high mass state $T_{(cn [\bar{s}\bar{n}])0^+}^{1}(3007)$ dominantly decays into the $DK$ and
$D_s\pi$, $D_s^*\rho$, and $D^*K^*$ channels with branching fractions about $10\%$, $10\%$, $40\%$,
and $41\%$, respectively. If the $T_{c\bar{s}0}^a(2900)$ corresponds to $T_{(cn\{\bar{s}\bar{n}\})0^+}^{1}(2922)$ indeed,
both the $T_{(cn [\bar{s}\bar{n}])0^+}^{1}(2527)$ and $T_{(cn [\bar{s}\bar{n}])0^+}^{1}(3007)$ may also have large potentials to be established
in the $D_s^+\pi^{\pm}$ and $D^{0,+}K^+$ final states in future experiments.

More states of the $cn\bar{s}\bar{n}$ system may have potentials to be observed in
some of their dominant decay channels as shown in Table~\ref{tab:table_Qnsn_decay_merged}, since their fall-apart decay widths are
predicted to be relatively narrow, $\Gamma\sim 10-100$~MeV. For example, the high-lying scalar state
with $I=0$, $T_{(cn\{\bar{s}\bar{n}\})0^+}^{0}(2873)$ as a narrow width state, is likely to be
observed in the $D^0K^+$ final state, the branching fraction can reach up to $\sim 72\%$.

The tensor states of the $cn\bar{s}\bar{n}$ system lie in the high mass range of $\sim 3.1-3.2$~GeV.
Their fall-apart decay widths are predicted to be very narrow; the values are only a few MeV.
They dominantly decay into the final states containing two vector mesons, such as $D_s^*\rho$.

Finally, it should be mentioned that the $T_{c\bar{s}}(2327)^{++,0}$ reported by LHCb
cannot be explained as tetraquarks, since our predicted mass for the lowest
state decaying into the $D_s\pi$ channel is notably ($\sim 200$~MeV) larger than the observation.

\subsubsection{$bn\bar{s}\bar{n}$ system}

The $bn\bar{s}\bar{n}$ is an interesting system which relates to
the exotic states $T_{b\bar{s}}(5568)^{\pm}$ observed in the $B_s^0\pi^{\pm}$ final states by the D0 collaboration~\cite{D0:2016mwd,D0:2017qqm}.
The $T_{b\bar{s}}(5568)^{\pm}$ may be candidates of the $I=1$ tetraquark states composed of $bu\bar{s}\bar{d}$ and $bd\bar{s}\bar{u}$.
For the $bn\bar{s}\bar{n}$ system, our predicted masses of various $1S$-wave states belonging to the $\mathbf{6}_F$ representation scatter in the range of $\sim 6200-6600$~MeV. While the predicted masses of the states belonging to $\bar{\mathbf{3}}_F$ scatter in a broader mass range
of $\sim 5800-6450$~MeV. The detailed behavior of the spectrum can be seen in Table~\ref{tab:table_Qnsn_radii_merged} and Fig.~\ref{fig:Mass_Qnsn_Qsnn}.

The lowest state with $I=1$ is the $J^P=0^+$ state, $T_{(bn[\bar{s}\bar{n}])0^+}^{1}(5910)$.
Its mass, 5910~MeV, is $\sim 350$~MeV larger than that of $T_{b\bar{s}}(5568)$.
Thus, the $T_{b\bar{s}}(5568)$ cannot be assigned to any state of the $bn\bar{s}\bar{n}$ system.
It should be mentioned that $T_{b\bar{s}}(5568)$ was not seen in the experiments carried out
at LHC~\cite{LHCb:2016dxl,CMS:2017hfy,ATLAS:2018udc} and CDF~\cite{CDF:2017dwr}.
According to our prediction, the $T_{(bn[\bar{s}\bar{n}])0^+}^{1}(5910)$ is a mixed state between $|(bn)_1^6 [\bar{s}\bar{n}]_1^{\bar{6}}\rangle_0$
and $|(bn)_0^{\bar{3}} [\bar{s}\bar{n}]_0^3\rangle_0$ with components $\sim 69\%$
and $\sim 31\%$, respectively. The low mass nature of $T_{(bn[\bar{s}\bar{n}])0^+}^{1}(5910)$ in the $I=1$ states is mainly due to the strong attractive color-magnetic interactions together with the effect of configuration mixing.
As shown in Table~\ref{tab:table_Qnsn_decay_merged}, the $T_{(bn[\bar{s}\bar{n}])0^+}^{1}(5910)$ has a
fall-apart width of $\sim 52$~MeV, which is saturated by the $B_s\pi$ and $BK$ channels.
The partial width ratio is predicted to be
\begin{eqnarray}
\frac{\Gamma[T_{(bn[\bar{s}\bar{n}])0^+}^{1}(5910)\to B_s\pi]}{\Gamma[T_{(bn[\bar{s}\bar{n}])0^+}^{1}(5910)\to BK]}\simeq 0.70.
\end{eqnarray}
The $T_{(bn[\bar{s}\bar{n}])0^+}^{1}(5910)$ may have good potentials to be observed in the $B_s^0\pi^{\pm}$
and $B^{0,-}K^+$ final states.

Another interesting state with $I=1$ is the scalar state, $T_{(bn\{\bar{s}\bar{n}\})0^+}^{1}(6263)$, which
may be the bottom partner of $T_{c\bar{s}0}(2900)$ according to our previous analysis in Sec.~\ref{Tcs2900}.
Our predicted mass, $\sim 6263$~MeV, is about $200$~MeV smaller than
the predictions only including the OGE potentials in the GI model~\cite{Lu:2020qmp}.
From Table~\ref{tab:table_Qnsn_decay_merged}, it is seen that the $T_{(bn\{\bar{s}\bar{n}\})0^+}^{1}(6263)$ has a
fall-apart width of $\sim 57$~MeV, and dominantly decays into the $B_s\pi$, $BK$, $B^*K^*$, and $B_s^*\rho$ channels
with branching fractions of $\sim 13\%$, $\sim 27\%$, $\sim 43\%$, and $\sim 17\%$, respectively.
Since the $T_{c\bar{s}0}(2900)$ has been observed in the $D_s^+\pi^{\pm}$ final state,
as its partner, the $T_{(bn\{\bar{s}\bar{n}\})0^+}^{1}(6263)$ is most likely to be observed
in the $B_s^0\pi^{\pm}$ channel. Furthermore, the $B^{0,-}K^+$
final states are also worth observing in future experiments.

For the $1S$-wave states with isospin $I=0$, their quantum numbers are the same as
the $1P$-wave $B_s$ mesons. The lowest one is the scalar state $T_{(bn[\bar{s}\bar{n}])0^+}^{0}(5850)$.
It highly overlaps with the orbitally excited $B_s$ meson state $B_s(1^3P_0)$ predicted
within the quark model~\cite{li:2021hss}.
The decay width can distinguish between the $T_{(bn[\bar{s}\bar{n}])0^+}^{0}(5850)$ and $B_s(1^3P_0)$, although they have similar masses and dominantly decay into $BK$ channel. According to our analysis, the $T_{(bn[\bar{s}\bar{n}])0^+}^{0}(5850)$ is a fairly narrow state with a width of $\sim23$ MeV, while the $B_s(1^3P_0)$ state is expected to be a very broad state with a width of $\sim200$ MeV~\cite{li:2021hss}.
It should be mentioned that the $B_s(1^3P_0)$ may lie $\sim 100$~MeV below the $T_{(bn[\bar{s}\bar{n}])0^+}^{0}(5850)$ according to
our recent unquenched quark model prediction~\cite{Ni:2023lvx}. In this case, the $B_s(1^3P_0)$ should be an extremely narrow state due to
the absence of strong decay processes. It may be interesting to search for the narrow scalar state
$T_{(bn[\bar{s}\bar{n}])0^+}^{0}(5850)$ in the $B^{-}K^+$ final states.

Furthermore, several high-lying tetraquark states may have good potentials to be discovered in their dominant
decay channels as shown in Table~\ref{tab:table_Qnsn_decay_merged}. For example, the scalar state $T_{(bn\{\bar{s}\bar{n}\})0^+}^{0}(6213)$ with a width of $\sim 26$~MeV and the axial state $T_{(bn[\bar{s}\bar{n}])1^+}^{0}(5988)$ with a width of $\sim 39$~MeV
may have good potentials to be observed in the $B^-K^+$ and $B^{*-}K^+$ final states, respectively.

\subsection{$Qs\bar{n}\bar{n}$ system}

The $Qs\bar{n}\bar{n}$ system is interesting because the states formed by it cannot mix with conventional meson states.
When they are produced in experiments, their exotic nature can be clearly distinguished through their decay channels.
For the $Qs\bar{n}\bar{n}$ system, there are six $1S$-wave configurations with isospin $I=1$ and six ones with isospin $I=0$.

From Table~\ref{tab:merged_csnn_bsnn} given in the appendix, one can see the dynamic roles of the OBE potentials,
which are summarized as follows: (i) the $\pi$ exchange contributes a significant attractive potential,
$\langle V^{\pi}\rangle\sim -160$~MeV, to the two $I=0$ configurations $\left| (Qs)^{\bar{\mathbf{3}}}_0 [\bar{n}\bar{n}]^{\mathbf{3}}_0 \right\rangle_{0}$ and $\left| (Qs)^{\bar{\mathbf{3}}}_1 [\bar{n}\bar{n}]^{\mathbf{3}}_0 \right\rangle_{1}$. For the other configurations, the contribution of the $\pi$ exchange is less important.
(ii) For the $I=1$ states, the interactions from both the $\omega$ and $\rho$ exchanges are weakly repulsive, their total contributions
are estimated to be a small value, $\langle V^{\rho}\rangle+\langle V^{\omega}\rangle\sim 20$~MeV.
(iii) For the $I=0$ states, the potentials from the $\omega$ and $\rho$ exchanges are repulsive and attractive, respectively.
The $\rho$ exchange plays an important role in the $\left| (Qs)^{\bar{\mathbf{3}}}_0 [\bar{n}\bar{n}]^{\mathbf{3}}_0 \right\rangle_{0}$ and
$\left| (Qs)^{\bar{\mathbf{3}}}_1 [\bar{n}\bar{n}]^{\mathbf{3}}_0 \right\rangle_{1}$ configurations,
it contributes a significant attractive potential, $\langle V^{\rho}\rangle\sim -170$~MeV.
However, for the other configurations the total contribution from the $\omega$ and $\rho$ exchanges is small,
$\langle V^{\rho}\rangle+\langle V^{\omega}\rangle\sim -20$~MeV.


\subsubsection{$cs\bar{n}\bar{n}$ system}\label{tcs2900}

The $cs\bar{n}\bar{n}$ (or its antiparticle $\bar{c}\bar{s}nn$) is an interesting system which relates to the exotic resonance
$T_{\bar{c}\bar{s}0}(2870)^0$ and $T_{\bar{c}\bar{s}1}(2900)^0$ observed at LHCb~\cite{LHCb:2020pxc,LHCb:2020bls,LHCb:2024vfz}.
These two resonances should be compelling candidates for tetraquark states composed of $\bar{c}\bar{s}ud$.
For the $cs\bar{n}\bar{n}/\bar{c}\bar{s}nn$ system, our predicted masses of various $1S$-wave states with $I=1$ scatter in the range
of $\sim 2980-3270$~MeV. While the predicted masses of the states with $I=0$ scatter in a broader mass range
of $\sim 2500-3200$~MeV. The detailed behavior of the spectrum can be seen in Table~\ref{tab:table_Qsnn_radii_merged} and Fig.~\ref{fig:Mass_Qnsn_Qsnn}.

It is found that the $IJ^P=00^+$ state $T_{(\bar{c}\bar{s}[ud])0^+}^0(2919)$ (as the antiparticle of $T_{(cs[\bar{u}\bar{d}])0^+}^0(2919)$)
predicted in the present work favors the assignment of the $T_{\bar{c}\bar{s}0}(2870)$ observed in the $D^-K^+$ final state at LHCb~\cite{LHCb:2020pxc,LHCb:2020bls,LHCb:2024vfz}.
The $T_{(\bar{c}\bar{s}[ud])0^+}^0(2919)$ is a mixed state between the $|(\bar{c}\bar{s})_1^6 [ud]_1^{\bar{6}}\rangle_0$
and $| (\bar{c}\bar{s})^{\bar{\mathbf{3}}}_0 [ud]^{\mathbf{3}}_0 \rangle_{0}$ configurations with components
around $65\%$ and $35\%$, respectively. Our predicted mass, $\sim 2919$~MeV, is very close to the average measured value $M_{exp}=2872\pm 16$~MeV
from the PDG~\cite{ParticleDataGroup:2024cfk}. We also analyze its fall-apart decays.
As shown in Table~\ref{tab:table_Qsnn_decay_merged}, the $T_{(\bar{c}\bar{s}[ud])0^+}^0(2919)$
has a width of $\Gamma\simeq 71$~MeV, which is also consistent with the average measured value
$\Gamma_{exp}=67\pm 24$~MeV from the PDG~\cite{ParticleDataGroup:2024cfk}.
The fall-apart decays of $T_{(\bar{c}\bar{s}[ud])0^+}^0(2919)$ are saturated by the
$DK$ and $D^*K^*$ channels. The partial width ratio is predicted to be
\begin{eqnarray}
\frac{\Gamma[T_{(\bar{c}\bar{s}[ud])0^+}^0(2919)\to DK]}{\Gamma[T_{(\bar{c}\bar{s}[ud])0^+}^0(2919)\to D^*K^*]}\simeq 	0.12.
\end{eqnarray}
The significant decay rate of $T_{(\bar{c}\bar{s}[ud])0^+}^0(2919)\to D^-K^+$, $\sim 10\%$, can explain why
the $T_{\bar{c}\bar{s}0}(2870)$ was first observed in the $D^-K^+$ final state at LHCb. To further confirm the nature of $T_{(\bar{c}\bar{s}[ud])0^+}^0(2919)$, the dominant decay channel $D^*K^*$ is worth observing in future experiments.

If the observed resonance $T_{\bar{c}\bar{s}0}(2870)$ corresponds to the $T_{(\bar{c}\bar{s}[ud])0^+}^0(2919)$ indeed,
its mixed partner, $T_{(\bar{c}\bar{s}[ud])0^+}^0(2510)$, may have discovery potentials as well.
It is the lowest state of the $\bar{c}\bar{s}nn$ system, and the dominant component is
$\left| (\bar{c}\bar{s})^{\bar{\mathbf{3}}}_0 [ud]^{\mathbf{3}}_0 \right\rangle_{0}$.
Our predicted mass, $\sim 2510$~MeV, is consistent with that obtained in a multiquark color flux-tube model~\cite{Wei:2022wtr}.
Due to the strong attractive interactions of the $\pi$ and $\rho$ exchanges, however, our predicted mass
for the lowest state is about 300~MeV smaller than that obtained within the OGE potential model in
our previous work~\cite{Liu:2022hbk}, where it was suggested as the assignment of $T_{\bar{c}\bar{s}0}(2870)$.
As shown in Table~\ref{tab:table_Qsnn_decay_merged}, the $T_{(\bar{c}\bar{s}[ud])0^+}^0(2510)$ has
a fall-apart width of $\sim 86$~MeV, which is saturated by the $DK$ channel.
To confirm our predictions, we suggest searching for this low-mass state around the mass range of $2.5$~GeV in the $D^-K^+$ final state in future experiments.

In the isospin $I=1$ sector, the two scalar states $T_{(\bar{c}\bar{s}\{ud\})0^+}^1(2982)$
and $T_{(\bar{c}\bar{s}\{ud\})0^+}^1(3263)$ are also worth exploring in experiments.
They have widths of $\sim 45$ and $\sim 26$~MeV, respectively.
The partial width ratios between their two allowed fall-apart channels $DK$ and $D^*K^*$ are predicted to be
\begin{eqnarray}
\frac{\Gamma[T_{(\bar{c}\bar{s}\{ud\})0^+}^1(2982)\to DK]}{\Gamma[T_{(\bar{c}\bar{s}\{ud\})0^+}^1(2982)\to D^*K^*]}\simeq 1.0,\\
\frac{\Gamma[T_{(\bar{c}\bar{s}\{ud\})0^+}^1(3263)\to DK]}{\Gamma[T_{(\bar{c}\bar{s}\{ud\})0^+}^1(3263)\to D^*K^*]}\simeq 0.02.
\end{eqnarray}
The $T_{(\bar{c}\bar{s}\{ud\})0^+}^1(2982)$ may have potentials to be observed in the $D^-K^+$ final state.
It should be mentioned that in Refs.~\cite{Wang:2020prk,Wei:2022wtr} the authors suggested the $T_{\bar{c}\bar{s}0}(2870)$
assigning as the low-lying state with $I=1$, however, our predicted mass, $\sim 2982$~MeV,
is about 100~MeV larger than the observation.

Furthermore, in the isospin $I=0$ sector, the axial vector states $T_{(\bar{c}\bar{s}[ud])1^+}^0(2658)$ and
$T_{(\bar{c}\bar{s}[ud])1^+}^0(2932)$ have discovery potentials as well.
The $T_{(\bar{c}\bar{s}[ud])1^+}^0(2658)$ has a fall-apart decay width of $\sim 64$~MeV, which is saturated
by the $D^*K$ channel. While the $T_{(\bar{c}\bar{s}[ud])1^+}^0(2932)$ has a relatively narrow width
of $\sim 49$~MeV, and dominantly decays into $DK^*$ and $D^*K^*$ channels with branching fractions
$\sim 72\%$ and $\sim 26\%$, respectively. The $T_{(\bar{c}\bar{s}[ud])1^+}^0(2658)$ and
$T_{(\bar{c}\bar{s}[ud])1^+}^0(2932)$ may be observed in the $D^{*-}K^+$ and $D^{-}K^{*+}$
final states, respectively. More details of the decay properties can be found in Table~\ref{tab:table_Qsnn_decay_merged}.
It should be mentioned that a hint of $T_{(\bar{c}\bar{s}[ud])1^+}^0(2658)$ might be observed in
the $D^{*-}K^+$ invariant mass spectrum by the LHCb Collaboration in the $B^+\to D^{*-}D^+K^+$ decay~\cite{LHCb:2024vfz}.

\subsubsection{$bs\bar{n}\bar{n}$ system}

For the $bs\bar{n}\bar{n}$ system, our predicted masses of various $1S$-wave states with $I=1$ scatter in the range
of $\sim 6300-6600$~MeV. While the predicted masses of the states with $I=0$ scatter in a broader mass range
of $\sim 5800-6500$~MeV. The detailed behavior of the spectrum can be seen in Table~\ref{tab:table_Qsnn_radii_merged} and Fig.~\ref{fig:Mass_Qnsn_Qsnn}.

In the $bs\bar{n}\bar{n}$ system, the lowest state is the scalar state with $I=0$,
$T_{(bs[\bar{u}\bar{d}])0^+}^{0}(5887)$, which is a $|(bs)_0^{\bar{3}} [\bar{u}\bar{d}]_0^3\rangle_0$
dominant state with a significant $|(bs)_1^6 [\bar{u}\bar{d}]_1^{\bar{6}}\rangle_0$ mixing.
The mass, $\sim 5887$~MeV, predicted in the present work is about $300$~MeV lower than
that of similar calculations with the GI model~\cite{Lu:2020qmp}, which is mainly caused by the strong
attractive interactions from the $\pi$ and $\rho$ exchanges.
It should be mentioned that the lowest state composed of $bs\bar{u}\bar{d}$ may be
a bound state below the $BK$ threshold discussed in the literature
~\cite{Chen:2018hts,Huang:2019otd,Chen:2023syh,Agaev:2019wkk,Tan:2020ldi}.
We also study the fall-apart decay properties, our results show that
the low-mass state $T_{(bs[\bar{u}\bar{d}])0^+}^{0}(5887)$ may be a narrow state with a fall-apart width of $\sim 74$~MeV,
which is saturated by the $BK$ channel. This state may have good potentials to be observed
in the $B^0K^-$ final state.


Another interesting state is $T_{(bs[\bar{u}\bar{d}])0^+}^{0}(6252)$,
as the mixed partner of $T_{(bs[\bar{u}\bar{d}])0^+}^{0}(5887)$,
whose dominant component is $|(bs)_1^6 [\bar{u}\bar{d}]_1^{\bar{6}}\rangle_0$.
Our predicted mass, $6252$~MeV, is about 200~MeV smaller than that obtained with the GI model~\cite{Lu:2020qmp},
since the OBE potentials are included in the present work.
According to our analysis in Sec.~\ref{tcs2900}, the $T_{(bs[\bar{u}\bar{d}])0^+}^{0}(6252)$ may be
the bottom partner of the antiparticle $T_{\bar{c}\bar{s}0}(2870)$ observed at LHCb.
As shown in Table~\ref{tab:table_Qsnn_decay_merged}, the $T_{(bs[\bar{u}\bar{d}])0^+}^{0}(6252)$ has
a fall-apart width of $\Gamma\simeq 86$~MeV, and dominantly decays
into the $BK$ and $B^*K^*$ channels with a partial width ratio of
\begin{eqnarray}
\frac{\Gamma[T_{(bs[\bar{u}\bar{d}])0^+}^{0}(6252)\to BK]}{\Gamma[T_{(bs[\bar{u}\bar{d}])0^+}^{0}(6252)\to B^*K^*]}\simeq 0.05.
\end{eqnarray}
The $B^0K^-$ may be an optimal final state for searching for
$T_{(bs[\bar{u}\bar{d}])0^+}^{0}(6252)$ in experiments.

Except for the two $IJ^P=00^+$ states mentioned above, several other states are also worth exploring in experiments.
For example, in the isospin $I=1$ sector, the scalar states $T_{(bs\{\bar{n}\bar{n}\})0^+}^1(6333)$
and $T_{(bs\{\bar{n}\bar{n}\})0^+}^1(6580)$ with widths of $\sim 35$ and $\sim 23$~MeV, respectively, may
have discovery potentials. Their fall-apart decays are saturated by the $BK$ and $B^*K^*$ channels.
The decay rates of $T_{(bs\{\bar{n}\bar{n}\})0^+}^1(6333)\to BK$ and $T_{(bs\{\bar{n}\bar{n}\})0^+}^1(6580)\to BK$
are predicted to be $\sim 45\%$ and $\sim 4\%$, respectively. To establish these two scalar states with $I=1$,
the $B^{0,-}K^-$ final states are worth observing. While in the isospin $I=0$ sector, the axial vector states
$T_{(bs[\bar{u}\bar{d}])1^+}^0(5954)$ and $T_{(bs[\bar{u}\bar{d}])1^+}^0(6276)$ with comparable widths of $\sim72$ MeV
may be likely to be observed in the $B^{0*}K^-$ (with a probability of $\sim 50\%$) and $B^0K^{*-}$ (with a probability of $\sim46\%$)
final states, respectively. More details of the decay properties can be found in Table~\ref{tab:table_Qsnn_decay_merged}.

\begin{figure}[htbp]
 \centering \epsfxsize=8.5 cm \epsfbox{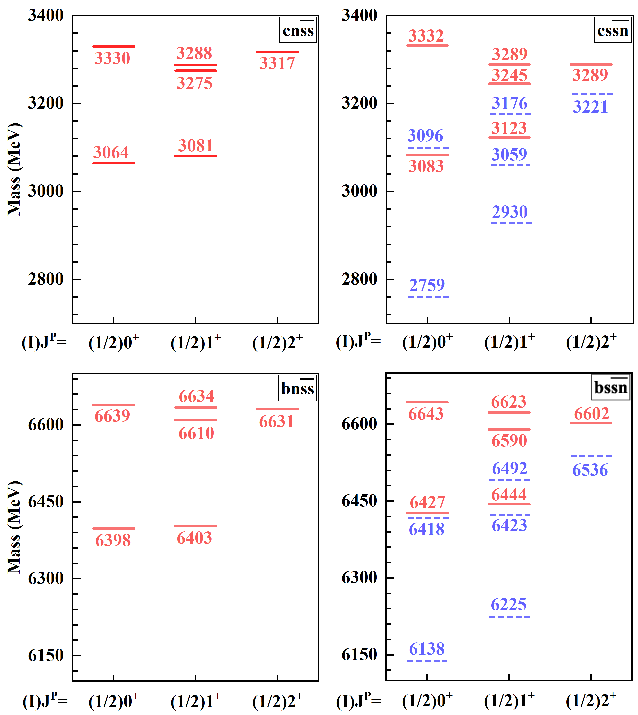}
 \caption{Mass spectra of the $1S$-wave states for the $Qs\bar{s}\bar{n}$ (left panel) and $Qn\bar{s}\bar{s}$ (right panel) systems, with the charmed ($c$) and bottom ($b$) states shown in the upper and lower rows, respectively. The red solid and blue dashed lines represent the states with flavor symmetries belonging to $\mathbf{6}_F$ and $\bar{\mathbf{3}}_F$, respectively.}\label{fig:Mass_Qssn_Qnss}
\end{figure}

\subsection{$Qs\bar{s}\bar{n}$ system}

For the $Qs\bar{s}\bar{n}$ system, there are six $1S$-wave configurations belonging to the $\mathbf{6}_F$ and $\bar{\mathbf{3}}_F$ representations, respectively. The flavor symmetry of $Qs\bar{s}\bar{n}$ is the same as that of $Qs\bar{n}\bar{n}$ in the SU(3) limit.

In the $Qs\bar{s}\bar{n}$ system, the $\pi$, $\rho$, and $\omega$ exchanges are absent.
Except for the $\eta$, $\eta'$, and $\sigma$ exchanges, there are $K$, $K^*$, and $\phi$ exchanges.
From Table~\ref{tab:merged_cssn_bssn} given in the appendix, it is found that
(i) for the two configurations belonging to $\bar{\mathbf{3}}_F$, $\left| (Qs)^{\bar{\mathbf{3}}}_0 [\bar{s}\bar{n}]^{\mathbf{3}}_0 \right\rangle_{0}$ and $\left| (Qs)^{\bar{\mathbf{3}}}_1 [\bar{s}\bar{n}]^{\mathbf{3}}_0 \right\rangle_{1}$, there is a sizable attractive $K$-exchange potential,
$\langle V^{K}\rangle\sim -30$~MeV. While for other configurations the $K$ exchange is less important, the magnitude of the potentials is often no more than 10~MeV. (ii) For the configurations belonging to $\mathbf{6}_F$, due to a nearly exact cancelation among the $K^*$- and $\phi$-exchange terms, the contributions from the vector meson exchanges are negligibly small. However, in the configurations belonging to $\bar{\mathbf{3}}_F$,
the $K^*$ and $\phi$ exchanges play important roles, there is a notable attractive potential
$\langle V^{K^*}\rangle+\langle V^{\phi}\rangle\sim (-100,-40)$~MeV.

\subsubsection{$cs\bar{s}\bar{n}$ system}

For the $cs\bar{s}\bar{n}$ system, our predicted masses for the six $1S$-wave states belonging to
the $\mathbf{6}_F$ representation scatter in the range
of $\sim 3080-3330$~MeV. While the predicted masses for the six states belonging to $\bar{\mathbf{3}}_F$ scatter in a broader mass range
of $\sim 2760-3220$~MeV. The detailed behavior of the spectrum can be seen in Table~\ref{tab:table_Qssn_radii_merged} and Fig.~\ref{fig:Mass_Qssn_Qnss}.
A few studies of the $cs\bar{s}\bar{n}$ system can be found in the
literature~\cite{Lu:2020qmp,Lu:2016zhe,Guo:2021mja,Jalili:2023kmw,Ebert:2010af,Zheng:2025uzy,Gerasyuta:2008ps}.
Our predicted mass spectrum lies about $100-500$~MeV above that predicted in Refs.~\cite{Jalili:2023kmw,Ebert:2010af,Guo:2021mja},
while about $100-200$~MeV below that predicted in Refs.~\cite{Lu:2020qmp,Lu:2016zhe}.

In the $cs\bar{s}\bar{n}$ system, the lowest state is the scalar state belonging to $\bar{\mathbf{3}}_F$,
$T_{(cs[\bar{s}\bar{n}])0^+}^{1/2}(2759)$, which is a mixed state between $|(cs)_1^6 [\bar{s}\bar{n}]_1^{\bar{6}}\rangle_0$
and $|(cs)_0^{\bar{3}} [\bar{s}\bar{n}]_0^3\rangle_0$ with components $\sim 59\%$ and $\sim 41\%$, respectively.
It should be mentioned that the $T_{(cs[\bar{s}\bar{n}])0^+}^{1/2}(2759)$ is close to the conventional excited $D$ meson state $D(2^3P_0)$
predicted in the quark model~\cite{Ni:2021pce}.
We also study the fall-apart decay properties, the results are given in Table~\ref{tab:table_Qssn_decay_merged}.
It is seen that the $T_{(cs[\bar{s}\bar{n}])0^+}^{1/2}(2759)$ may have a narrow width of $\Gamma\simeq 44$~MeV,
and dominantly decays into $D_sK$ and $D\eta$ channels with a partial width ratio of
\begin{eqnarray}
\frac{\Gamma[T_{(cs[\bar{s}\bar{n}])0^+}^{1/2}(2759)\to D_sK]}{\Gamma[T_{(cs[\bar{s}\bar{n}])0^+}^{1/2}(2759)\to D\eta]}\simeq 1.8.
\end{eqnarray}
The $T_{(cs[\bar{s}\bar{n}])0^+}^{1/2}(2759)$ can be distinguished from the
$D(2^3P_0)$ through the decay properties, because the $D(2^3P_0)$ is predicted to be a
very broad state ($\Gamma\sim 1000$~MeV) dominantly decaying into the $D\pi$ channel~\cite{Ni:2021pce}.
Thus, if a narrow structure is observed in the $D_s^+K^-$ and/or $D\eta$ final states around the mass
region of $2.76$~GeV, it should correspond to the tetraquark state $T_{(cs[\bar{s}\bar{n}])0^+}^{1/2}(2759)$.

The high-mass mixed state $T_{(cs[\bar{s}\bar{n}])0^+}^{1/2}(3096)$, as the partner of
$T_{(cs[\bar{s}\bar{n}])0^+}^{1/2}(2759)$, may have discovery potentials in future experiments as well.
The $T_{(cs[\bar{s}\bar{n}])0^+}^{1/2}(3096)$ is also the strange partner of the antiparticle
of $T_{\bar{c}\bar{s}0}(2870)$ observed in the $D^-K^+$ final state at LHCb according to our
previous analysis in Sec.~\ref{tcs2900}. As shown in Table~\ref{tab:table_Qssn_decay_merged}, this state is a relatively broad state with
a width of $\Gamma\simeq 102$~MeV, and mainly decays into the $D_sK$ ($\sim 10\%$), $D_s^*K^*$ ($\sim 44\%$),
and $D^*\phi$ ($\sim 39\%$) channels. The $D_s^+K^-$ and $D^{*}\phi$ may be optimal
modes for searching for the high-mass state $T_{(cs[\bar{s}\bar{n}])0^+}^{1/2}(3096)$.

Furthermore, several other states may have potentials to be observed in
their dominant decay channels. Such as, the axial vector states
$T_{(cs[\bar{s}\bar{n}])1^+}^{1/2}(2930)$ and $T_{(cs[\bar{s}\bar{n}])1^+}^{1/2}(3059)$ have large
decay rates into the optimal observation channels $D_s^{*+}K^-$ and $D\phi$, respectively.
The details of their decay properties can be seen in Table~\ref{tab:table_Qssn_decay_merged}.

\subsubsection{$bs\bar{s}\bar{n}$ system}

For the $bs\bar{s}\bar{n}$ system, our predicted masses of various $1S$-wave states belonging to
the $\mathbf{6}_F$ representation scatter in the range
of $\sim 6400-6650$~MeV. While the predicted masses of the states belonging to $\bar{\mathbf{3}}_F$ scatter in a broader mass range
of $\sim 6140-6540$~MeV. The detailed behavior of the spectrum can be seen in Table~\ref{tab:table_Qssn_radii_merged} and Fig.~\ref{fig:Mass_Qssn_Qnss}.
A few studies of the $bs\bar{s}\bar{n}$ system can be found in the
literature~\cite{Lu:2020qmp,Lu:2016zhe,Guo:2021mja,Jalili:2023kmw,Ebert:2010af,Zheng:2025uzy}.
Our predicted mass spectrum lies about $100-200$~MeV below that predicted in Refs.~\cite{Lu:2020qmp,Lu:2016zhe},
while about $100-400$~MeV above that predicted in Refs.~\cite{Jalili:2023kmw,Guo:2021mja,Ebert:2010af}.

In the $bs\bar{s}\bar{n}$ system, the lowest state is the scalar state belonging to $\bar{\mathbf{3}}_F$,
$T_{(bs[\bar{s}\bar{n}])0^+}^{1/2}(6138)$, which is a $|(bs)_1^6 [\bar{s}\bar{n}]_1^{\bar{6}}\rangle_0$
dominant state with significant mixing of $|(bs)_0^{\bar{3}} [\bar{s}\bar{n}]_0^3\rangle_0$ ($\sim 45\%$).
It is found that the $T_{(bs[\bar{s}\bar{n}])0^+}^{1/2}(6138)$ is close to the conventional excited $B$ meson state $B(2^3P_0)$
predicted in the quark models~\cite{DiPierro:2001dwf,Ebert:2009ua}.
As shown in Table~\ref{tab:table_Qssn_decay_merged}, the $T_{(bs[\bar{s}\bar{n}])0^+}^{1/2}(6138)$ has a narrow width of $\Gamma\simeq 39$~MeV,
and dominantly decays into $B_sK$ and $B\eta$ channels with branching fractions $\sim 60\%$ and $\sim 40\%$,
respectively. It may be easily distinguished from the $B(2^3P_0)$ through their decay properties, because the $B(2^3P_0)$ is predicted to be a
very broad state ($\Gamma\sim 400-800$~MeV) dominantly decaying into the $B\pi$ channel~\cite{Ni:2021pce}.
Thus, if a narrow structure is observed in the $B_s^0K^-$ and/or $B\eta$ channels around the mass
region of $6.14$~GeV, it should correspond to the $T_{(bs[\bar{s}\bar{n}])0^+}^{1/2}(6138)$.

The other high-mass scalar state belonging to $\bar{\mathbf{3}}_F$, $T_{(bs[\bar{s}\bar{n}])0^+}^{1/2}(6418)$, as the mixing partner of $T_{(bs[\bar{s}\bar{n}])0^+}^{1/2}(6138)$,
has a broader width of $\sim 100$~MeV, and dominantly decays into the $B_sK$, $B_s^*K^*$, and $B^*\phi$ channels
with branching fractions $\sim 8\%$, $\sim 44\%$, and $\sim 42\%$, respectively.
The $T_{(bs[\bar{s}\bar{n}])0^+}^{1/2}(6418)$ may have potentials to be observed in the
$B_s^0K^-$ and $B^*\phi$ final states. Furthermore, from Table~\ref{tab:table_Qssn_decay_merged} one can
see that the narrow scalar state belonging to $\mathbf{6}_F$, $T_{(bs\{\bar{s}\bar{n}\})0^+}^{1/2}(6427)$,
may be likely to be observed in the $B^*\phi$ final state, while two axial vector
states belonging to $\bar{\mathbf{3}}_F$, $T_{(bs[\bar{s}\bar{n}])1^+}^{1/2}(6225)$ and
$T_{(bs[\bar{s}\bar{n}])1^+}^{1/2}(6423)$, may have potentials to be observed in the $B_s^{0*}K^-$ and
$B_s^{0*}K^{*-}$ final states, respectively. The details of their decay properties can be seen in Table~\ref{tab:table_Qssn_decay_merged}.

\subsection{$Qn\bar{s}\bar{s}$ system}

The $Qn\bar{s}\bar{s}$ system is interesting because the exotic nature of the physical states it forms
can be clearly distinguished through their fall-apart decay channels.
For the $Qn\bar{s}\bar{s}$ system, due to the constraint of symmetry there are only six $1S$-wave configurations
belonging to the $\mathbf{6}_F$.

For the $Qn\bar{s}\bar{s}$ system, in the meson exchanges considered in the present work only the $\eta$, $\eta'$, $\sigma$, and $\phi$ exchanges remain in the OBE potentials. Their dynamical roles can be found in Table~\ref{tab:merged_cnss_bnss} of the appendix.
The contributions from the $\eta$ and $\eta'$ exchanges are negligibly small. The $\phi$-meson exchange
provides a sizeable repulsive interaction, $\langle V^{\phi}\rangle\sim 20$~MeV, for each configuration. While the $\sigma$-meson exchange
provides a moderate attractive interaction, $\langle V^{\sigma}\rangle\sim -60$~MeV.

\subsubsection{$cn\bar{s}\bar{s}$ system}

For the $cn\bar{s}\bar{s}$ system, our predicted masses for the six $1S$-wave states scatter in the range
of $\sim 3060-3330$~MeV, which highly overlaps with that predicted for the $cs\bar{s}\bar{n}$ system.
The detailed behavior of the spectrum can be seen in Table~\ref{tab:table_Qnss_radii_merged} and Fig.~\ref{fig:Mass_Qssn_Qnss}.
There are a few studies of the $cn\bar{s}\bar{s}$ system in the literature~\cite{Lu:2020qmp,Guo:2021mja,Zheng:2025uzy}.
The mass spectrum predicted in the present work lies about 100-200~MeV below that obtained within the GI model~\cite{Lu:2020qmp},
while about 500-700~MeV above that predicted within an improved ICMI model~\cite{Guo:2021mja}.

The two scalar states, $T_{(cn\{\bar{s}\bar{s}\})0^+}^{1/2}(3064)$ and $T_{(cn\{\bar{s}\bar{s}\})0^+}^{1/2}(3330)$,
are mixed states between $|(cn)_0^6 \{\bar{s}\bar{s}\}_0^{\bar{6}}\rangle_0$
and $|(cn)_1^{\bar{3}} \{\bar{s}\bar{s}\}_1^3\rangle_0$ with comparable components.
The large mass splitting and significant mixing of the physical states are due to the strong color-magnetic spin-spin interaction
between the two different color configurations.  As shown in Table~\ref{tab:table_Qnss_decay_merged},
both $T_{(cn\{\bar{s}\bar{s}\})0^+}^{1/2}(3064)$ and $T_{(cn\{\bar{s}\bar{s}\})0^+}^{1/2}(3330)$
have relatively narrow fall-apart decay widths of $\sim 56$~MeV, and $\sim 20$~MeV, respectively.
Both of them dominantly decay into the $D_sK$ and $D_s^*K^*$ channels with partial width ratios
\begin{eqnarray}
\frac{\Gamma[T_{(cn\{\bar{s}\bar{s}\})0^+}^{1/2}(3064)\to D_sK]}{\Gamma[T_{(cn\{\bar{s}\bar{s}\})0^+}^{1/2}(3064)\to D_s^*K^*]}\simeq 0.56,\\
\frac{\Gamma[T_{(cn\{\bar{s}\bar{s}\})0^+}^{1/2}(3330)\to D_sK]}{\Gamma[T_{(cn\{\bar{s}\bar{s}\})0^+}^{1/2}(3330)\to D_s^*K^*]}\simeq 0.10.
\end{eqnarray}
The two scalar states composed of $cu\bar{s}\bar{s}$ may have good potentials to be observed in the $D_s^+K^+$ final state.

For the axial vector sector, there are physical states $T_{(cn\{\bar{s}\bar{s}\})1^+}^{1/2}(3081)$,
$T_{(cn\{\bar{s}\bar{s}\})1^+}^{1/2}(3275)$, and $T_{(cn\{\bar{s}\bar{s}\})1^+}^{1/2}(3288)$,
predicted in the present work. These states are admixtures of different configurations.
The mixing eigenvectors have been given in Table~\ref{tab:table_Qnss_radii_merged}. The two high-lying states
highly overlap with each other. The low-lying state $T_{(cn\{\bar{s}\bar{s}\})1^+}^{1/2}(3081)$ has a narrow width
of $\sim 32$~MeV, and dominantly decays into the $D_s^*K$, $D_sK^*$, and $D_s^*K^*$ channels with branching fractions
$\sim 8\%$, $\sim 52\%$, and $\sim 40\%$, respectively. The two high-lying states $T_{(cn\{\bar{s}\bar{s}\})1^+}^{1/2}(3275)$ and $T_{(cn\{\bar{s}\bar{s}\})1^+}^{1/2}(3288)$ have comparable narrow widths
of $\sim 10$~MeV, and dominantly decay into the $D_sK^*$ ($\sim 91\%$) and $D_s^*K$ ($\sim 72\%$), respectively.
These axial vector states composed of $cu\bar{s}\bar{s}$ may have discovery potentials
in the $D_s^{*+}K^+$ and $D_s^{+}K^{*+}$ final states.

There is one tensor state, $T_{(cn\{\bar{s}\bar{s}\})2^+}^{1/2}(3317)$, which is a pure $|(cn)_1^{\bar{3}} \{\bar{s}\bar{s}\}_1^3\rangle_2$ configuration. The $T_{(cn\{\bar{s}\bar{s}\})2^+}^{1/2}(3317)$ has a very narrow width of $\sim 0.02$~MeV, which is saturated by the
$D_s^*K^*$ channel. It may be a challenge to search for this tensor state in the $D_s^*K^*$ channel.

Finally, it should be mentioned that recently the $cn\bar{s}\bar{s}$
system was analyzed within an OGE model by combining a complex scaling method~\cite{Zheng:2025uzy}, no resonances
with $J^P=0^+$ and $1^+$ were found, while two narrow resonances with $J^P=2^+$ in the
mass range of $3.7-3.9$~GeV were found, whose masses are notably larger than that
of $T_{(cn\{\bar{s}\bar{s}\})2^+}^{1/2}(3317)$ predicted in the present work.

\subsubsection{$bn\bar{s}\bar{s}$ system}

For the $bn\bar{s}\bar{s}$ system, our predicted masses for the six $1S$-wave states scatter in the range
of $\sim 6400-6650$~MeV, which highly overlaps with that predicted for the $bs\bar{s}\bar{n}$ system.
The detailed behavior of the spectrum can be seen in Table~\ref{tab:table_Qnss_radii_merged} and Fig.~\ref{fig:Mass_Qssn_Qnss}.
There are only several studies of the $bn\bar{s}\bar{s}$ system in the literature~\cite{Lu:2020qmp,Guo:2021mja,Zheng:2025uzy}.
The mass spectrum predicted in the present work lies about 100-200~MeV below that obtained within the GI model~\cite{Lu:2020qmp},
while about 400-600~MeV above that predicted within an improved ICMI~\cite{Guo:2021mja}.

The two scalar states, $T_{(bn\{\bar{s}\bar{s}\})0^+}^{1/2}(6398)$ and $T_{(bn\{\bar{s}\bar{s}\})0^+}^{1/2}(6639)$,
as the partners of $T_{(cn\{\bar{s}\bar{s}\})0^+}^{1/2}(3064)$ and $T_{(cn\{\bar{s}\bar{s}\})0^+}^{1/2}(3330)$ in the charmed sector,
have fall-apart widths of $\Gamma\sim 56$~MeV, and $\sim 16$~MeV, respectively.
As shown in Table~\ref{tab:table_Qnss_decay_merged}, their widths are saturated by the $B_sK$ and $B_s^*K^*$ channels with partial width ratios
\begin{eqnarray}
		\frac{\Gamma[T_{(bn\{\bar{s}\bar{s}\})0^+}^{1/2}(6398)\to B_sK]}{\Gamma[T_{(bn\{\bar{s}\bar{s}\})0^+}^{1/2}(6398)\to B_s^*K^*]}\simeq 	 0.36,\\
		\frac{\Gamma[T_{(bn\{\bar{s}\bar{s}\})0^+}^{1/2}(6639)\to B_sK]}{\Gamma[T_{(bn\{\bar{s}\bar{s}\})0^+}^{1/2}(6639)\to B_s^*K^*]}\simeq 	 0.18.
\end{eqnarray}
To establish these two scalar states, the $B_s^0K^+$ final state is worth observing in future experiments.

In the three axial vector states, the low-lying state $T_{(bn\{\bar{s}\bar{s}\})1^+}^{1/2}(6403)$ has a
relatively large fall-apart width of $\sim 45$~MeV. It mainly decays into the $B_s^*K$, $B_sK^*$, and $B_s^*K^*$ channels
with fractions $\sim 18\%$, $\sim 34\%$, and $\sim 49\%$, respectively.
The two high-lying states $T_{(bn\{\bar{s}\bar{s}\})1^+}^{1/2}(6610)$ and $T_{(bn\{\bar{s}\bar{s}\})1^+}^{1/2}(6634)$
highly overlap with each other. The $T_{(bn\{\bar{s}\bar{s}\})1^+}^{1/2}(6610)$ has a very narrow width of $\sim 7$~MeV,
and dominantly decays into the $B_s^*K$ and $B_sK^*$ channels with a partial width ratio
\begin{eqnarray}
\frac{\Gamma[T_{(bn\{\bar{s}\bar{s}\})1^+}^{1/2}(6610)\to B_s^*K]}{\Gamma[T_{(bn\{\bar{s}\bar{s}\})1^+}^{1/2}(6610)\to B_sK^*]}\simeq 	0.34,
\end{eqnarray}
While the $T_{(bn\{\bar{s}\bar{s}\})1^+}^{1/2}(6634)$ has a narrow width of $\sim 3$~MeV, and dominantly decays into
the $B_s^*K$, $B_sK^*$, and $B_s^*K^*$ channels with branching fractions $\sim 58\%$, $\sim 23\%$, and $\sim 20\%$, respectively.
These axial vector states composed of $bu\bar{s}\bar{s}$ may have discovery potentials
in the $B_s^{*0}K^+$ and $B_s^{0}K^{*+}$ final states.

There is one tensor state, $T_{(bn\{\bar{s}\bar{s}\})2^+}^{1/2}(6631)$, which highly overlaps with the two
high-lying axial vector states. The $T_{(bn\{\bar{s}\bar{s}\})2^+}^{1/2}(6631)$ has a very narrow
width of $\sim 0.08$~MeV, which is saturated by the
$B_s^*K^*$ channel. It may be a challenge to search for this tensor state in experiments.

Finally, it should be mentioned that recently the $bn\bar{s}\bar{s}$
system was analyzed within an OGE model by combining a complex scaling method~\cite{Zheng:2025uzy}, no resonances
with $J^P=0^+$ and $1^+$ were found, while two narrow resonances with $J^P=2^+$ in the
mass range of $7.0-7.2$~GeV were found, whose masses are notably larger than that
of $T_{(bn\{\bar{s}\bar{s}\})2^+}^{1/2}(6631)$ predicted in the present work.

\begin{figure}[htbp]
 \centering \epsfxsize=8.5 cm \epsfbox{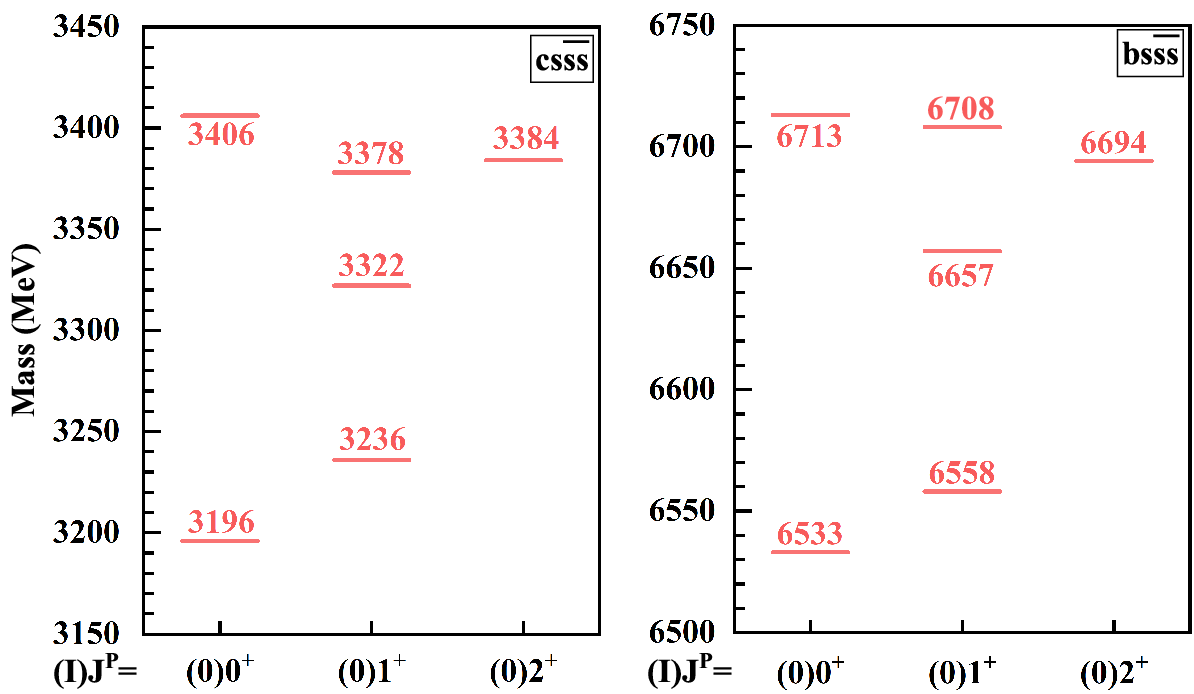}
 \caption{Mass spectra of the $1S$-wave states for the $cs\bar{s}\bar{s}$ (left panel) and $bs\bar{s}\bar{s}$ (right panel) systems.}\label{fig:Mass_Qsss}
\end{figure}

\subsection{$Qs\bar{s}\bar{s}$ system}

For the $Qs\bar{s}\bar{s}$ system, due to the constraint of symmetry there are only six $1S$-wave configurations
belonging to the $\mathbf{6}_F$. The flavor symmetry is the same as that of $Qn\bar{s}\bar{s}$.
In the meson-exchange potentials considered in the present work, only those from the $\eta$, $\eta'$, $\sigma$, and $\phi$ exchanges remain.
The $\eta$ and $\eta'$ exchanges are less important, the magnitude of the potentials is in the order of several MeV.
The $\sigma$-exchange term contributes a moderate attractive potential $\langle V^{\sigma}\rangle\sim -60$~MeV
for each configuration. While the $\phi$-meson exchange
provide a sizeable attractive potential $\langle V^{\phi}\rangle\sim -25$~MeV for each configuration.

\subsubsection{$cs\bar{s}\bar{s}$ system}

For the $cs\bar{s}\bar{s}$ system, our predicted masses for the six $1S$-wave states scatter in the range
of $\sim 3200-3410$~MeV. The detailed behavior of the spectrum can be seen in Table~\ref{tab:table_Qsss_radii_merged} and Fig.~\ref{fig:Mass_Qsss}.
A few studies of the $cs\bar{s}\bar{s}$ system can be found in the
literature~\cite{Zhang:2006hv,Lu:2020qmp,Guo:2021mja,Jalili:2023kmw,Ebert:2010af,Zheng:2025uzy}.
Our predicted mass spectrum of the $1S$-wave states lies about $100$~MeV below that predicted in Ref.~\cite{Lu:2020qmp},
while about $100-600$~MeV above that predicted in Refs.~\cite{Zhang:2006hv,Lu:2016zhe,Jalili:2023kmw,Guo:2021mja,Ebert:2010af}.


The two scalar states, $T_{(cs\{\bar{s}\bar{s}\})0^+}^{0}(3196)$ and $T_{(cs\{\bar{s}\bar{s}\})0^+}^{0}(3406)$,
are mixed states between $|(cs)_0^6 \{\bar{s}\bar{s}\}_0^{\bar{6}}\rangle_0$
and $|(cs)_1^{\bar{3}} \{\bar{s}\bar{s}\}_1^3\rangle_0$ with comparable components.
From Table~\ref{tab:table_Qsss_decay_merged}, it is seen that the low-mass state $T_{(cs\{\bar{s}\bar{s}\})0^+}^{0}(3196)$ has a narrow width of $\sim 25$~MeV, and dominantly decays into $D_s\eta$, $D_s\eta'$, and $D_s^*\phi$ channels with branching
fractions $\sim 16\%$, $\sim 26\%$, and $\sim 58\%$, respectively. The high-mass state
$T_{(cs\{\bar{s}\bar{s}\})0^+}^{0}(3406)$ has a narrow width of $\sim 11$~MeV,
which is nearly saturated by the $D_s^*\phi$ channel. The $D_s\eta$, $D_s\eta'$, and $D_s^*\phi$
may be optimal channels for searching for these scalar states.

The three axial vector states may be very narrow states with comparable fall-apart widths of $\sim 7$~MeV.
The lowest state $T_{(cs\{\bar{s}\bar{s}\})1^+}^{0}(3236)$ mainly decays into
the $D_s\phi$, and $D_s^*\phi$ channels with a partial width ratio
\begin{eqnarray}
\frac{\Gamma[T_{(cs\{\bar{s}\bar{s}\})1^+}^{0}(3236)\to D_s\phi]}{\Gamma[T_{(cs\{\bar{s}\bar{s}\})1^+}^{0}(3236)\to D_s^*\phi]}\simeq 2.3.
\end{eqnarray}
There is only a small splitting, $\Delta M\sim 56$~MeV, between the two high-lying states $T_{(cs\{\bar{s}\bar{s}\})1^+}^{0}(3322)$ and $T_{(cs\{\bar{s}\bar{s}\})1^+}^{0}(3378)$. The $T_{(cs\{\bar{s}\bar{s}\})1^+}^{0}(3322)$
dominantly decays into the $D_s^*\eta$, $D_s^*\eta'$, and $D_s^*\phi$ channels with branching
fractions $\sim 27\%$, $\sim 39\%$, and $\sim 34\%$, respectively.
While the fall-apart decays of the highest state $T_{(cs\{\bar{s}\bar{s}\})1^+}^{0}(3378)$
is governed by the $D_s\phi$ channel. More details of the decay properties can be found in Table~\ref{tab:table_Qsss_decay_merged}.
To establish these axial vector states,
the $D_s\eta$, $D_s\eta'$, and $D_s\phi$ channels are worth observing in experiments.

There is one tensor state, $T_{(cs\{\bar{s}\bar{s}\})2^+}^{0}(3384)$, which nearly degenerates with the
highest axial vector state $T_{(cs\{\bar{s}\bar{s}\})1^+}^{0}(3378)$.
However, their decay properties are very different. As shown in Table~\ref{tab:table_Qsss_decay_merged},
the $T_{(cs\{\bar{s}\bar{s}\})2^+}^{0}(3384)$ has a very narrow
width of $\sim 2$~MeV, which is saturated by the
$D_s^*\phi$ channel. It may be interesting to search for this narrow tensor state in the $D_s^*\phi$ channel.

\subsubsection{$bs\bar{s}\bar{s}$ system}

For the $bs\bar{s}\bar{s}$ system, our predicted masses of various $1S$-wave states scatter in the range
of $\sim 6530-6715$~MeV. The detailed behavior of the spectrum can be seen in Table~\ref{tab:table_Qsss_radii_merged} and Fig.~\ref{fig:Mass_Qsss}.
A few studies of the $bs\bar{s}\bar{s}$ system can be found in the
literature~\cite{Lu:2020qmp,Guo:2021mja,Jalili:2023kmw,Ebert:2010af,Zheng:2025uzy}.
Our predicted mass spectrum of the $1S$-wave states lies about $150$~MeV below that predicted in Ref.~\cite{Lu:2020qmp},
while about $200-500$~MeV above that predicted in Refs.~\cite{Lu:2016zhe,Jalili:2023kmw,Guo:2021mja,Ebert:2010af}.


The two scalar states, $T_{(bs\{\bar{s}\bar{s}\})0^+}^{0}(6533)$ and $T_{(bs\{\bar{s}\bar{s}\})0^+}^{0}(6713)$,
as the bottom partners of $T_{(cs\{\bar{s}\bar{s}\})0^+}^{0}(3196)$ and $T_{(cs\{\bar{s}\bar{s}\})0^+}^{0}(3406)$ in the charmed sector,
have relatively narrow fall-apart widths of $\sim 20$~MeV, and $\sim 8$~MeV, respectively.
From Table~\ref{tab:table_Qsss_decay_merged}, it is seen that the low-mass state
$T_{(bs\{\bar{s}\bar{s}\})0^+}^{0}(6533)$ dominantly decays
into $B_s\eta$, $B_s\eta'$, and $B_s^*\phi$ channels with branching
fractions $\sim 13\%$, $\sim 21\%$, and $\sim 66\%$, respectively.
While for the high-mass state
$T_{(bs\{\bar{s}\bar{s}\})0^+}^{0}(6713)$, the fall-apart decays are governed by the $B_s^*\phi$ channel.
The $B_s\eta$, $B_s\eta'$, and $B_s^*\phi$ may be optimal channels for searching for these scalar states.

For the axial vector sector, as shown in Table~\ref{tab:table_Qsss_decay_merged} the lowest state $T_{(bs\{\bar{s}\bar{s}\})1^+}^{0}(6558)$
has a narrow fall-apart width of $\sim 12$~MeV, and dominantly decays
into $B_s^*\eta'$, $B_s\phi$, and $B_s^*\phi$ channels with branching
fractions $\sim 15\%$, $\sim 35\%$, and $\sim 43\%$, respectively.
The two high-lying states $T_{(bs\{\bar{s}\bar{s}\})1^+}^{0}(6657)$ and
$T_{(bs\{\bar{s}\bar{s}\})1^+}^{0}(6708)$ with a small mass splitting of
$\Delta M\sim 51$~MeV lie about 120~MeV above the $T_{(bs\{\bar{s}\bar{s}\})1^+}^{0}(6558)$.
The $T_{(bs\{\bar{s}\bar{s}\})1^+}^{0}(6657)$ has a very narrow fall-apart width of $\sim 3$~MeV,
and dominantly decays into the $B_s^*\phi$ channel with a branching fraction $\sim 58\%$.
While the $T_{(bs\{\bar{s}\bar{s}\})1^+}^{0}(6708)$ has a narrow fall-apart width of $\sim 6$~MeV,
and dominantly decays into the $B_s\phi$ and $B_s^*\phi$ channel with
branching fractions $\sim 74\%$ and $\sim 26\%$, respectively.
The $B_s\phi$ and $B_s^*\phi$ may be optimal channels for searching for these axial vector states.

The tensor state $T_{(bs\{\bar{s}\bar{s}\})2^+}^{0}(6694)$ highly overlaps with the axial
vector state $T_{(bs\{\bar{s}\bar{s}\})1^+}^{0}(6708)$.
However, their decay properties are very different. As shown in Table~\ref{tab:table_Qsss_decay_merged},
the $T_{(bs\{\bar{s}\bar{s}\})2^+}^{0}(6694)$ has an extremely narrow
width of $\sim 2$~MeV, which is saturated by the
$B_s^*\phi$ channel. It may be interesting to search for this narrow tensor state in the $B_s^*\phi$ channel.

\section{Summary}\label{sec:summary}

In this work, we have carried out a systematic study of the mass spectra of the $1S$-wave states for the whole singly-heavy tetraquark systems $Qn\bar{n}\bar{n}$, $Qs\bar{n}\bar{n}$, $Qn\bar{s}\bar{n}$, $Qs\bar{s}\bar{n}$, $Qn\bar{s}\bar{s}$, and $Qs\bar{s}\bar{s}$ within a hybrid quark potential model, in which both the OGE and OBE interactions are included. Then, we further evaluated the fall-apart decays by
combining the obtained spectra within the quark exchange model.
Based on the obtained mass spectra and decay properties, the observation strategies of some interesting
states have been discussed and suggested, which may be helpful for the future experimental
explorations at LHC, Belle II, and BESIII. Several key findings are emphasized as follows.

Besides the OGE potentials, the OBE potentials are also crucial for describing the spectrum.
(i) Important roles of the $\pi$-meson exchange are found in the $Qn\bar{n}\bar{n}$ system. For the $\bar{3} 3$ configurations, $\left| (Qn)^{\bar{\mathbf{3}}}_1 [\bar{n}\bar{n}]^{\mathbf{3}}_1 \right\rangle_0$, $\left| (Qn)^{\bar{\mathbf{3}}}_0 [\bar{n}\bar{n}]^{\mathbf{3}}_0 \right\rangle_0$ and $\left| (Qn)^{\bar{\mathbf{3}}}_1 [\bar{n}\bar{n}]^{\mathbf{3}}_0 \right\rangle_1$, the contribution can reach up to $|\langle V^{\pi}\rangle|\sim 160$~MeV.
(ii) Moreover, important roles of the $\omega$- and $\rho$-meson exchanges are found in the $Qn\bar{n}\bar{n}$
and $Qn\bar{s}\bar{n}$ systems. For most of the $Qn\bar{n}\bar{n}$ states with $I=1/2$
and $Qn\bar{s}\bar{n}$ states with $I=0$, there are large attractive interactions
$\langle V^{\rho}\rangle+\langle V^{\omega}\rangle\sim (-200,-100)$~MeV.
(iii) Important roles of the $\rho$-meson exchange are also found in the $Qs\bar{n}\bar{n}$ system.
For the $I=0$ configurations, $\left| (Qs)^{\bar{\mathbf{3}}}_0 [\bar{n}\bar{n}]^{\mathbf{3}}_0 \right\rangle_{0}$ and $\left| (Qs)^{\bar{\mathbf{3}}}_1 [\bar{n}\bar{n}]^{\mathbf{3}}_0 \right\rangle_{1}$,
the contribution can reach up to $\langle V^{\rho}\rangle\sim -170$~MeV.

The obtained $1S$-wave states are compact and lie far above the lowest dissociation meson-meson threshold.
The $J^P=2^+$ states have extremely narrow fall-apart widths of $\mathcal{O}(1)$~MeV,
while most of the $J^P=0^+,1^+$ states also have narrow widths of a few tens of MeV,
which indicates that they may be stable enough to be observed in experiments.

For the $Qn\bar{n}\bar{n}$ sector, there are two narrow scalar states with $I=1/2$,
$T_{(cn[\bar{u}\bar{d}])0^+}^{1/2}(2212)$ and $T_{(bn[\bar{u}\bar{d}])0^+}^{1/2}(5606)$,
whose masses highly overlap with the broad $D(1^3P_0)$ and $B(1^3P_0)$ states
predicted within the unquenched quark model~\cite{Ni:2023lvx}, respectively.
The $T_{(cn[\bar{u}\bar{d}])0^+}^{1/2}(2212)$ and $T_{(bn[\bar{u}\bar{d}])0^+}^{1/2}(5606)$
may have potentials to be observed in their dominant decay modes $D\pi$ and $B\pi$, respectively.

For the $cn\bar{s}\bar{n}$ system, the mass of the lowest state
is predicted to be $\sim 2527$~MeV, which is far larger
than the observed masses of $D_{s0}(2317)$, $D_{s1}(2460)$, and $T_{c\bar{s}}(2327)$.
The $T_{c\bar{s}0}^a(2900)$ resonance favors the lowest $I=1$ scalar state belonging to $\mathbf{6}_F$,
$T_{(cn\{\bar{s}\bar{n}\})0^+}^{1}(2922)$, which should be observed in $D^{0,+}K^+$ as well.
Furthermore, another $IJ^P=10^+$ state belonging to $\mathbf{\bar{3}}_F$,
$T_{(cn [\bar{s}\bar{n}])0^+}^{1}(2527)$, may have large potentials to be established
in the $D_s^+\pi^{\pm}$ and $D^{0,+}K^+$ final states in future experiments.

For the $bn\bar{s}\bar{n}$ system, the mass of the lowest state
is predicted to be $\sim 5850$~MeV, which is far larger
than that of the $T_{b\bar{s}}(5568)$ reported by the D0 collaboration.
The $T_{(bn\{\bar{s}\bar{n}\})0^+}^{1}(6263)$ with a width of $\sim 57$~MeV may be the
bottom partner of $T_{c\bar{s}0}(2900)$. This state is most likely to be observed
in the $B_s^0\pi^{\pm}$ and/or $B^{0,-}K^+$ final states.

For the $cs\bar{n}\bar{n}/\bar{c}\bar{s}nn$ system, the $IJ^P=00^+$ state $T_{(\bar{c}\bar{s}[ud])0^+}^0(2919)$ (as the antiparticle of $T_{(cs[\bar{u}\bar{d}])0^+}^0(2919)$) favors the assignment of the $T_{\bar{c}\bar{s}0}(2870)$ newly observed in the $D^-K^+$ final state at LHCb.
Its low-mass mixed partner, $T_{(\bar{c}\bar{s}[ud])0^+}^0(2510)$ with a width of $\Gamma\sim 86$~MeV,
should have discovery potentials in the $D^-K^+$ final state as well.

\begin{table*}[!htbp]
\caption{\label{tab:table_Qnnn_radii_merged}
The spectral properties of the $1S$-wave states for the $cn\bar{n}\bar{n}$ and $bn\bar{n}\bar{n}$ systems. For each physical states mixed between different configurations, we present the Hamiltonian matrix elements, mixing coefficients (eigenvectors), predicted mass, and root-mean-square radii between different quark pairs, $R_{ij} \equiv  \sqrt{\langle r_{ij}^2 \rangle}$ (fm).}
	\centering
	\setlength{\tabcolsep}{0.32cm}
	\renewcommand\arraystretch{1.40}
\begin{tabular}{clcccc}
		\hline\hline
		\multicolumn{6}{c}{$cn\bar{n}\bar{n}$ system} \\
		\hline
		$(I)J^{P}$ & Configuration  &$ \langle H \rangle$(MeV) & Eigenvector & Mass(MeV) & $R_{12}/R_{34}/R_{13}/R_{24}/R_{14}/R_{23}$  \\
		\hline
			$(\frac{3}{2})0^+$& $|(cn)^{6}_{0} \{\bar{n}\bar{n}\}^{\bar{6}}_{0}\rangle_{0}$ &\multirow{2}{*}{$\begin{pmatrix}3049&184\\184&2916\end{pmatrix}$}& \multirow{2}{*}{$\begin{pmatrix}0.575&-0.818\\-0.818&-0.575\end{pmatrix}$} & 2787 & 0.56/0.64/0.56/0.59/0.56/0.59  \\
			& $|(cn)^{\bar{3}}_{1} \{\bar{n}\bar{n}\}^{3}_{1}\rangle_{0}$ & & & 3179 & 0.62/0.70/0.58/0.62/0.58/0.62  \\
			$(\frac{3}{2})1^+$& $|(cn)^{6}_{1} \{\bar{n}\bar{n}\}^{\bar{6}}_{0}\rangle_{1}$ &\multirow{3}{*}{$\begin{pmatrix}3035&90&104\\90&2994&-97\\104&-97&3006\end{pmatrix}$}& \multirow{3}{*}{$\begin{pmatrix}-0.526&-0.351&0.775\\0.603&-0.797&0.049\\0.600&0.492&0.631\end{pmatrix}$} & 2818 & 0.56/0.64/0.57/0.62/0.57/0.62  \\
			& $|(cn)^{\bar{3}}_{1} \{\bar{n}\bar{n}\}^{3}_{1}\rangle_{1}$ & & & 3094 & 0.54/0.61/0.56/0.61/0.56/0.61  \\
			& $|(cn)^{\bar{3}}_{0} \{\bar{n}\bar{n}\}^{3}_{1}\rangle_{1}$ & & & 3123 & 0.60/0.70/0.58/0.63/0.58/0.63  \\
			$(\frac{3}{2})2^+$& $|(cn)^{\bar{3}}_{1} \{\bar{n}\bar{n}\}^{3}_{1}\rangle_{2}$ & 3126 & $\begin{pmatrix}1\end{pmatrix}$ & 3126 & 0.56/0.63/0.60/0.66/0.60/0.66  \\
\hline\hline
	\end{tabular}
\end{table*}

\begin{table*}[!htbp]
	\caption{\label{tab:table_Qnsn_radii_merged} The spectral properties of $1S$-wave states for the $cn\bar{s}\bar{n}$ and $bn\bar{s}\bar{n}$ systems. The caption is the same as that of Table~\ref{tab:table_Qnnn_radii_merged}.}
	\centering
	\setlength{\tabcolsep}{0.32cm}
	\renewcommand\arraystretch{1.30}
	\begin{tabular}{clcccc}
		\hline\hline
		\multicolumn{6}{c}{$cn\bar{s}\bar{n}$ system} \\
		\hline
		$(I)J^{P}$ & Configuration  &$ \langle H \rangle$(MeV) & Eigenvector & Mass(MeV) & $R_{12}/R_{34}/R_{13}/R_{24}/R_{14}/R_{23}$  \\
		\hline
			$(1)0^+$& $|(cn)^{6}_{0} \{\bar{s}\bar{n}\}^{\bar{6}}_{0}\rangle_{0}$ &\multirow{2}{*}{$\begin{pmatrix}3116&-162\\-162&3056\end{pmatrix}$}& \multirow{2}{*}{$\begin{pmatrix}-0.640&-0.768\\-0.768&0.640\end{pmatrix}$} & 2922 & 0.58/0.63/0.56/0.60/0.56/0.60  \\
			& $|(cn)^{\bar{3}}_{1} \{\bar{s}\bar{n}\}^{3}_{1}\rangle_{0}$ & & & 3251 & 0.60/0.66/0.56/0.61/0.56/0.61  \\
			$(1)1^+$& $|(cn)^{6}_{1} \{\bar{s}\bar{n}\}^{\bar{6}}_{0}\rangle_{1}$ &\multirow{3}{*}{$\begin{pmatrix}3101&77&88\\77&3114&-76\\88&-76&3110\end{pmatrix}$}& \multirow{3}{*}{$\begin{pmatrix}-0.609&-0.457&0.649\\0.539&-0.838&-0.084\\0.582&0.299&0.756\end{pmatrix}$} & 2947 & 0.57/0.64/0.56/0.61/0.56/0.61  \\
			& $|(cn)^{\bar{3}}_{1} \{\bar{s}\bar{n}\}^{3}_{1}\rangle_{1}$ & & & 3183 & 0.55/0.61/0.56/0.61/0.56/0.61  \\
			& $|(cn)^{\bar{3}}_{0} \{\bar{s}\bar{n}\}^{3}_{1}\rangle_{1}$ & & & 3196 & 0.57/0.65/0.57/0.62/0.57/0.62  \\
			$(1)2^+$& $|(cn)^{\bar{3}}_{1} \{\bar{s}\bar{n}\}^{3}_{1}\rangle_{2}$ & 3216 & $\begin{pmatrix}1\end{pmatrix}$ & 3216 & 0.56/0.62/0.59/0.65/0.59/0.65  \\
\hline\hline
	\end{tabular}
\end{table*}

\begin{table*}[!htbp]
	\caption{\label{tab:table_Qsnn_radii_merged}The spectral properties of $1S$-wave states for the $cs\bar{n}\bar{n}$ and $bs\bar{n}\bar{n}$ systems. The caption is the same as that of Table~\ref{tab:table_Qnnn_radii_merged}.}
	\centering
	\setlength{\tabcolsep}{0.35cm}
	\renewcommand\arraystretch{1.40}
	\begin{tabular}{clcccc}
		\hline\hline
		\multicolumn{6}{c}{$cs\bar{n}\bar{n}$ system} \\
		\hline
		$(I)J^{P}$ & Configuration  &$ \langle H \rangle$(MeV) & Eigenvector & Mass(MeV) & $R_{12}/R_{34}/R_{13}/R_{24}/R_{14}/R_{23}$  \\
		\hline
			$(1)0^+$& $|(cs)^{6}_{0} \{\bar{n}\bar{n}\}^{\bar{6}}_{0}\rangle_{0}$ &\multirow{2}{*}{$\begin{pmatrix}3167&133\\133&3078\end{pmatrix}$}& \multirow{2}{*}{$\begin{pmatrix}0.584&-0.812\\-0.812&-0.584\end{pmatrix}$} & 2982 & 0.56/0.65/0.57/0.60/0.57/0.60  \\
			& $|(cs)^{\bar{3}}_{1} \{\bar{n}\bar{n}\}^{3}_{1}\rangle_{0}$ & & & 3263 & 0.61/0.70/0.58/0.62/0.58/0.62  \\
			$(1)1^+$& $|(cs)^{6}_{1} \{\bar{n}\bar{n}\}^{\bar{6}}_{0}\rangle_{1}$ &\multirow{3}{*}{$\begin{pmatrix}3157&-49&-76\\-49&3123&-57\\-76&-57&3124\end{pmatrix}$}& \multirow{3}{*}{$\begin{pmatrix}-0.518&-0.331&0.789\\-0.566&0.824&-0.026\\-0.641&-0.459&-0.614\end{pmatrix}$} & 3012 & 0.55/0.65/0.58/0.61/0.58/0.61  \\
			& $|(cs)^{\bar{3}}_{1} \{\bar{n}\bar{n}\}^{3}_{1}\rangle_{1}$ & & & 3175 & 0.53/0.62/0.57/0.61/0.57/0.61  \\
			& $|(cs)^{\bar{3}}_{0} \{\bar{n}\bar{n}\}^{3}_{1}\rangle_{1}$ & & & 3217 & 0.59/0.70/0.59/0.63/0.59/0.63  \\
			$(1)2^+$& $|(cs)^{\bar{3}}_{1} \{\bar{n}\bar{n}\}^{3}_{1}\rangle_{2}$ & 3204 & $\begin{pmatrix}1\end{pmatrix}$ & 3204 & 0.54/0.62/0.59/0.64/0.59/0.64  \\
\hline\hline
	\end{tabular}
\end{table*}

\begin{table*}[!htbp]
	\caption{\label{tab:table_Qssn_radii_merged}The spectral properties of $1S$-wave states for the $cs\bar{s}\bar{n}$ and $bs\bar{s}\bar{n}$ systems. The caption is the same as that of Table~\ref{tab:table_Qnnn_radii_merged}.}
	\centering
	\setlength{\tabcolsep}{0.32cm}
	\renewcommand\arraystretch{1.40}
	\begin{tabular}{clcccc}
		\hline\hline
		\multicolumn{6}{c}{$cs\bar{s}\bar{n}$ system} \\
		\hline
		$(I)J^{P}$ & Configuration  &$ \langle H \rangle$(MeV) & Eigenvector & Mass(MeV) & $R_{12}/R_{34}/R_{13}/R_{24}/R_{14}/R_{23}$  \\
		\hline
			$(\frac{1}{2})0^+$& $|(cs)^{6}_{0} \{\bar{s}\bar{n}\}^{\bar{6}}_{0}\rangle_{0}$ &\multirow{2}{*}{$\begin{pmatrix}3232&122\\122&3184\end{pmatrix}$}& \multirow{2}{*}{$\begin{pmatrix}0.635&-0.773\\-0.773&-0.635\end{pmatrix}$} & 3083 & 0.56/0.63/0.56/0.59/0.56/0.59  \\
			& $|(cs)^{\bar{3}}_{1} \{\bar{s}\bar{n}\}^{3}_{1}\rangle_{0}$ & & & 3332 & 0.59/0.66/0.56/0.60/0.56/0.60  \\
			$(\frac{1}{2})1^+$& $|(cs)^{6}_{1} \{\bar{s}\bar{n}\}^{\bar{6}}_{0}\rangle_{1}$ &\multirow{3}{*}{$\begin{pmatrix}3221&-44&69\\-44&3221&28\\69&28&3215\end{pmatrix}$}& \multirow{3}{*}{$\begin{pmatrix}-0.637&-0.131&-0.760\\-0.464&0.852&0.242\\0.616&0.507&-0.604\end{pmatrix}$} & 3123 & 0.56/0.64/0.56/0.60/0.56/0.60  \\
			& $|(cs)^{\bar{3}}_{1} \{\bar{s}\bar{n}\}^{3}_{1}\rangle_{1}$ & & & 3245 & 0.51/0.59/0.56/0.59/0.56/0.59  \\
			& $|(cs)^{\bar{3}}_{0} \{\bar{s}\bar{n}\}^{3}_{1}\rangle_{1}$ & & & 3289 & 0.58/0.66/0.57/0.60/0.57/0.60  \\
			$(\frac{1}{2})2^+$& $|(cs)^{\bar{3}}_{1} \{\bar{s}\bar{n}\}^{3}_{1}\rangle_{2}$ & 3289 & $\begin{pmatrix}1\end{pmatrix}$ & 3289 & 0.54/0.61/0.58/0.63/0.58/0.63  \\
\hline\hline
	\end{tabular}
\end{table*}

\begin{table*}[!htbp]
	\caption{\label{tab:table_Qnss_radii_merged}The spectral properties of $1S$-wave states for the $cn\bar{s}\bar{s}$ and $bn\bar{s}\bar{s}$ systems. The caption is the same as that of Table~\ref{tab:table_Qnnn_radii_merged}.}
	\centering
	\setlength{\tabcolsep}{0.35cm}
	\renewcommand\arraystretch{1.30}
	\begin{tabular}{clcccc}
		\hline\hline
		\multicolumn{6}{c}{$cn\bar{s}\bar{s}$ system} \\
		\hline
		$(I)J^{P}$ & Configuration  &$ \langle H \rangle$(MeV) & Eigenvector & Mass(MeV) & $R_{12}/R_{34}/R_{13}/R_{24}/R_{14}/R_{23}$  \\
		\hline
			$(\frac{1}{2})0^+$& $|(cn)^{6}_{0} \{\bar{s}\bar{s}\}^{\bar{6}}_{0}\rangle_{0}$ &\multirow{2}{*}{$\begin{pmatrix}3193&133\\133&3201\end{pmatrix}$}& \multirow{2}{*}{$\begin{pmatrix}-0.717&0.697\\0.697&0.717\end{pmatrix}$} & 3064 & 0.59/0.63/0.55/0.60/0.55/0.60  \\
			& $|(cn)^{\bar{3}}_{1} \{\bar{s}\bar{s}\}^{3}_{1}\rangle_{0}$ & & & 3330 & 0.59/0.63/0.55/0.60/0.55/0.60  \\
			$(\frac{1}{2})1^+$& $|(cn)^{6}_{1} \{\bar{s}\bar{s}\}^{\bar{6}}_{0}\rangle_{1}$ &\multirow{3}{*}{$\begin{pmatrix}3178&-61&-75\\-61&3242&-54\\-75&-54&3224\end{pmatrix}$}& \multirow{3}{*}{$\begin{pmatrix}0.707&-0.707&-0.030\\0.455&0.486&-0.746\\0.542&0.514&0.665\end{pmatrix}$} & 3081 & 0.58/0.63/0.56/0.61/0.56/0.61  \\
			& $|(cn)^{\bar{3}}_{1} \{\bar{s}\bar{s}\}^{3}_{1}\rangle_{1}$ & & & 3275 & 0.58/0.63/0.56/0.61/0.56/0.61  \\
			& $|(cn)^{\bar{3}}_{0} \{\bar{s}\bar{s}\}^{3}_{1}\rangle_{1}$ & & & 3288 & 0.52/0.58/0.55/0.60/0.55/0.60  \\
			$(\frac{1}{2})2^+$& $|(cn)^{\bar{3}}_{1} \{\bar{s}\bar{s}\}^{3}_{1}\rangle_{2}$ & 3317 & $\begin{pmatrix}1\end{pmatrix}$ & 3317 & 0.55/0.60/0.58/0.64/0.58/0.64  \\
\hline\hline
	\end{tabular}
\end{table*}

\begin{table*}[!htbp]
	\caption{\label{tab:table_Qsss_radii_merged} The spectral properties of $1S$-wave states for the $cs\bar{s}\bar{s}$ and $bs\bar{s}\bar{s}$ systems. The caption is the same as that of Table~\ref{tab:table_Qnnn_radii_merged}.}
	\centering
	\setlength{\tabcolsep}{0.35cm}
	\renewcommand\arraystretch{1.30}
	\begin{tabular}{clcccc}
		\hline\hline
		\multicolumn{6}{c}{$cs\bar{s}\bar{s}$ system} \\
		\hline
		$(I)J^{P}$ & Configuration  &$ \langle H \rangle$(MeV) & Eigenvector & Mass(MeV) & $R_{12}/R_{34}/R_{13}/R_{24}/R_{14}/R_{23}$ \\
		\hline
			$(0)0^+$& $|(cs)^{6}_{0} \{\bar{s}\bar{s}\}^{\bar{6}}_{0}\rangle_{0}$ &\multirow{2}{*}{$\begin{pmatrix}3305&105\\105&3297\end{pmatrix}$}& \multirow{2}{*}{$\begin{pmatrix}0.694&-0.720\\-0.720&-0.694\end{pmatrix}$} & 3196 & 0.57/0.62/0.55/0.58/0.55/0.58  \\
			& $|(cs)^{\bar{3}}_{1} \{\bar{s}\bar{s}\}^{3}_{1}\rangle_{0}$ & & & 3406 & 0.57/0.62/0.55/0.58/0.55/0.58  \\
			$(0)1^+$& $|(cs)^{6}_{1} \{\bar{s}\bar{s}\}^{\bar{6}}_{0}\rangle_{1}$ &\multirow{3}{*}{$\begin{pmatrix}3294&38&-60\\38&3328&2\\-60&2&3314\end{pmatrix}$}& \multirow{3}{*}{$\begin{pmatrix}-0.767&0.099&-0.634\\0.303&-0.815&-0.494\\-0.565&-0.572&0.595\end{pmatrix}$} & 3236 & 0.57/0.63/0.55/0.59/0.55/0.59  \\
			& $|(cs)^{\bar{3}}_{1} \{\bar{s}\bar{s}\}^{3}_{1}\rangle_{1}$ & & & 3322 & 0.51/0.57/0.55/0.58/0.55/0.58  \\
			& $|(cs)^{\bar{3}}_{0} \{\bar{s}\bar{s}\}^{3}_{1}\rangle_{1}$ & & & 3378 & 0.55/0.62/0.55/0.58/0.55/0.58  \\
			$(0)2^+$& $|(cs)^{\bar{3}}_{1} \{\bar{s}\bar{s}\}^{3}_{1}\rangle_{2}$ & 3384 & $\begin{pmatrix}1\end{pmatrix}$ & 3384 & 0.53/0.59/0.57/0.61/0.57/0.61  \\
\hline\hline
	\end{tabular}
\end{table*}

\begin{table*}[!htbp]
	\begin{center}
		\caption{\label{tab:table_Qnnn_decay_merged}Partial and total fall-apart decay widths (MeV) of the $1S$-wave states for the $cn\bar{n}\bar{n}$ and $bn\bar{n}\bar{n}$ systems. The symbol $\cdots$ indicates that the corresponding decay channel is forbidden by kinematics or symmetry. }
		\tabcolsep=0.35cm
		\renewcommand\arraystretch{1.50}
		\begin{tabular}{cccccccccc}
			\hline\hline
			\multicolumn{10}{c}{$cn\bar{n}\bar{n}$ system} \\
			\hline
			\textrm{State}&	
			\textrm{$\Gamma _{T\to D \pi }$}&
			\textrm{$\Gamma _{T\to D \eta/D\eta' }$}&
			\textrm{$\Gamma _{T\to D^\ast \pi }$}&
			\textrm{$\Gamma _{T\to D^\ast \eta/D^\ast\eta' }$}&
			\textrm{$\Gamma _{T\to D  \rho }$}&
			\textrm{$\Gamma _{T\to D \omega }$} &
			\textrm{$\Gamma _{T\to D^\ast \rho }$}&
			\textrm{$\Gamma _{T\to D^\ast \omega  }$}&
			\textrm{$\Gamma _{Sum }$}\\
			\hline
			$T_{(cn\{\bar{n}\bar{n}\})0^+}^{3/2}(2787)$ & 40.51 & $\cdots$ & $\cdots$ & $\cdots$ & $\cdots$ & $\cdots$ & 8.55 & $\cdots$ & 49.06 \\
			$T_{(cn\{\bar{n}\bar{n}\})0^+}^{3/2}(3179)$ & 0.01 & $\cdots$ & $\cdots$ & $\cdots$ & $\cdots$ & $\cdots$ & 36.22 & $\cdots$ & 36.23 \\
			$T_{(cn\{\bar{n}\bar{n}\})1^+}^{3/2}(2818)$ & $\cdots$ & $\cdots$ & 9.38 & $\cdots$ & 21.99 & $\cdots$ & 7.86 & $\cdots$ & 39.23 \\
			$T_{(cn\{\bar{n}\bar{n}\})1^+}^{3/2}(3094)$ & $\cdots$ & $\cdots$ & 6.06 & $\cdots$ & 2.11 & $\cdots$ & 5.19 & $\cdots$ & 13.36 \\
			$T_{(cn\{\bar{n}\bar{n}\})1^+}^{3/2}(3123)$ & $\cdots$ & $\cdots$ & 3.75 & $\cdots$ & 10.68 & $\cdots$ & 0.00 & $\cdots$ & 14.43 \\
			$T_{(cn\{\bar{n}\bar{n}\})2^+}^{3/2}(3126)$ & $\cdots$ & $\cdots$ & $\cdots$ & $\cdots$ & $\cdots$ & $\cdots$ & 2.29 & $\cdots$ & 2.29 \\
\hline\hline
		\end{tabular}
	\end{center}
\end{table*}

\begin{table*}[!htbp]
	\begin{center}
		\caption{\label{tab:table_Qnsn_decay_merged}Partial and total fall-apart decay widths (MeV) of the $1S$-wave states for the $Qn\bar{s}\bar{n}$ systems.}
		\tabcolsep=0.31cm
		\renewcommand\arraystretch{1.0}
		\begin{tabular}{cccccccccc}
			\hline\hline
			\multicolumn{10}{c}{$cn\bar{s}\bar{n}$ system} \\
			\hline
			\textrm{State}&
			\textrm{$\Gamma _{T\to D_s \pi }$}&
			\textrm{$\Gamma _{T\to D_s^\ast  \pi }$}&
			\textrm{$\Gamma _{T\to D_s  \rho }$}&
			\textrm{$\Gamma _{T\to D_s^\ast  \rho }$}&
			\textrm{$\Gamma _{T\to DK }$} &
			\textrm{$\Gamma _{T\to D^\ast K }$}&
			\textrm{$\Gamma _{T\to D K^\ast }$}&
			\textrm{$\Gamma _{T\to D^\ast K^\ast }$}&
			\textrm{$\Gamma _{Sum }$}\\
			\hline
			$T_{(cn\{\bar{s}\bar{n}\})0^+}^{1}(2922)$ & 10.81 & $\cdots$ & $\cdots$ & 8.71 & 17.56 & $\cdots$ & $\cdots$ & 15.66 & 52.74 \\
			$T_{(cn\{\bar{s}\bar{n}\})0^+}^{1}(3251)$ & 0.00 & $\cdots$ & $\cdots$ & 10.89 & 0.19 & $\cdots$ & $\cdots$ & 15.10 & 26.18 \\
			$T_{(cn\{\bar{s}\bar{n}\})1^+}^{1}(2947)$ & $\cdots$ & 0.96 & 5.72 & 1.58 & $\cdots$ & 4.51 & 13.01 & 7.09 & 32.87 \\
			$T_{(cn\{\bar{s}\bar{n}\})1^+}^{1}(3183)$ & $\cdots$ & 2.03 & 1.33 & 0.78 & $\cdots$ & 2.66 & 2.07 & 3.03 & 11.90 \\
			$T_{(cn\{\bar{s}\bar{n}\})1^+}^{1}(3196)$ & $\cdots$ & 2.21 & 3.26 & 0.64 & $\cdots$ & 1.40 & 3.54 & 0.39 & 11.44 \\
			$T_{(cn\{\bar{s}\bar{n}\})2^+}^{1}(3216)$ & $\cdots$ & $\cdots$ & $\cdots$ & 1.05 & $\cdots$ & $\cdots$ & $\cdots$ & 0.59 & 1.64 \\
\hline\hline
		\end{tabular}
	\end{center}
\end{table*}

\begin{table*}[!htbp]
	\caption{\label{tab:table_Qsnn_decay_merged}Partial and total fall-apart decay widths (MeV) of the $1S$-wave states for the $cs\bar{n}\bar{n}$ and $bs\bar{n}\bar{n}$ systems.}
	\tabcolsep=0.18cm
	\renewcommand\arraystretch{1.50}
	\begin{tabular}{ccccccccccccc}
		\hline\hline
		\multicolumn{6}{c}{$cs\bar{n}\bar{n}$ system} & & \multicolumn{6}{c}{$bs\bar{n}\bar{n}$ system}\\
		\cline{1-6} \cline{8-13}
		\textrm{State}&	
		\textrm{$\Gamma _{T\to DK }$} &
		\textrm{$\Gamma _{T\to D^\ast K }$}&
		\textrm{$\Gamma _{T\to D K^\ast }$}&
		\textrm{$\Gamma _{T\to D^\ast K^\ast }$}&
		\textrm{$\Gamma _{Sum }$}&
		&
		\textrm{State}&	
		\textrm{$\Gamma _{T\to BK }$} &
		\textrm{$\Gamma _{T\to B^\ast K }$}&
		\textrm{$\Gamma _{T\to B K^\ast }$}&
		\textrm{$\Gamma _{T\to B^\ast K^\ast }$}&
		\textrm{$\Gamma _{Sum }$}\\
		\hline
		$T_{(cs\{\bar{n}\bar{n}\})0^+}^{1}(2982)$ & 22.30 & $\cdots$ & $\cdots$ & 22.36 & 44.66 & &
		$T_{(bs\{\bar{n}\bar{n}\})0^+}^{1}(6333)$ & 15.71 & $\cdots$ & $\cdots$ & 19.36 & 35.07 \\
		$T_{(cs\{\bar{n}\bar{n}\})0^+}^{1}(3263)$ & 0.40 & $\cdots$ & $\cdots$ & 25.62 & 26.02 & &
		$T_{(bs\{\bar{n}\bar{n}\})0^+}^{1}(6580)$ & 1.03 & $\cdots$ & $\cdots$ & 21.90 & 22.93 \\
		$T_{(cs\{\bar{n}\bar{n}\})1^+}^{1}(3012)$ & $\cdots$ & 2.76 & 9.44 & 2.74 & 14.94 & &
		$T_{(bs\{\bar{n}\bar{n}\})1^+}^{1}(6333)$ & $\cdots$ & 7.40 & 7.58 & 7.12 & 22.10 \\
		$T_{(cs\{\bar{n}\bar{n}\})1^+}^{1}(3175)$ & $\cdots$ & 3.26 & 1.61 & 2.82 & 7.69 & &
		$T_{(bs\{\bar{n}\bar{n}\})1^+}^{1}(6531)$ & $\cdots$ & 0.86 & 0.50 & 2.56 & 3.92 \\
		$T_{(cs\{\bar{n}\bar{n}\})1^+}^{1}(3217)$ & $\cdots$ & 4.52 & 8.09 & 0.06 & 12.67 & &
		$T_{(bs\{\bar{n}\bar{n}\})1^+}^{1}(6554)$ & $\cdots$ & 2.12 & 7.41 & 5.52 & 15.05 \\
		$T_{(cs\{\bar{n}\bar{n}\})2^+}^{1}(3204)$ & $\cdots$ & $\cdots$ & $\cdots$ & 3.16 & 3.16 & &
		$T_{(bs\{\bar{n}\bar{n}\})2^+}^{1}(6519)$ & $\cdots$ & $\cdots$ & $\cdots$ & 3.52 & 3.52 \\
\hline\hline
		\end{tabular}
	\end{table*}

\begin{table*}[!htbp]
	\caption{\label{tab:table_Qnss_decay_merged}Partial and total fall-apart decay widths (MeV) of the $1S$-wave states for the $cn\bar{s}\bar{s}$ and $bn\bar{s}\bar{s}$ systems.}
	\tabcolsep=0.17cm
	\renewcommand\arraystretch{1.50}
	\begin{tabular}{ccccccccccccc}
		\hline\hline
		\multicolumn{6}{c}{$cn\bar{s}\bar{s}$ system} & &\multicolumn{6}{c}{$bn\bar{s}\bar{s}$ system}\\
		\cline{1-6}\cline{8-13}
		\textrm{State}&	
		\textrm{$\Gamma _{T\to D_s K }$}&
		\textrm{$\Gamma _{T\to D_s^\ast  K }$}&
		\textrm{$\Gamma _{T\to D_s  K^\ast}$}&
		\textrm{$\Gamma _{T\to D_s^\ast  K^\ast }$}&
		\textrm{$\Gamma _{Sum }$}&
		&
		\textrm{State}&	
		\textrm{$\Gamma _{T\to B_s K }$}&
		\textrm{$\Gamma _{T\to B_s^\ast  K }$}&
		\textrm{$\Gamma _{T\to B_s  K^\ast}$}&
		\textrm{$\Gamma _{T\to B_s^\ast  K^\ast }$}&
		\textrm{$\Gamma _{Sum }$}\\
		\hline
		$T_{(cn\{\bar{s}\bar{s}\})0^+}^{1/2}(3064)$ & 20.06 & $\cdots$ & $\cdots$ & 35.58 & 55.64 & &
		$T_{(bn\{\bar{s}\bar{s}\})0^+}^{1/2}(6398)$ & 15.03 & $\cdots$ & $\cdots$ & 41.63 & 56.66 \\
		$T_{(cn\{\bar{s}\bar{s}\})0^+}^{1/2}(3330)$ & 1.85 & $\cdots$ & $\cdots$ & 17.94 & 19.79 & &
		$T_{(bn\{\bar{s}\bar{s}\})0^+}^{1/2}(6639)$ & 2.41 & $\cdots$ & $\cdots$ & 13.54 & 15.95 \\
		$T_{(cn\{\bar{s}\bar{s}\})1^+}^{1/2}(3081)$ & $\cdots$ & 2.54 & 16.44 & 12.69 & 31.67 & &
		$T_{(bn\{\bar{s}\bar{s}\})1^+}^{1/2}(6403)$ & $\cdots$ & 8.21 & 15.35 & 22.31 & 45.87 \\
		$T_{(cn\{\bar{s}\bar{s}\})1^+}^{1/2}(3275)$ & $\cdots$ & 0.68 & 7.67 & 0.06 & 8.41 & &
		$T_{(bn\{\bar{s}\bar{s}\})1^+}^{1/2}(6610)$ & $\cdots$ & 1.37 & 4.09 & 1.90 & 7.36 \\
		$T_{(cn\{\bar{s}\bar{s}\})1^+}^{1/2}(3288)$ & $\cdots$ & 7.78 & 0.61 & 2.39 & 10.78 & &
		$T_{(bn\{\bar{s}\bar{s}\})1^+}^{1/2}(6634)$ & $\cdots$ & 1.59 & 0.63 & 0.54 & 2.76 \\
		$T_{(cn\{\bar{s}\bar{s}\})2^+}^{1/2}(3317)$ & $\cdots$ & $\cdots$ & $\cdots$ & 0.02 & 0.02 & &
		$T_{(bn\{\bar{s}\bar{s}\})2^+}^{1/2}(6631)$ & $\cdots$ & $\cdots$ & $\cdots$ & 0.08 & 0.08 \\
		\hline\hline
	\end{tabular}
\end{table*}

\begin{table*}[!htbp]
	\begin{center}
		\caption{\label{tab:table_Qssn_decay_merged}Partial and total fall-apart decay widths (MeV) of the $1S$-wave states for the $cs\bar{s}\bar{n}$ and $bs\bar{s}\bar{n}$ systems.}
		\tabcolsep=0.31cm
		\renewcommand\arraystretch{1.50}
		\begin{tabular}{cccccccccc}
			\hline\hline
			\multicolumn{10}{c}{$cs\bar{s}\bar{n}$ system} \\
			\hline
			\textrm{State}&	
			\textrm{$\Gamma _{T\to D_s K  }$}&
			\textrm{$\Gamma _{T\to D_s^\ast K }$}&
			\textrm{$\Gamma _{T\to D_s  K^* }$}&
			\textrm{$\Gamma _{T\to D_s^\ast K^* }$}&
			\textrm{$\Gamma _{T\to  D \eta/D\eta' }$} &
			\textrm{$\Gamma _{T\to  D^\ast \eta/D^\ast\eta'}$}&
			\textrm{$\Gamma _{T\to D \phi }$}&
			\textrm{$\Gamma _{T\to D^\ast \phi }$}&
			\textrm{$\Gamma _{Sum }$}\\
			\hline
			$T_{(cs\{\bar{s}\bar{n}\})0^+}^{1/2}(3083)$ & 4.96 & $\cdots$ & $\cdots$ & 3.20 & 3.85/6.43 & 0.00/0.00 & $\cdots$ & 11.50 & 29.94 \\
			$T_{(cs\{\bar{s}\bar{n}\})0^+}^{1/2}(3332)$ & 0.44 & $\cdots$ & $\cdots$ & 8.11 & 0.00/0.00 & 0.00/0.00 & $\cdots$ & 9.63 & 18.18 \\
			$T_{(cs\{\bar{s}\bar{n}\})1^+}^{1/2}(3123)$ & $\cdots$ & 0.00 & 1.59 & 0.03 & 0.00/0.00 & 0.51/0.98 & 5.32 & 2.73 & 11.16 \\
			$T_{(cs\{\bar{s}\bar{n}\})1^+}^{1/2}(3245)$ & $\cdots$ & 1.56 & 0.05 & 1.21 & 0.00/0.00 & 1.02/1.56 & 0.38 & 2.05 & 7.83 \\
			$T_{(cs\{\bar{s}\bar{n}\})1^+}^{1/2}(3289)$ & $\cdots$ & 2.02 & 4.65 & 0.08 & 0.00/0.00 & 0.19/0.33 & 4.10 & 0.02 & 11.39 \\
			$T_{(cs\{\bar{s}\bar{n}\})2^+}^{1/2}(3289)$ & $\cdots$ & $\cdots$ & $\cdots$ & 4.22 & 0.00/0.00 & 0.00/0.00 & $\cdots$ & 0.91 & 5.13 \\
\hline\hline
		\end{tabular}
	\end{center}
\end{table*}

\begin{table*}[!htbp]
	\caption{\label{tab:table_Qsss_decay_merged} Partial and total fall-apart decay widths (MeV) of the $1S$-wave states for the $cs\bar{s}\bar{s}$ and $bs\bar{s}\bar{s}$ systems.}
	\tabcolsep=0.12cm
	\renewcommand\arraystretch{1.50}
	\begin{tabular}{ccccccccccccc}
		\hline\hline
		\multicolumn{6}{c}{$cs\bar{s}\bar{s}$ system} & &\multicolumn{6}{c}{$bs\bar{s}\bar{s}$ system}\\
		\cline{1-6}\cline{8-13}
		\textrm{State}&				
		\textrm{$\Gamma _{T\to D_s\eta/D_s\eta'  }$}&
		\textrm{$\Gamma _{T\to D_s^\ast\eta/D_s^\ast\eta'}$}&
		\textrm{$\Gamma _{T\to D_s \phi}$}&
		\textrm{$\Gamma _{T\to D_s^\ast  \phi }$}&
		\textrm{$\Gamma _{Sum }$}&
		&
		\textrm{State}&				
		\textrm{$\Gamma _{T\to B_s\eta/B_s\eta'  }$}&
		\textrm{$\Gamma _{T\to B_s^\ast\eta/B_s^\ast\eta'}$}&
		\textrm{$\Gamma _{T\to B_s \phi}$}&
		\textrm{$\Gamma _{T\to B_s^\ast  \phi }$}&
		\textrm{$\Gamma _{Sum }$}\\
		\hline
		$T_{(cs\{\bar{s}\bar{s}\})0^+}^{0}(3196)$ & 4.13/6.40 & 0.00/0.00 & $\cdots$ & 13.75 & 24.28 & &
		$T_{(bs\{\bar{s}\bar{s}\})0^+}^{0}(6533)$ & 2.57/4.37 & 0.00/0.00 & $\cdots$ & 13.46 & 20.40 \\
		$T_{(cs\{\bar{s}\bar{s}\})0^+}^{0}(3406)$ & 0.02/0.03 & 0.00/0.00 & $\cdots$ & 11.03 & 11.08 & &
		$T_{(bs\{\bar{s}\bar{s}\})0^+}^{0}(6713)$ & 0.00/0.00 & 0.00/0.00 & $\cdots$ & 8.35 & 8.35 \\
		$T_{(cs\{\bar{s}\bar{s}\})1^+}^{0}(3236)$ & 0.00/0.00 & 0.05/0.10 & 4.84 & 2.12 & 7.11 & &
		$T_{(bs\{\bar{s}\bar{s}\})1^+}^{0}(6558)$ & 0.00/0.00 & 0.98/1.77 & 4.21 & 5.18 & 12.14 \\
		$T_{(cs\{\bar{s}\bar{s}\})1^+}^{0}(3322)$ & 0.00/0.00 & 1.98/2.88 & 0.03 & 2.56 & 7.45 & &
		$T_{(bs\{\bar{s}\bar{s}\})1^+}^{0}(6657)$ & 0.00/0.00 & 0.30/0.47 & 0.30 & 1.50 & 2.57 \\
		$T_{(cs\{\bar{s}\bar{s}\})1^+}^{0}(3378)$ & 0.00/0.00 & 0.04/0.05 & 7.23 & 0.04 & 7.36 & &
		$T_{(bs\{\bar{s}\bar{s}\})1^+}^{0}(6708)$ & 0.00/0.00 & 0.00/0.00 & 4.37 & 1.56 & 5.93 \\
		$T_{(cs\{\bar{s}\bar{s}\})2^+}^{0}(3384)$ & 0.00/0.00 & 0.00/0.00 & $\cdots$ & 1.60 & 1.60 & &
		$T_{(bs\{\bar{s}\bar{s}\})2^+}^{0}(6694)$ & 0.00/0.00 & 0.00/0.00 & $\cdots$ & 1.68 & 1.68 \\
		\hline\hline
	\end{tabular}
\end{table*}

\section*{Acknowledgements}

This work is supported by the National Natural Science Foundation of China under Grants Nos. 12235018 and 12175065.

\begin{appendix}


\section*{Appendix}

The contributions of each part of the Hamiltonian for each configuration of the singly-heavy tetraquarks, $Qn\bar{n}\bar{n}$, $Qs\bar{n}\bar{n}$, $Qn\bar{s}\bar{n}$, $Qs\bar{s}\bar{n}$, $Qn\bar{s}\bar{s}$, and $Qs\bar{s}\bar{s}$, are presented in Tables~\ref{tab:merged_cnnn_bnnn}-\ref{tab:merged_csss_bsss} as an appendix.

\begin{table*}[!htbp]
	\caption{\label{tab:merged_cnnn_bnnn}
		The expectation values of each term of the Hamiltonian for each configuration of the $Qn\bar{n}\bar{n}$ ($Q=c,b$) systems.
		Here $\langle T\rangle$, $\langle V^{Conf}\rangle$, $\langle V^{Coul}\rangle$, and $\langle V^{SS}\rangle$ stand for the contributions from the relativistic kinetic energy term, the linear confinement potential, color Coulomb potential, the color-magnetic spin-spin potential, respectively. The $\langle V^{\pi}\rangle$, $\langle V^{K}\rangle$, $\langle V^{\eta}\rangle$, $\langle V^{\sigma}\rangle$, $\langle V^{\rho}\rangle$, and $\langle V^{\omega}\rangle$ stand for the contributions from the $\pi$, $K$, $\eta$, $\sigma$, $\rho$, and $\omega$ meson-exchange potentials, respectively.
The unit for all of the values is MeV.}
	\tabcolsep=0.15cm
	\renewcommand\arraystretch{1.30}
\begin{tabular}{lccccccccccccccccc}
		\hline\hline
		\multicolumn{16}{c}{$cn\bar{n}\bar{n}$ system} \\
		\hline
		\rule{0pt}{12pt}$(I)~J^{P}$ & Configuration  & Mass  & $\left \langle T \right \rangle $&$\left \langle V^{Conf} \right \rangle $ & $\left \langle V^{Coul} \right \rangle $&$\left \langle V^{SS} \right \rangle $& $\left \langle V^{\pi} \right \rangle $  & $\left \langle  V^{K}  \right \rangle $ & $\left \langle V^{\eta} \right \rangle $ & $\left \langle V^{\eta'} \right \rangle $ & $\left \langle V^{\sigma} \right \rangle $ & $\left \langle V^{\rho} \right \rangle $ & $\left \langle V^{\omega} \right \rangle $ & $\left \langle V^{K^*} \right \rangle $ & $\left \langle V^{\phi} \right \rangle $ \\
		\hline
			$(\frac{3}{2})~0^+$& $\left|(cn)_{0}^{6}\{\bar{n}\bar{n}\}_{0}^{\bar{6}}\right\rangle _{0}$ &3049& 3493& 92& -572& 44& 17& 0& 2& 0& -52& 39& -14& 0& 0 \\
			& $\left|(cn)_{1}^{\bar{3}}\{\bar{n}\bar{n}\}_{1}^{3}\right\rangle _{0}$ &2916& 3757& -33& -653& -62& -57& 0& 7& 2& -66& 65& -43& 0& 0 \\
		\hline
			$(\frac{3}{2})~1^+$& $\left|(cn)_{1}^{6}\{\bar{n}\bar{n}\}_{0}^{\bar{6}}\right\rangle _{1}$ &3035& 3511& 79& -579& 31& 17& 0& 2& 0& -53& 40& -14& 0& 0 \\
			& $\left|(cn)_{1}^{\bar{3}}\{\bar{n}\bar{n}\}_{1}^{3}\right\rangle _{1}$ &2994& 3666& 16& -622& 5& -31& 0& 2& 1& -63& 50& -28& 0& 0 \\
			& $\left|(cn)_{0}^{\bar{3}}\{\bar{n}\bar{n}\}_{1}^{3}\right\rangle _{1}$ &3006& 3654& 25& -619& -1& -10& 0& -2& -1& -61& 39& -18& 0& 0 \\
		\hline
			$(\frac{3}{2})~2^+$& $\left|(cn)_{1}^{\bar{3}}\{\bar{n}\bar{n}\}_{1}^{3}\right\rangle _{2}$ &3126& 3519& 110& -569& 102& 7& 0& -5& -1& -56& 29& -9& 0& 0 \\
\hline\hline
	\end{tabular}
\end{table*}

\begin{table*}[!htbp]
	\caption{\label{tab:merged_cnsn_bnsn}
		The expectation values of each term of the Hamiltonian for each configuration of the $Qn\bar{s}\bar{n}$ ($Q=c,b$) systems.}
	\tabcolsep=0.15cm
	\renewcommand\arraystretch{1.10}
	\begin{tabular}{lccccccccccccccccc}
		\hline\hline
		\multicolumn{16}{c}{$cn\bar{s}\bar{n}$ system} \\
		\hline
		\rule{0pt}{12pt}$(I)~J^{P}$ & Configuration  & Mass  & $\left \langle T \right \rangle $&$\left \langle V^{Conf} \right \rangle $ & $\left \langle V^{Coul} \right \rangle $&$\left \langle V^{SS} \right \rangle $& $\left \langle V^{\pi} \right \rangle $  & $\left \langle  V^{K}  \right \rangle $ & $\left \langle V^{\eta} \right \rangle $ & $\left \langle V^{\eta'} \right \rangle $ & $\left \langle V^{\sigma} \right \rangle $ & $\left \langle V^{\rho} \right \rangle $ & $\left \langle V^{\omega} \right \rangle $ & $\left \langle V^{K^*} \right \rangle $ & $\left \langle V^{\phi} \right \rangle $ \\
		\hline
		$(1)~0^+$& $\left|(cn)_{0}^{6}\{\bar{s}\bar{n}\}_{0}^{\bar{6}}\right\rangle _{0}$ &3116& 3637& 29& -559& 37& 0& 8& -3& 0& -55& 14& -13& 21& 0 \\
		& $\left|(cn)_{1}^{\bar{3}}\{\bar{s}\bar{n}\}_{1}^{3}\right\rangle _{0}$ &3056& 3820& -39& -604& -56& -18& -6& 1& 2& -65& 25& -24& 21& 0 \\
		$(1)~1^+$& $\left|(cn)_{1}^{6}\{\bar{s}\bar{n}\}_{0}^{\bar{6}}\right\rangle _{1}$ &3101& 3656& 16& -566& 23& 0& 9& -3& 0& -56& 14& -14& 21& 0 \\
		& $\left|(cn)_{1}^{\bar{3}}\{\bar{s}\bar{n}\}_{1}^{3}\right\rangle _{1}$ &3114& 3751& 0& -581& -2& -9& -5& 1& 1& -62& 19& -18& 20& 0 \\
		& $\left|(cn)_{0}^{\bar{3}}\{\bar{s}\bar{n}\}_{1}^{3}\right\rangle _{1}$ &3110& 3757& -1& -585& -16& 0& -5& 2& -1& -61& 14& -14& 20& 0 \\
		$(1)~2^+$& $\left|(cn)_{1}^{\bar{3}}\{\bar{s}\bar{n}\}_{1}^{3}\right\rangle _{2}$ &3216& 3638& 74& -542& 82& 7& -5& 2& -1& -57& 10& -10& 19& 0 \\
\hline\hline
	\end{tabular}
\end{table*}

\begin{table*}[!htbp]
	\caption{\label{tab:merged_csnn_bsnn}
		The expectation values of each term of the Hamiltonian for each configuration of the $Qs\bar{n}\bar{n}$ tetraquark systems.}
	\tabcolsep=0.15cm
	\renewcommand\arraystretch{1.40}
	\begin{tabular}{lccccccccccccccccc}
		\hline\hline
		\multicolumn{16}{c}{$cs\bar{n}\bar{n}$ system} \\
		\hline
		\rule{0pt}{12pt}$(I)~J^{P}$ & Configuration  & Mass  & $\left \langle T \right \rangle $&$\left \langle V^{Conf} \right \rangle $ & $\left \langle V^{Coul} \right \rangle $&$\left \langle V^{SS} \right \rangle $& $\left \langle V^{\pi} \right \rangle $  & $\left \langle  V^{K}  \right \rangle $ & $\left \langle V^{\eta} \right \rangle $ & $\left \langle V^{\eta'} \right \rangle $ & $\left \langle V^{\sigma} \right \rangle $ & $\left \langle V^{\rho} \right \rangle $ & $\left \langle V^{\omega} \right \rangle $ & $\left \langle V^{K^*} \right \rangle $ & $\left \langle V^{\phi} \right \rangle $ \\
		\hline
		$(1)~0^+$& $\left|(cs)_{0}^{6}\{\bar{n}\bar{n}\}_{0}^{\bar{6}}\right\rangle _{0}$ &3167& 3601& 73& -538& 40& 16& 0& 2& 0& -52& 12& 12& 0& 0 \\
		& $\left|(cs)_{1}^{\bar{3}}\{\bar{n}\bar{n}\}_{1}^{3}\right\rangle _{0}$ &3078& 3797& -31& -598& -29& -11& 0& -9& 2& -64& 10& 10& 0& 0 \\
		$(1)~1^+$& $\left|(cs)_{1}^{6}\{\bar{n}\bar{n}\}_{0}^{\bar{6}}\right\rangle _{1}$ &3157& 3613& 63& -542& 31& 17& 0& 2& 0& -53& 13& 12& 0& 0 \\
		& $\left|(cs)_{1}^{\bar{3}}\{\bar{n}\bar{n}\}_{1}^{3}\right\rangle _{1}$ &3123& 3741& 2& -579& 16& -11& 0& -5& 1& -62& 10& 10& 0& 0 \\
		& $\left|(cs)_{0}^{\bar{3}}\{\bar{n}\bar{n}\}_{1}^{3}\right\rangle _{1}$ &3124& 3740& 4& -580& 13& -10& 0& -2& -1& -61& 10& 10& 0& 0 \\
		$(1)~2^+$& $\left|(cs)_{1}^{\bar{3}}\{\bar{n}\bar{n}\}_{1}^{3}\right\rangle _{2}$ &3204& 3646& 64& -546& 89& -10& 0& 1& -1& -58& 10& 10& 0& 0 \\
\hline\hline
	\end{tabular}
\end{table*}

\begin{table*}[!htbp]
	\caption{\label{tab:merged_cssn_bssn}
		The expectation values of each term of the Hamiltonian for each configuration of the $Qs\bar{s}\bar{n}$ tetraquark systems.}
	\tabcolsep=0.15cm
	\renewcommand\arraystretch{1.40}
	\begin{tabular}{lccccccccccccccccc}
		\hline\hline
		\multicolumn{16}{c}{$cs\bar{s}\bar{n}$ system} \\
		\hline
		\rule{0pt}{12pt}$(I)~J^{P}$ & Configuration  & Mass  & $\left \langle T \right \rangle $&$\left \langle V^{Conf} \right \rangle $ & $\left \langle V^{Coul} \right \rangle $&$\left \langle V^{SS} \right \rangle $& $\left \langle V^{\pi} \right \rangle $  & $\left \langle  V^{K}  \right \rangle $ & $\left \langle V^{\eta} \right \rangle $ & $\left \langle V^{\eta'} \right \rangle $ & $\left \langle V^{\sigma} \right \rangle $ & $\left \langle V^{\rho} \right \rangle $ & $\left \langle V^{\omega} \right \rangle $ & $\left \langle V^{K^*} \right \rangle $ & $\left \langle V^{\phi} \right \rangle $ \\
		\hline
		$(\frac{1}{2})~0^+$& $\left|(cs)_{0}^{6}\{\bar{s}\bar{n}\}_{0}^{\bar{6}}\right\rangle _{0}$ &3232& 3758& 18& -531& 34& 0& 9& -3& 0& -56& 0& 0& 21& -19 \\
		& $\left|(cs)_{1}^{\bar{3}}\{\bar{s}\bar{n}\}_{1}^{3}\right\rangle _{0}$ &3184& 3904& -52& -565& -32& 0& -6& 3& 2& -65& 0& 0& 21& -27 \\
		$(\frac{1}{2})~1^+$& $\left|(cs)_{1}^{6}\{\bar{s}\bar{n}\}_{0}^{\bar{6}}\right\rangle _{1}$ &3221& 3772& 8& -536& 24& 0& 9& -3& 0& -57& 0& 0& 22& -19 \\
		& $\left|(cs)_{1}^{\bar{3}}\{\bar{s}\bar{n}\}_{1}^{3}\right\rangle _{1}$ &3221& 3855& -24& -549& 7& 0& -5& 2& 1& -63& 0& 0& 20& -23 \\
		& $\left|(cs)_{0}^{\bar{3}}\{\bar{s}\bar{n}\}_{1}^{3}\right\rangle _{1}$ &3215& 3861& -26& -552& -2& 0& -5& 2& -1& -62& 0& 0& 20& -20 \\
		$(\frac{1}{2})~2^+$& $\left|(cs)_{1}^{\bar{3}}\{\bar{s}\bar{n}\}_{1}^{3}\right\rangle _{2}$ &3289& 3772& 30& -521& 71& 0& -5& 1& -2& -59& 0& 0& 20& -17 \\
\hline\hline
	\end{tabular}
\end{table*}

\begin{table*}[!htbp]
	\caption{\label{tab:merged_cnss_bnss}
		The expectation values of each term of the Hamiltonian for each configuration of the $Qn\bar{s}\bar{s}$ tetraquark systems.}
	\tabcolsep=0.15cm
	\renewcommand\arraystretch{1.40}
	\begin{tabular}{lccccccccccccccccc}
		\hline\hline
		\multicolumn{16}{c}{$cn\bar{s}\bar{s}$ system} \\
		\hline
		\rule{0pt}{12pt}$(I)~J^{P}$ & Configuration  & Mass  & $\left \langle T \right \rangle $&$\left \langle V^{Conf} \right \rangle $ & $\left \langle V^{Coul} \right \rangle $&$\left \langle V^{SS} \right \rangle $& $\left \langle V^{\pi} \right \rangle $  & $\left \langle  V^{K}  \right \rangle $ & $\left \langle V^{\eta} \right \rangle $ & $\left \langle V^{\eta'} \right \rangle $ & $\left \langle V^{\sigma} \right \rangle $ & $\left \langle V^{\rho} \right \rangle $ & $\left \langle V^{\omega} \right \rangle $ & $\left \langle V^{K^*} \right \rangle $ & $\left \langle V^{\phi} \right \rangle $ \\
		\hline
		$(\frac{1}{2})~0^+$& $\left|(cn)_{0}^{6}\{\bar{s}\bar{s}\}_{0}^{\bar{6}}\right\rangle _{0}$ &3193& 3762& -26& -536& 30& 0& 0& 3& 1& -57& 0& 0& 0& 17 \\
		& $\left|(cn)_{1}^{\bar{3}}\{\bar{s}\bar{s}\}_{1}^{3}\right\rangle _{0}$ &3201& 3895& -36& -562& -45& 0& 0& -8& 2& -65& 0& 0& 0& 19 \\
		$(\frac{1}{2})~1^+$& $\left|(cn)_{1}^{6}\{\bar{s}\bar{s}\}_{0}^{\bar{6}}\right\rangle _{1}$ &3178& 3781& -39& -543& 15& 0& 0& 3& 1& -57& 0& 0& 0& 17 \\
		& $\left|(cn)_{1}^{\bar{3}}\{\bar{s}\bar{s}\}_{1}^{3}\right\rangle _{1}$ &3242& 3844& -6& -545& -2& 0& 0& -5& 0& -62& 0& 0& 0& 19 \\
		& $\left|(cn)_{0}^{\bar{3}}\{\bar{s}\bar{s}\}_{1}^{3}\right\rangle _{1}$ &3224& 3864& -15& -554& -25& 0& 0& -1& -1& -62& 0& 0& 0& 19 \\
		$(\frac{1}{2})~2^+$& $\left|(cn)_{1}^{\bar{3}}\{\bar{s}\bar{s}\}_{1}^{3}\right\rangle _{2}$ &3317& 3758& 51& -516& 66& 0& 0& 1& -2& -59& 0& 0& 0& 18 \\
\hline\hline
	\end{tabular}
\end{table*}

\begin{table*}[!htbp]
	\caption{\label{tab:merged_csss_bsss}
		The expectation values of each term of the Hamiltonian for each configuration of the $Qs\bar{s}\bar{s}$ tetraquark systems.}
	\tabcolsep=0.15cm
	\renewcommand\arraystretch{1.40}
	\begin{tabular}{lccccccccccccccccc}
		\hline\hline
		\multicolumn{16}{c}{$cs\bar{s}\bar{s}$ system} \\
		\hline
		\rule{0pt}{12pt}$(I)~J^{P}$ & Configuration  & Mass  & $\left \langle T \right \rangle $&$\left \langle V^{Conf} \right \rangle $ & $\left \langle V^{Coul} \right \rangle $&$\left \langle V^{SS} \right \rangle $& $\left \langle V^{\pi} \right \rangle $  & $\left \langle  V^{K}  \right \rangle $ & $\left \langle V^{\eta} \right \rangle $ & $\left \langle V^{\eta'} \right \rangle $ & $\left \langle V^{\sigma} \right \rangle $ & $\left \langle V^{\rho} \right \rangle $ & $\left \langle V^{\omega} \right \rangle $ & $\left \langle V^{K^*} \right \rangle $ & $\left \langle V^{\phi} \right \rangle $ \\
		\hline
		$(0)~0^+$& $\left|(cs)_{0}^{6}\{\bar{s}\bar{s}\}_{0}^{\bar{6}}\right\rangle _{0}$ &3305& 3898& -29& -514& 28& 0& 0& 3& 1& -59& 0& 0& 0& -23 \\
		& $\left|(cs)_{1}^{\bar{3}}\{\bar{s}\bar{s}\}_{1}^{3}\right\rangle _{0}$ &3297& 4014& -61& -534& -29& 0& 0& 5& 2& -66& 0& 0& 0& -33 \\
		$(0)~1^+$& $\left|(cs)_{1}^{6}\{\bar{s}\bar{s}\}_{0}^{\bar{6}}\right\rangle _{1}$ &3294& 3913& -39& -519& 17& 0& 0& 3& 1& -59& 0& 0& 0& -23 \\
		& $\left|(cs)_{1}^{\bar{3}}\{\bar{s}\bar{s}\}_{1}^{3}\right\rangle _{1}$ &3328& 3972& -37& -521& 3& 0& 0& 1& 1& -64& 0& 0& 0& -27 \\
		& $\left|(cs)_{0}^{\bar{3}}\{\bar{s}\bar{s}\}_{1}^{3}\right\rangle _{1}$ &3314& 3985& -43& -527& -12& 0& 0& -1& -1& -64& 0& 0& 0& -22 \\
		$(0)~2^+$& $\left|(cs)_{1}^{\bar{3}}\{\bar{s}\bar{s}\}_{1}^{3}\right\rangle _{2}$ &3384& 3900& 8& -499& 57& 0& 0& -4& -2& -61& 0& 0& 0& -17 \\
\hline\hline
	\end{tabular}
\end{table*}

	\clearpage

\end{appendix}

\bibliographystyle{unsrt}

\end{document}